\documentclass[aps,prd,onecolumn,nofootinbib,preprintnumbers,superscriptaddress]{revtex4}
\usepackage[mathscr]{eucal}
\usepackage{color}
\usepackage{bm}
\usepackage{hyperref}

\usepackage{dcolumn}
\usepackage{bm}
\usepackage{amsmath}
\usepackage{amssymb}
\usepackage{amsfonts}
\usepackage{amsthm}
\usepackage{amscd}
\usepackage{graphicx}
\graphicspath{{./picsExc/}}

\def\slashchar#1{\setbox0=\hbox{$#1$}           
   \dimen0=\wd0                                     
   \setbox1=\hbox{/} \dimen1=\wd1                   
   \ifdim\dimen0>\dimen1                            
      \rlap{\hbox to \dimen0{\hfil/\hfil}}          
      #1                                            
   \else                                            
      \rlap{\hbox to \dimen1{\hfil$#1$\hfil}}       
      /                                             
   \fi}

\begin{document}

\title{Exciton-driven quantum phase transitions in holography}

\author{E. Gubankova}

\affiliation{Institute for Theoretical Physics, J. W. Goethe-University,\\
D-60438 Frankfurt am Main, Germany\footnote{Also at ITEP, Moscow, Russia}}
\email{elena1@mit.edu}

\author{M. \v{C}ubrovi\'{c}}

\author{J. Zaanen}
\affiliation{Instituut Lorentz, Leiden University, Niels Bohrweg 2,\\
2300 RA Leiden, Netherlands}
\email{cubrovic, jan@lorentz.leidenuniv.nl}

\begin{abstract}
We study phase transitions driven by fermionic double-trace
deformations in gauge-gravity duality. Both the strength of the
double trace deformation and the infrared conformal
dimension/self-energy scaling of the quasiparticle can be used to
decrease the critical temperature to zero, leading to a line of
quantum critical points. The self-energy scaling is controlled
indirectly through an applied magnetic field and the quantum phase
transition naturally involves the condensation of a fermion
bilinear which models the spin density wave in an antiferromagnetic
state. The nature of the quantum critical points depends on the
parameters and we find either a Berezinskii-Kosterlitz-Thouless-type transition or one of two
distinct second order transitions with non-mean field exponents.
One of these is an anomalous branch where the order parameter of
constituent non-Fermi liquid quasiparticles is enhanced by the
magnetic field. Stabilization of ordered non-Fermi liquids by a
strong magnetic field is observed in experiments with highly
oriented pyrolytic graphite.

\vspace{2cm}

\end{abstract}

\maketitle

\section{Introduction}\label{section:0}

The anti-de Sitter/conformal Field Theory correspondence (AdS/CFT) or gauge/gravity duality is a new proving ground to describe strongly
correlated systems, and its application to unresolved questions in
condensed matter is an exciting new direction. It is especially compelling, as conventional methods, such as large-$N$ 
Ref.\cite{largeN} and $(4-\epsilon)$-type Ref.\cite{epsilon} expansions
fail to describe quantum critical behavior in $2+1$-dimensional
systems. 
The primary examples of such are the strange metal
states in the high $T_c$ cuprates and heavy fermion systems. Both
systems are characterized by anomalous behavior of transport and
thermodynamic quantities. 
In heavy fermions,
the Sommerfeld coefficient 
grows as the temperature is
lowered, meaning that the effective mass of the electrons on the Fermi surface diverges
or the Fermi energy of the electrons vanishes Ref.\cite{Schofield:2005}. In
the strange metal phase of the high $T_c$ superconductors as well as
in heavy fermions near a quantum phase transition, the resistivity
is linear with temperature $\rho\sim T$. These anomalous behaviors are  partly explained by the phenomenological
marginal Fermi liquid model Ref.\cite{Varma:1989zz}, and it
is an early success of AdS/CFT that the marginal Fermi liquid can be seen to emerge as the low-energy dynamics of a consistent theory.

A particularly simple gravity 
description for strongly interacting finite density matter is the
planar AdS-Reissner-Nordstr\"{o}m (AdS-RN) black hole (BH), which is
dual to a system at finite chemical potential. While the
AdS-RN black hole is a natural starting point to study the universal
aspects of finite charge density systems, the
universality of a black hole makes it difficult to explain
experiments that are keen on the nature of the charge carriers,
such as transport properties (e.~g. conductivity). In particular the dominance of Pauli blocking for observed physics, requires that at the minimum one
needs to add free Dirac fermions to the AdS-RN background. A
self-consistent treatment shows that this system is unstable to a quasi-Lifshitz geometry in the bulk Refs.\cite{Review:2010,Hartnoll:es,czs2010}, that encodes for a deconfined Fermi liquid system Refs.\cite{Hartnoll:2010xj,Sachdev:2010um,Huijse:2011hp,Sachdev:2011ze}. Here we shall initiate the study of instabilities in the unstable metallic AdS-RN phase that are driven by Fermi bilinears.

The essential low-energy property of the metallic system dual to
the AdS-RN black hole background is the emergence of Fermi surfaces Refs.\cite{Leiden:2009,MIT} where the notion of a quasiparticle needs not
be well defined, i.e. stable Ref.\cite{Faulkner:2009}. In Ref.\cite{Leiden:2010}, we used
the magnetic field as an external probe to change the characteristics of the Fermi surface excitations and as a consequence the transport properties of the system. It strongly suggested that a quantum phase transition should occur when the underlying quasiparticle becomes (un)stable as a function of the magnetic field. The study in this article of the influence on stability of Fermi bilinears allows us to show that there is a phase transition between the two regimes and that for a specific set of parameters the critical temperature vanishes. Our work is therefore also a fermionic companion to Ref.\cite{Faulkner:2011}.

Continuing the connection of AdS models to actual observations, the results we find resemble other experimental findings in quantum-critical systems.
At low temperatures and in high magnetic fields, the
resistance of single-layer graphene at the Dirac point undergoes a
thousandfold increase within a narrow interval of field strengths Ref.\cite{Novoselov-Geim:2007}.
The abruptness of the increase suggests that a transition to a
field-induced insulating, ordered state occurs at the critical field
$h_c$ Ref.\cite{Checkelsky:2009}. In bilayer graphene, measurements
taken at the filling factor $\nu=0$ point show that, similar to
single layer graphene, the bilayer becomes insulating at strong
magnetic field Ref.\cite{Cadden-Zimansky:2009}. In these systems, the
divergent resistivity in strong magnetic fields was analyzed in
terms of Kosterlitz-Thouless localization Ref.\cite{Checkelsky:2009} and
the gap opening in the zeroth Landau level Ref.\cite{Novoselov:2009}.
However, it remains a theoretical challenge to explain a highly
unusual approach to the insulating state. Despite the steep
divergence of resistivity, the profile of $\rho$ vs. $T$ at fixed
$h$ saturates to a $T$-independent value at low temperatures, which
is consistent with gapless charge-carrying excitations
Ref.\cite{Checkelsky:2009}. Moreover, in highly oriented pyrolytic
graphite in the magnetic field, the temperature of the
metal-insulator phase transition $T_c(h)$ increases with increasing
field strength, contrary to the $T_c(h)$ dependence in the classical
low field limit Ref.\cite{Kopelevich:1999}. The anomalous $T_c(h)$
behavior has been successfully modeled within a dynamical gap picture
Ref.\cite{Shovkovy:2d}.
The available data suggest that by tuning the magnetic field
graphene approaches a quantum critical point, beyond which a new
insulating phase develops with anomalous behavior $T_c(h)$. This
picture is in agreement with expectations of quantum critical
behavior, where e.~g. in heavy fermion metal a new magnetically
ordered state (antiferromagnet) emerges when tuned through the
quantum critical point Ref.\cite{Schofield:2005}.

We shall see that the same qualitative physics emerges with our use of the
the magnetic field as a knob to tune to the IR fixed point  to gain
some insight into the quantum critical behavior driven by fermion bilinears. In our gauge/gravity dual
prescription, the unusual properties characteristic for quantum
criticality can be understood as being controlled by the scaling
dimension of the fermion operator in the emergent IR fixed point.
The novel insight of AdS/CFT is that the low-energy behavior of a strongly coupled quantum critical system is governed by a
nontrivial unstable fixed point which exhibits nonanalytic scaling
behavior in the temporal direction only (the retarded Green's
function of the IR CFT is $G_{IR}^{R}\sim\omega^{2\nu}$)
Ref.\cite{Faulkner:2009}. This
fixed point manifests itself as a near-horizon region of the black
hole with AdS${}_2$ geometry which is (presumably) dual to a
one-dimensional IR CFT. Building on the semilocal description of the quasiparticle characteristics by simple Dyson summation in a Fermi gas coupled to this 1+1-dimensional IR CFT Ref.\cite{FaulknerPol} 
an appealing picture arises that quantum critical fermionic fluctuations in the IR
CFT generate relevant order parameter perturbations of the Fermi liquid theory.
Whether this is truly what is driving the physics is an open question.
Regardless, quantum critical matter is
universal in the sense that no information about the microscopic
nature of the material enters. Qualitatively our study should apply to any bilinear instability in the strange metal phase of unconventional
superconductors, heavy fermions as well as for a critical point in
graphene. Universality makes applications of AdS/CFT to quantum
critical phenomena justifiable and appealing.

The paper is organized as follows. In Sec. II, we review the AdS-RN black hole solution in AdS-Einstein-Maxwell gravity coupled to charged fermions and the dual interpretation as a quantum critical fermion system at finite density.  
In Sec. III we use the bilinear formalism put forward in Ref.\cite{czs2010} to explore an instability of a quantum system towards a quantum phase transition
using the AdS dual description. We study a quantum phase transition to an insulating phase as a function of the magnetic field.
For completeness we test the various phases by a spectral analysis in Sec. IV.  We
conclude by discussing a phase space in $(h,T)$ variables for a
quantum critical matter at nonzero temperatures.

\section{Holographic fermions in the background of a dyonic black hole}

The gravity dual to a $2+1$-dimensional CFT at finite density in the
presence of a magnetic field 
starts with
the Einstein-Maxwell action
describing an asymptotically AdS${}_4$ geometry
\begin{equation}
S_g=\frac{1}{2\kappa^2}\int d^4x \sqrt{-g}\left( {\mathcal R}
+6 -\frac{1}{g_F^2}F_{MN}F^{MN}\right).
\label{action-g}
\end{equation}
Here $A_M$ is the gauge field, $g_F^2$ is an effective
dimensionless gauge coupling and the curvature radius of
AdS${}_4$ is set to unity. The equations of motion following from eq.(\ref{action-g}) 
are solved by a dyonic AdS black hole, having both
electric and magnetic charge
\begin{equation}
ds^2=\frac{1}{(1-z)^2}\left(-fdt^2+dx^2+dy^2+\frac{dz^2}{f}\right)
\label{geometry}
\end{equation}
where the redshift factor $f$ and the vector field $A_M$ are given
by
\begin{eqnarray}
f&=&z\left(3-3z+z^2-(Q^2+H^2)(1-z)^3\right),\nonumber\\
A_t &=& \mu z,\;\;A_y=hx,\;\mathrm{with}~~ \mu=g_FQ,\;\;  h=g_F H
. \label{gauge}
\end{eqnarray}
The AdS boundary is reached for $z\rightarrow 1$, the black hole horizon is at $z\rightarrow 0$
and the electric and magnetic charge of the black hole $Q$ and $H$, encoding the chemical potential $\mu$ and magnetic field $h$ of the dual CFT, are scaled such that the black hole temperature equals\footnote{The independent black hole mass parameter is restored after rescaling $t\rightarrow M t~, x\rightarrow M x, y \rightarrow My$ and $h\rightarrow M^{-2}h$.}
\begin{equation}
T=\frac{1}{4\pi}\left(3-(Q^2+H^2)\right).
\end{equation}
In these units, the extremal $T=0$ black hole corresponds to
$Q^2+H^2=3$ and in this case the red shift factor develops a double zero at the
horizon
\begin{eqnarray}
f=3z^2(z-z_*)(z-\bar{z}_*), \;\; z_*=(4+i\sqrt{2})/3.
\label{double-zero}
\end{eqnarray}

To include the bulk fermions, we consider a spinor field $\psi$ in
the AdS${}_4$ of charge $q$ and mass $m$, which is dual to a fermionic
operator ${\mathcal O}$ in the boundary CFT${}_3$ of charge $q$ and
dimension
\begin{equation}
\Delta_\Psi = \frac{3}{2} + m, \label{conformal-dimension}
\end{equation}
with $m\geq -\frac{1}{2}$ (in units of the AdS radius). The quadratic
action for $\psi$ reads
\begin{equation}
S_{\psi} = \int d^4x \sqrt{-g}\left(\bar{\psi}\Gamma^M{\mathcal
D}_{M}\psi-m\bar{\psi}\psi\right), \label{action-psi}
\end{equation}
where $\bar{\psi}=\psi^\dagger i\Gamma^{\hat{t}}$, and
\begin{equation}
{\mathcal D}_M=\partial_M
+\frac{1}{4}\omega_{Mab}\Gamma^{ab}-iqA_M,
\label{spin-connection}
\end{equation}
where $\omega_{Mab}$ is the spin connection, and
$\Gamma^{ab}=\frac{1}{2}[\Gamma^a,\Gamma^b]$. Here, $M$ and $a,b$
denote the bulk space-time and tangent-space indices respectively,
while $\mu,\nu$ are indices along the boundary directions, i.~e.
$M=(z,\mu)$. 
The Dirac equation in the dyonic AdS-black hole background becomes
\begin{eqnarray}
&&
\left(\Gamma^{\hat{z}}\sqrt{f}\partial_z+\Gamma^{\hat{z}}\frac{\sqrt{f}}{2(1-z)}(3+\frac{(1-z)f^{\prime}}{2f})
-\Gamma^{\hat{t}}\frac{i(\omega+q\mu z)}{\sqrt{f}}-\frac{1}{(1-z)}m
+\Gamma^{\hat{x}}\partial_x+\Gamma^{\hat{y}}i(k_y-qhx)\phantom{\sqrt{f}}\hspace{-0.5cm}\right)\psi=0
\end{eqnarray}
where $\psi$ is the Fourier transform in the $y$ directions
and time. The $z$ and $x$ dependences can
be separated as in Refs.\cite{Albash:2009wz,Albash:2010yr,Leiden:2010}. Define
\begin{eqnarray}
P &=&
\Gamma^{\hat{z}}\sqrt{f}\left(\partial_z+\frac{1}{2(1-z)}(3+\frac{(1-z)f^{\prime}}{2f})\right)
-\Gamma^{\hat{t}}\frac{i(\omega+q\mu z)}{\sqrt{f}}-\frac{1}{(1-z)}m \nonumber\\
Q &=&
\Gamma^{\hat{x}}\partial_x+\Gamma^{\hat{y}}(ik_y-iqhx),
\end{eqnarray}
in terms of which the Dirac equation is $(P+Q)\psi=0$. In order to separate the variables, we can proceed by finding the matrix
$U$ such that $UP\psi=-UQ\psi=\lambda\psi$. The idea is that, although $P$ and $Q$ do not commute, we can find $U$ so that $[UP,UQ]$ commute and can be diagonalized simultaneously Ref.\cite{Leiden:2010}.\footnote{Rather the part in $P$ not proportional to the identity {\em anticommutes} with $Q$. This realization shows why the relations in the next sentence are the solution.} To this end, $U$ must satisfy the relations
$\{U,\Gamma^z\}=0$, $\{U,\Gamma^t\}=0$, $[U,\Gamma^x]=0$,
$[U,\Gamma^y]=0$. 
A clear solution is $U=[\Gamma^z,\Gamma^t]$.

In a convenient gamma matrix basis (Minkowski signature)
Ref.\cite{Faulkner:2009}
\begin{eqnarray}
&& \Gamma^{\hat{z}}= \left(\begin{array}{cc}
-\sigma^3 & 0 \\
0 & -\sigma^3
\end{array}
\right),\;\; \Gamma^{\hat{t}}= \left(\begin{array}{cc}
i\sigma^1 & 0 \\
0 & i\sigma^1
\end{array}
\right),\;\; \Gamma^{\hat{x}}= \left(\begin{array}{cc}
-\sigma^2 & 0 \\
0 & \sigma^2
\end{array}
\right),\;\;
\nonumber\\
&& \Gamma^{\hat{y}}= \left(\begin{array}{cc}
0 & \sigma^2 \\
\sigma^2 & 0
\end{array}
\right),\;\; \Gamma^{\hat{5}}= \left(\begin{array}{cc}
0 & i\sigma^2 \\
-i\sigma^2 & 0
\end{array}
\right) \equiv i\Gamma^{\hat{t}}\Gamma^{\hat{x}}\Gamma^{\hat{y}}\Gamma^{\hat{z}}. \label{matrices}
\end{eqnarray}
the matrix $U$ equals
\begin{eqnarray}
U=\left(\begin{array}{cc}
-i\sigma^2 & 0 \\
0 & -i\sigma^2
\end{array}
\right). \label{transform}
\end{eqnarray}
This choice of the basis allows one to obtain $k_y=0$ spectral functions in a simple way. In the absence of a magnetic field one can use rotational invariance to rotate to a frame where this is so.
The gauge choice for the a magnetic field obviously breaks the
isotropy, but the physical isotropy still ensures that the spectral functions simplify in this basis Ref.\cite{Leiden:2010}.
The $x$-dependent part of the Dirac equation can be solved
analytically in terms of Gaussian-damped Hermite polynomials $H_n(\sqrt{qh}(x+\frac{k_y}{qh}))$ with eigenvalues $\lambda_n=\sqrt{|qh|n}$ quantized in terms of the Landau index $n=0,1,\dots$ Refs.\cite{Albash:2009wz,Albash:2010yr,Leiden:2010}. The
 Dirac equation $(P-U^{-1}\lambda)\psi=0$, where $\lambda$ is a diagonal matrix in terms of $\lambda_n$ and whose square is proportional to the identity, then reduces to
\begin{eqnarray}
\left((\partial_z+\frac{1}{2(1-z)}(3+\frac{(1-z)f^{\prime}}{2f}))\Gamma^{\hat{z}}
-\frac{i(\omega+q\mu z)}{f}\Gamma^{\hat{t}}-\frac{m}{\sqrt{f}(1-z)}
-U^{-1}\frac{\lambda_n}{\sqrt{f}}\right)\psi=0.
\label{general-Dirac}
\end{eqnarray}
We introduce now the projectors $\Pi_\alpha$ that split the
four-component bispinors into two two-component spinors
$\Psi=(\psi_1,\psi_2)^T$ where the index
$\alpha=1,2$ is the Dirac index of the boundary theory
\begin{equation}
\Pi_\alpha=\frac{1}{2}(1-(-1)^\alpha\Gamma^{\hat{z}}\Gamma^{\hat{t}}\frac{1}{|\lambda|}Q),\;\;
\alpha=1,2,\;\; \Pi_1+\Pi_2=1. \label{projection}
\end{equation}
The projectors commute with both $P$ and $Q$ (recall that $Q^2=\lambda^21\!\!1$). At zero magnetic field
projectors are given by $\Pi_\alpha=\frac{1}{2}(1-(-1)^\alpha\Gamma^{\hat{z}}\Gamma^{\hat{t}}\hat{k}_i\Gamma^i)$ with unit vector $\hat{k}_i=\vec{k}/|\vec{k}|$. The projections
$\psi_\alpha=\Pi_\alpha\psi$ with $\alpha=1,2$ therefore decouple from
each other and one finds two independent copies of the two-component Dirac equation
\begin{equation}
\left(\partial_z+\frac{1}{2}(\frac{3}{1-z}+\frac{f^{\prime}}{2f})-\frac{i(\omega+\mu_q z)}{f}\sigma^2
+\frac{m}{\sqrt{f}(1-z)}\sigma^3+\frac{\lambda_n}{\sqrt{f}}\sigma^1 \right)\psi_{1;2}=0,
\label{Dirac}
\end{equation}
where the magnetic momentum $\lambda_n=\sqrt{2|qh|n}$ is Landau quantized with integer values $n=0,1,\dots$ and
$\mu_q\equiv \mu q$. 
It is identical to the AdS-Dirac equation for an AdS-RN black hole with zero magnetic charge when the discrete eigenvalue $\lambda$ is identified with the (size of the) momentum $k$.

As we have shown in Ref.\cite{Leiden:2010},
solving eq.(\ref{Dirac}) is equivalent to solving
the Dirac equation at zero magnetic field but with a rescaled
chemical potential and fermion charge. At $T=0$ the mapping is
given by Ref.\cite{Leiden:2010}
\begin{equation}
(\mu_q,h,q)\mapsto (\mu_{q,eff},\,h_{eff},\,q_{eff})=(\sqrt{3}q\sqrt{1-\frac{h^2}{3}},\,0,\,q\sqrt{1-\frac{h^2}{3}}),
\label{mapping}
\end{equation}
which we will use further.

\section{Bilinear approach to particle-hole pairing}

The objective of this paper is to use the magnetic field as a tool to
probe our unstable quantum critical system dual to the dyonic AdS-RN geometry. We show that the instability is manifest in the appearance of ordering in the system: the
magnetic field acts as a catalyzer for the particle-hole pairing. In
particular, we will find an unusual behavior for the critical temperature
of the normal to paired phase transition as the dialing of the magnetic field drives
the system to a quantum crtitical point: for a critical magnetic field the critical temperature vanishes indicating a new emergent quantum critical point.

We will
identify the bulk quantities in the bilinear approach which are dual to the sought-for
quantities on the CFT side. 
We have given the setup of the bilinear formalism in
Ref.\cite{czs2010}. Here, we will first give a concise review with the
focus on the transport properties and the influence of magnetic
fields, and then derive the bilinear equations relevant for
computing the pairing gap.

\subsection{Bulk propagators and currents}

A controlled method for calculating the expectation value of
some composite operator $J$ with the structure of a fermion
bilinear ($J\sim\psi^\dagger\psi$) has been put forward in Ref.\cite{czs2010} and it is based on a relation between the bulk and the boundary propagator in the isotropic single-particle approximation. This allows us to identify the familiar quantities at the
boundary by matching the resulting expression to known
thermodynamic relations. The crucial object was identified in
Ref.\cite{czs2010}
\begin{equation}
\label{jeq} J^\mu(E,p,z)=\int d\omega\int d^2k
\bar{\psi}(\omega,k,z)\Gamma^\mu\psi(E-\omega,p-k,z)
\end{equation}
and it is the spatial average of the $U(1)$ current four-vector in the bulk\footnote{As shown in Ref.\cite{czs2010}, even though the current is defined as spatial average, the only mode that contributes at the leading order (tree level) is the quasinormal mode at $k=k_F$.}. The metric then assumes the form given in
the first section by eq.(\ref{geometry}) (so that the horizon is
located at $z_H=0$ and the boundary is at $z_0\to 1$).
Having defined the radial projection of the bulk Dirac equation
in eq.(\ref{projection}) we can also define the
radial projections of the current as
\begin{eqnarray}
J^\mu_\alpha(E,p,z)=\int d\omega\int d^2k
\bar{\psi}_\alpha(\omega,k,z)\gamma^\mu\psi_\alpha(E-\omega,p-k,z),
\label{psipm}
\end{eqnarray}
where $\alpha=1,2$ and $\gamma^\mu$ is a Pauli matrix acting in the
boundary frame.

The boundary interpretation of this current is, however,
subtler than the simple $U(1)$ conserved current which it is in the bulk
Ref.\cite{czs2010}: it expresses the Migdal theorem, i.e. the density of
quasiparticles in the vicinity of the Fermi surface. To see this,
express the bulk spinors $\psi_\alpha(z)$ at an arbitrary value of
$z$ through the bulk-to-boundary propagators
$\mathcal{G}_\alpha(z,z^\prime)$ and the boundary spinors
$\psi_\alpha(z_0)$ as
\begin{eqnarray}
\label{bulk-psi}
\psi_\alpha(z)=\mathcal{G}_\alpha(z_H,z)\mathcal{G}_\alpha^{-1}(z_H,z_0)\psi_\alpha(z_0).
\end{eqnarray}
The meaning of the above expressions is clear: the spinors evolve
from their horizon values toward the values in the bulk at some $z$,
under the action of the bulk-to-boundary propagator $\mathcal{G}_\alpha(z,z^\prime)$ acting upon them (normalized by its value at the boundary). To find the
relation with the boundary Green's function we need to know the
asymptotics of the solutions of the Dirac equation (\ref{Dirac}) at
the boundary, see eq.(\ref{bndpsir}) in Appendix A
\begin{eqnarray}
\psi_1 &\sim&
a_1(1-z)^{3/2-m}\psi_{0,+}^{in}+b_1(1-z)^{3/2+m}\psi_{0,-}^{in} \nonumber\\
\psi_2 &\sim&
a_2(1-z)^{5/2-m}\psi_{0,+}^{in}+b_2(1-z)^{5/2+m}
\psi_{0,-}^{in}. \label{asymptotics}
\end{eqnarray}
On the other hand, the boundary retarded propagator is given by
the dictionary entry Ref.\cite{Vegh:2009}, eq.(\ref{gfinish}),
where $\gamma^0=i\sigma^1$.

The bulk-to-boundary Green's function (in dimensionless units) can
be constructed from the solutions to the Dirac equation Ref.\cite{Hartman:2010}
as in eq.(\ref{bulk-boundary}). 
Using eq.(\ref{bndpsir}) and the expression for the
Wronskian, we arrive at the following relation between the
boundary asymptotics of the solutions $\psi^{in}$ and $\psi^{bdy}$
\begin{eqnarray}
\label{bndasymprel}
\psi^{in}_\alpha(z_0)=\left(\frac{(1-z)^{-2m}}{G_\alpha}(-i\gamma^0)+1\right)\psi^{bdy}_\alpha(z_0).
\end{eqnarray}
Taking into account the dictionary entry for the boundary
propagator from eq.(\ref{gfinish}) and the representation
eq.(\ref{bulk-psi}) for $\psi^{in}$ and $\psi^{bdy}$, the retarded
propagator at the boundary is
\begin{equation}
\label{gret} G_\alpha=\lim_{z_0\to
1}(1-z_0)^{-2m}\psi^{bdy}_\alpha(z_0)(\psi^{in}_\alpha(z_0))^{-1}=
\lim _{z_0\to
1}\mathcal{G}_\alpha(z_H,z_0)\gamma^0\mathcal{G}_\alpha(z_H,z_0)
\end{equation}
with $z_H=0$. Using eq.(\ref{gret}) and the definition for the
current in eq.(\ref{psipm}) it can now be shown that the current
$J^\mu_1\sim\int\bar{G}_1\gamma^\mu G_1$ \emph{for an on-shell solution}
becomes at the boundary Ref.\cite{czs2010}
\begin{equation}
\label{gretj} J^\mu_1(\omega=0,k=k_F,z_0\to 1)=\frac{1+2m}{\mu}\int
d\omega\gamma^\mu G_1(\omega,k_F).
\end{equation}
It is well known Ref.\cite{Landau9} that the integral of the propagator
is related to the charge density. In particular, for
$\gamma^\mu=\gamma^0$ and for the horizon boundary conditions chosen
so that $G=G_F$ (Feynman propagator), we obtain
\begin{eqnarray}
\label{noj} J^0_1\equiv\int d\omega
\psi_1^\dagger\psi_1=\frac{1+2m}{\mu}n_F,
\end{eqnarray}
i.e. the bilinear $J^0$ directly expresses the charge density
$n_F={\rm tr}\left(i\gamma^0G\right)|_{on-shell}\sim
|b_1(k_F)|^2$. Notice that to achieve this we need to set
$\omega=k-k_F=0$, i.e. look at the location of the Fermi surface.
By analogy, we can now see that the components $J^{1,2}$
correspond to current densities. In particular, the ratio of the
spatial components $J^i_1/E^j$ in the external electric field
$\mathbf{E}$ readily gives the expression for the conductivity
tensor $\sigma_{ij}$. Finally, the formalism outlined above allows
us to define an arbitrary bilinear $J^A=\int\bar{\psi}\hat{A}\psi$
and to compute its expectation value. By choosing the matrix
$\hat{A}$ appropriately we are able to model particle-hole,
particle-particle or any other current. Notice however that all bilinears $J^A$ are proportional on shell, as can be seen from eqs.(\ref{gret}-\ref{gretj}), which hold also for any other matrix $\hat{A}$ sandwiched between the two bulk propagators. The proportionality is at fixed parameters ($\mu$, $T$, etc) so the dependences of the form $J^A(\mu)$ and $J^A(T)$ will be different for different choices of $\hat{A}$.

To introduce another crucial current, we will study the form of the
action. (We will define our action to model the quantum phase transition and to define the pairing excitonic gap in section \ref{pairing}.) 
We pick a gauge, eq.(\ref{gauge}), so that the Maxwell field
is $A_\mu=\left(\Phi(z),0, h(z)x,0,0\right)$, meaning that the
non-zero components of $F^{\mu\nu}$ are $F^{z0}=\partial_z\Phi$,
$F^{z2}=x\partial_z h$, $F^{12}=h$ and their antisymmetric pairs.
The total action eqs.(\ref{action-g},\ref{action-psi}) is now
\begin{eqnarray}
S &=&\int dzd^3x \sqrt{-g}\left( \frac{1}{2\kappa^2}
\left({\mathcal R} +6-\frac{1}{4g_F^2}F_{MN}F^{MN}\right)
+\bar{\psi}\Gamma^M{\mathcal D}_M\psi-m\bar{\psi}\psi\right)\nonumber\\
&+& \int d^3x\sqrt{-h}\left({\mathcal R}_{bnd}A_\mu n_\nu F^{\mu\nu}
+\sum_\alpha\bar{\psi}_\alpha(-i\sigma^3)\psi_\alpha\right),
\label{action}
\end{eqnarray}
where $\bar{\psi}_\alpha=i\psi_\alpha^\dagger\sigma^1$. The second
integral is the boundary term added to regularize the bulk action,
for which the fermion part vanishes on shell.
Knowing the metric
eq.(\ref{geometry}) and the form of $A^\mu$, we find that the total
action (free energy, from the dictionary) can be expressed as
Ref.\cite{czs2010}
\begin{equation}
\label{free-act}
\mathcal{F}=\mathcal{F}_\mathrm{hor}-\frac{1}{2}(\mu\rho+h\mathcal{M})+\frac{3}{2}K
\end{equation}
where $\mathcal{F}_{\mathrm{hor}}$ is the free energy at the
horizon, which does not depend on the physical quantities on the
boundary as long as the metric is fixed Ref.\cite{czs2010} so we can
disregard it here. In eq.(\ref{free-act}), $\mu,\rho$ and
$h,\mathcal{M}$ are the leading and subleading terms in the
electric and magnetic field
\begin{eqnarray}
&& \Phi(z\to z_0)=\mu,\;\; \partial_z\Phi(z\to z_0)=\rho,\nonumber\\
&& h(z\to z_0)=h,\;\; \partial_z h(z\to z_0)=\mathcal{M},
\end{eqnarray}
and the fermionic contribution is
proportional to
\begin{equation}
\label{i0def} K=\int d\omega\int d^2k \sum_\alpha
\bar{\psi}_\alpha(\omega,k,z)\psi_\alpha(E-\omega,p-k,z)
\end{equation}
which brings us to the second crucial bilinear. Along the lines of
the derivation eqs.(\ref{psipm}-\ref{gretj}), we see that the fermionic
contribution to the boundary action eq.(\ref{action}) is proportional
to
\begin{equation}
\label{i0eq} K=2\sum_\alpha{\rm Re} G_\alpha,
\end{equation}
i.~e. it is the real part of the boundary propagator\footnote{In Ref.\cite{czs2010} this bilinear is denoted by $I$. In the present paper a different bilinear is called $I_\pm$.}. The bulk fermionic
term does not contribute, being proportional to the equation
of motion, while the boundary terms include the holographic
factors of the form $(1-z_0)^n$. In accordance with our earlier conclusion that the on-shell bilinears are all proportional, we can reexpress the free energy in eq.(\ref{free-act}) as
\begin{equation}
\label{free-acti}
\mathcal{F}=\mathcal{F}_\mathrm{hor}-\frac{1}{2}(\mu\rho+h\mathcal{M})+\frac{3}{4m+2}\mu J^0_1
\end{equation}
where the chemical potential reappears in the prefactor and the fermionic term becomes of the form $\mu J^0_1$, confirming again that $J^0_1$ can be associated with the number density.

\subsection{Pairing currents}\label{pairing}

Now we will put to work our bilinear approach in order to explicitly
compute the particle-hole (excitonic) pairing operator.
We add a scalar field which interacts with fermions by the Yukawa coupling as done in Ref.\cite{Faulkner_photoemission:2009}.
Both scalar and fermion fields are dynamical.
The matter action is given by
\begin{eqnarray}
S_{\psi} &=& i\int dzd^3x \sqrt{-g}\left(\bar{\psi}\Gamma^MD_M^{\psi}\psi - m_{\psi}\bar{\psi}\psi-\lambda|\phi|^2\bar{\psi}\psi\right)
\nonumber\\
S_{G} &=& \int dzd^3x \sqrt{-g}\frac{1}{2}G_{int}\left(\phi\bar{\psi}\Gamma\psi+\phi^{\star}\bar{\psi}\bar{\Gamma}\psi\right)
\nonumber\\
S_{\phi} &=& -\int dzd^3x \sqrt{-g}\left(|D_M^{\phi}\phi|^2+V(|\phi|)\right)
\label{Yukawa_action}
\end{eqnarray}
where the covariant derivatives are $D_M^{\psi}=\nabla_M+\frac{1}{4}\omega_{Mab}\Gamma^{ab}-iq_{\psi}A_M$, 
$D_M^{\phi}=\nabla_M-iq_{\phi}A_M$, and $\bar{\psi}=\psi^{\dagger}i\Gamma^t$. The gamma-matrix structure of the Yukawa interaction 
is specified further. 
Matter action is supplemented by the gauge-gravity action
\begin{equation}
S_{A}=\frac{1}{2\kappa^2}\int dzd^3x\sqrt{-g}\left(R+\frac{6}{L^2}-\frac{1}{4g_F^2}F_{MN}F^{MN}\right).
\end{equation}
we take the AdS radius $L=1$ and $g_F=1$. 
The gauge field components $A_0$ and $A_2$ are responsible for the chemical potential and magnetic field, respectively,
in the boundary theory.  
As in Ref.\cite{Faulkner_photoemission:2009}, we assume $\lambda=0$ and $V(|\phi|)=m_{\phi}^2|\phi|^2$ and the scalar is real 
$\phi^{\star}=\phi$. For the particle-hole sector, the scalar field is neutral $q_{\phi}=0$.

The Yukawa coupling 
$G_{int}$ is allowed to be positive and negative. When the coupling is positive
$G_{int}>0$, a repulsive interaction makes it harder to form the particle-hole condesate. Therefore it lowers
the critical temperature and can be used as a knob to tune to a vanishing critical temperature $T_c=0$ at a critical value $G_{int}^c$
which defines a quantum critical point.
When the coupling is negative $G_{int}<0$,
an attractive interaction facilitates pairing
and helps to form the condensate. 

Both situations can be described when the interaction term is viewed as a dynamical mass of either sign due to the fact that it is
in $\bar\psi\psi$ channel.
For $G_{int}>0$, interaction $G_{int}\phi$ introduces a new massive pole: massless free fermion field aquires a mass which
makes it harder to condense. For $G_{int}<0$, there is a tachyonic instability.   
The exponentially growing tachyonic mode is resolved by a condensate formation, a new stable ground state. 
It can be shown that we do not need  a nonzero chemical potential to form a condensate in this case. 
A similar situation was considered in Ref.\cite{Faulkner:2011} for the superconducting instability where  
the spontaneous symmetry breaking of $U(1)$ was achieved by the boundary double-trace deformation. In our case for the electron-hole pairing, 
$Z_2$ symmetry is spontaneously broken by a neutral order parameter.
Next we discuss the choice for the gamma-matrix structure $\Gamma$ of the Yukawa interaction eq.(\ref{Yukawa_action}) and 
the corresponding pairing parameter $\Delta$ 
\begin{equation}
\label{deltanew}
\Delta=G_{int}\langle\bar{\psi}\Gamma\psi\rangle.
\end{equation}
Now we explain our choice of the pairing operator and give a  rigorous
justification for this choice.

In principle, any operator that creates a particle and a hole with
the same quantum numbers could be taken to define $\Delta$. This
translates into the requirements
\begin{equation}
\label{deltacand}\left[\Gamma,\Gamma^i\right]=0,\lbrace\Gamma,\Gamma^0\rbrace=0,\left[\Gamma,\hat{C}\right]=0.
\end{equation}
(Anti) commutation with (time) space gamma matrices is required for
the preservation of homogeneity and isotropy, and the last one is there
to preserve the particle-hole symmetry. In the basis we have adopted, eq.(\ref{matrices}),
$\left(\Gamma^{t}\right)^\star=-\Gamma^t$ and $\Gamma^{z\star}=\Gamma^z$, and therefore
the charge conjugation is represented as
\begin{equation}
\label{ctrans}\hat{C}:\; \psi\rightarrow \Gamma^0\Gamma^3\psi^{\star}.
\end{equation}
We will also consider the parity of the order parameter. As defined in Ref.\cite{Stefano-Bolognesi}, parity in the presence of the AdS boundary acts as $x^1\rightarrow -x^1$ with $x^2,z$ unchanged, while the transformation of the spinor is given by
\begin{equation}
\label{ptrans}\hat{P}:\; \psi\rightarrow \Gamma^1\Gamma^5\psi.
\end{equation}
We can now expand $\Gamma$ in the usual basis
\begin{equation}
\label{basis}\mathbb{B}=\lbrace I,\Gamma^\mu,\Gamma^5,\Gamma^5\Gamma^\mu,\left[\Gamma^\mu,\Gamma^\nu\right]\rbrace
\end{equation}
where the indices in the commutators $[\Gamma^\mu,\Gamma^\nu]$ run along the six different combinations, and check directly that the conditions eq.(\ref{deltacand}) can only be satisfied by the matrices $I$, $\Gamma^5\Gamma^i$ and $[\Gamma^0,\Gamma^z]$. This gives three candidate bilinears
\begin{itemize}
  \item For $\Gamma=I$ we get the bulk current $\bar{\psi}\psi=-(\psi_1^{\dagger}\sigma^1\psi_1+\psi_2^{\dagger}\sigma^1\psi_2)$,
   i.e. the mass operator in the bulk. As noted in this section and in more detail in Ref.\cite{czs2010}, it can be identified as proportional to the bulk mass term. As such, it describes the free energy per particle,
as can be seen from the expression for the free energy eq.(\ref{free-act}). The equation of motion for $K=\langle\bar{\psi}\psi\rangle$ eq.(\ref{i0def}) exclusively depends on the $U(1)$ current and thus cannot encapsulate the density of the neutral particle-hole pairs: indeed, we directly see that the right-hand side equals zero if the total charge current vanishes.
  \item For $\Gamma=i\Gamma^y\Gamma^5$, the bulk current is $\bar{\psi}i\Gamma^y\Gamma^5\psi=-(\psi_1^{\dagger}\sigma^1\psi_1-\psi_2^{\dagger}\sigma^1\psi_2)$. The crucial difference with respect to the first case is the relative minus sign. It is due to this sign that the current \emph{couples to itself}, i.e. it is a response to a nonzero parameter $G_{int}$, as we will see soon.
  \item For $\Gamma=\Gamma^z$, the resulting bulk current is $\bar{\psi}\Gamma^3\psi=-i(\psi_1^{\dagger}\sigma^2\psi_1+\psi_2^{\dagger}\sigma^2\psi_2)$. It sources the radial gauge field $A_z$ which is believed to be equal to zero in all meaningful holographic setups, as the radial direction corresponds to the renormalization group (RG) scale. Thus, this operator is again not the response to the attractive pairing interaction.
\end{itemize}
We are therefore left with one possibility only: $\Gamma=i\Gamma^2\Gamma^5$ which is also consistent with the choice of our gauge at nonzero magnetic field. We will therefore work with the channel
\begin{equation}
\label{deltanewmat}\Gamma\equiv i\Gamma^y\Gamma^5=\left(\begin{array}{cc}
1&0\\
0&-1
\end{array}\right).
\end{equation}
As we have discussed earlier, the isotropy in the $x-y$ plane remains unbroken by the radial magnetic field, and hence the expectation value should in fact be ascribed to the current $I^\mu=i\bar{\psi}\Gamma^\mu\Gamma^5\psi$ with $\mu=1,2$. We show the equivalence of the $i\Gamma^x\Gamma^5$ and $i\Gamma^y\Gamma^5$ order parameters below.  
The choice of the $y$ channel is motivated by technical simplicity due to the form of the projection operator and the fermion basis we use, eq.(\ref{projection}):
$\Pi_{\alpha}=\frac{1}{2}\left(1-(-1)^{\alpha}\Gamma^3\Gamma^0\Gamma^1\right)$ with $\alpha=1,2$, since
$\Gamma^3\Gamma^0\Gamma^1=-i\Gamma^2\Gamma^5$ with $\Gamma^5=i\Gamma^0\Gamma^1\Gamma^2\Gamma^3$. Finally, we note that the structure of the currents defined in eqs.(\ref{jeq},\ref{psipm}) depends on the basis choice and that the currents as such have no physical interpretation in the boundary theory: physical meaning can only be ascribed to the expectation values Ref.\cite{czs2010}. It is exactly the expectation values that encode for the condensation (order) on the field theory side Refs.\cite{czs2010}, \cite{Stefano-Bolognesi}.

The AdS/CFT correspondence does not provide a straightforward
way to match a double-trace condensate to a boundary operator, though only single-trace fields are easy to identify with the operators at the boundary.
Indeed, in holographic superconductors a superconducting condensate is modeled by
a charged scalar field $\langle\Phi\rangle$ (see e.g. Ref.\cite{Horowitz:2009}). 
As in Ref.\cite{Stefano-Bolognesi}, we argue by matching discrete symmetries
on the gravity and field theory sides, that the expectation of the bulk current $\langle\bar{\psi}i\Gamma^2\Gamma^5\psi\rangle$
is gravity dual of the pairing particle-hole gap. 
Let us consider properties of the corresponding condensates with respect to discrete symmetries, parity and charge conjugation, in the AdS 
four-dimensional space.
According to eq.(\ref{ptrans}), $\langle\bar{\psi}\psi\rangle$ and $\langle\bar{\psi}\Gamma^3\psi\rangle$ are scalars and parity even, while
$\langle\bar{\psi}i\Gamma^2\Gamma^5\psi\rangle$ is a pseudoscalar and parity odd. As for the charge conjugation, we easily find that $\langle\bar{\psi}\psi\rangle$ and $\langle\bar{\psi}i\Gamma^2\Gamma^5\psi\rangle$ commute with $\hat{C}$,
while $\langle\bar{\psi}\Gamma^3\psi\rangle$ anticommutes. Since the latter is the component of a vector current while the former two are (pseudo)scalars, we find that all operators preserve the particle number, as promised. The magnetic field $H$ is odd under both parity and charge conjugation, and therefore it is unaffected by $\hat{C}\hat{P}$. The condensate
$\langle\bar{\psi}\psi\rangle$ is also unaffected by $\hat{C}\hat{P}$, however $\langle\bar{\psi}i\Gamma^2\Gamma^5\psi\rangle$ and
$\langle\bar{\psi}\Gamma^3\psi\rangle$ spontaneously break
the $\hat{C}\hat{P}$ symmetry.

In the three-dimensional boundary theory, gamma matrices can be deduced from the four-dimensional bulk gamma matrices; and the four component Dirac spinor $\psi$ is dual to a two-component
spinor operator $\Psi$. As has been also found in Ref.\cite{Stefano-Bolognesi},
the three-dimensional condensate $\bar{\Psi}\Psi$ is odd under parity and even under charge conjugation, and therefore it is odd under $\hat{C}\hat{P}$.
We summarize the transformation properties of the four- and three-dimensional condensates together with the magnetic field
\begin{equation}
\begin{array}{|c|c|c|c|c|c|}
\hline & \langle\bar{\psi}\psi\rangle_{4d} & \langle\bar{\psi}\Gamma^3\psi\rangle_{4d} & \langle\bar{\psi}i\Gamma^2\Gamma^5\psi\rangle_{4d} &
 \langle\bar{\Psi}\Psi\rangle_{3d}& H \\
\hline \hat{P} & + & + & -& -& - \\
 \hat{C} & + & - & +& + & -\\
 \hat{C}\hat{P} & + & - & -& -& +\\
 \hline
 \end{array}
 \label{discrete_symmetry}
\end{equation}
which shows that the symmetry properties are matched between $\langle\bar{\psi}i\Gamma^2\Gamma^5\psi\rangle_{4d}$ and
$\langle\bar{\Psi}\Psi\rangle_{3d}$ condensates: they spontaneously break the CP symmetry while the magnetic field leaves it intact.
Therefore our AdS/CFT dictionary between the bulk and boundary quantities is
$\psi \leftrightarrow \Psi$ and   
$\langle\bar{\psi}i\Gamma^2\Gamma^5\psi\rangle \leftrightarrow \langle\bar{\Psi}\Psi\rangle$, with the corresponding conformal dimensions
of boundary operators given by eq.(\ref{IRnu-psi}) and eq.(\ref{IRnu-I}).

The natural bulk extension is now the current
\begin{equation}
\label{jdelta}
I=(-i)\int d\omega\int d^2k\bar{\psi}(\omega,k,z)\Gamma\psi(E-\omega,p-k,z)
\end{equation}
and it is understood that in nonzero magnetic field the
integration over $k$ degenerates into the sum over Landau levels
(this holds for all currents in this section). 
We will soon show that a complete set of bulk equations of motion for
the operator eq.(\ref{jdelta}) requires a set of currents that we
label $J_\pm$, $I_\pm$ and $K_\pm$. In the representation
eq.(\ref{matrices}) we introduce the following bilinears of the
fermion field
\begin{eqnarray}
\nonumber J_\pm(E,p,z) &=& (-i)\int d\omega \int d^2 k
\left(\bar{\psi}_1(\omega,k,z)\sigma^1\psi_1(E-\omega,p-k,z)\pm
\bar{\psi}_2(\omega,k,z)\sigma^1\psi_2(E-\omega,p-k,z)\right)\nonumber\\
& \equiv & J_1(E,p,z)\pm J_2(E,p,z),\nonumber\\
I_\pm(E,p,z) &=& (-i)\int d\omega \int d^2 k
\left(\bar{\psi}_1(\omega,k,z)\psi_1(E-\omega,p-k,z)\pm
\bar{\psi}_2(\omega,k,z)\psi_2(E-\omega,p-k,z)\right)\nonumber\\
& \equiv & I_1(E,p,z)\pm I_2(E,p,z), \nonumber\\
K_\pm(E,p,z) &=& -\int d\omega \int d^2 k
\left(\bar{\psi}_1(\omega,k,z)\sigma^2\psi_1(E-\omega,p-k,z)\pm
\bar{\psi}_2(\omega,k,z)\sigma^2\psi_2(E-\omega,p-k,z)\right)\nonumber\\
\label{jdelta2} & \equiv & K_1(E,p,z)\pm K_2(E,p,z),
\end{eqnarray}
where the pairing parameter $\langle\bar{\psi}\Gamma\psi\rangle$ in eq.(\ref{jdelta}) is
$I\equiv I_{-}$, the index $0$ for the zeroth
component is omitted in $J_\pm$, and $\bar{\psi}_\alpha=i\psi^\dagger_\alpha\sigma^1$.

Let us now study the dynamics of the system. We need to know the
evolution equations for the currents and the scalar field and to complement them with
the Maxwell equations. We will show that the equations of motion
for all currents generically have nonzero solutions. This suggests
that, due to the coupling with the UV CFT, the pairing can occur
spontaneously, without explicitly adding new terms to the action
(there is no need to add an interaction for fermions
in the bulk). Nevertheless, we will also analyze the
situation with nonzero $G_{int}$ and show what new phenomena it
brings as compared to UV CFT-only coupling (i.e. no bulk
coupling).

Let us start from the equations of motion. 
The Dirac and Klein-Gordon equations are to be
complemented with the Maxwell equation
\begin{eqnarray}
\nabla^MF_{MN}=iq_{\phi}\left(\hspace{-0.3cm}\phantom{1\over1}\phi^{\star}(\nabla_N-iq_{\phi}A_N)\phi - \phi(\nabla_N+iq_{\phi}A_n)\phi^{\star}
\phantom{1\over 1}\hspace{-0.3cm}\right)
-iq_{\psi}\bar{\psi}\Gamma_N\psi
\end{eqnarray}
which is reduced when the scalar is real, $\phi^{\star}=\phi$, to
\begin{eqnarray}
\nabla^MF_{MN}=2q_{\phi}^2\phi^2A_N
-iq_{\psi}\bar{\psi}\Gamma_N\psi
\end{eqnarray}
In the background of a dyonic black hole with the metric
\begin{equation}
ds^2 = \frac{1}{(1-z)^2}\left(-fdt^2 +\frac{dz^2}{f} +dx^2 +dy^2\right)
\end{equation}
the Maxwell equation for the component $A_0$ is
\begin{eqnarray}
\partial_z^2A_0-\frac{2q_{\phi}^2\phi^2}{(1-z)^2f}A_0-\frac{iq_{\psi}J_{+}}{(1-z)^2f}=0
\label{Maxwell}
\end{eqnarray}
where we have used $\bar{\psi}\Gamma_0\psi\rightarrow -J_{+}$.

In our setup we ignore the
backreaction to $A_2=Hx$, treating it as a fixed external field.
The justification comes from the physics on the field theory side:
we consider a stationary nonmagnetic system with zero current and
magnetization density. In the bulk, this means that the currents
sourced by --- and backreacting to --- the magnetic field arise as
corrections of higher order that can be neglected to a good
approximation.\footnote{To see this, consider the corresponding
Maxwell equation
\begin{eqnarray}
\partial_z^2A_2+\frac{\partial_zf}{f}\partial_zA_2=\frac{2q_{\phi}^2\phi^2}{\sqrt{f}(1-z)^3}A_2
+\frac{iq_{\psi}}{\sqrt{f}(1-z)^3}K_+,
\label{Maxwell2}
\end{eqnarray}
and insert the ansatz $A_2=Hx+\delta(z,x)$. The resulting relation for the neutral scalar $q_{\phi}=0$
$\partial_z(f\partial_z\delta)=-q/\left(\sqrt{f}\left(1-z\right)^3\right)$
predicts $K\sim \bar{\psi}_{\alpha}\sigma^3\psi_{\alpha}\sim \delta$,
compared to the analogous estimate for
the electrostatic backreaction $J\sim \bar{\psi}_{\alpha}\sigma^1\psi_{\alpha}\sim \mu$.
Thus the spatial
current is of order of the small correction to the field,
$\delta$. The reason obviously lies in the fact that the magnetic
monopole sources a $z$-independent field.} Inclusion of the second
Maxwell equation for $A_2$ would likely only lead to a
renormalization of the magnetic field $H\mapsto H+\delta H$
without quantitative changes of the physics.

The equations of motion for the matter fields read
\begin{eqnarray}
&& e_A^M\Gamma^M\left(\tilde{D}_M^{\psi}-iq_{\psi}A_M\right)\psi - m_{\psi}\psi -iG_{int}\phi\Gamma\psi = 0
\nonumber\\
&& -(\partial_M-iq_{\phi}A_M)(\partial^M-iq_{\phi}A^M)\phi +\frac{1}{2}\frac{\phi}{|\phi|}V^{\prime}(|\phi|)
-\frac{1}{2}G_{int}\bar{\psi}\Gamma\psi =0
\end{eqnarray}
where we included the connection to the definition $\tilde{D}_M=\nabla_M+\frac{1}{4}\omega_{Mab}\Gamma^{ab}$. 
In the dyonic black hole background, the Dirac equation is
\begin{equation}
\left[(\partial_z+{\mathcal A})\Gamma^z -\frac{i(\omega+q A_0)}{f}\Gamma^t-\frac{m}{\sqrt{f}(1-z)}
\mp \frac{iG_{int}\phi}{\sqrt{f}(1-z)}-U^{-1}\frac{\lambda_n}{\sqrt{f}}
\right]\psi=0
\end{equation}
where $q_{\psi}\equiv q$, the scalar is neutral $q_{\phi}=0$, $m_{\psi}\equiv m$ and
\begin{eqnarray}
&& {\mathcal A} = \frac{1}{2}\left(\frac{3}{(1-z)}+\frac{f^{\prime}}{2f}\right), 
\;\; A_0 = \mu z,
\;\; \lambda_n=\sqrt{2qhn},
\;\; U^{-1} = \left(\begin{array}{cc}
i\sigma^2 & 0 \\
0 & i\sigma^2
\end{array}\right)
\end{eqnarray}
with $f^{\prime}\equiv \partial_z f$.
In the limit $\omega=0$ it is written as follows
\begin{equation}
\left(\partial_z+{\mathcal A} -\frac{iqA_0}{f}\sigma^2+\frac{m}{\sqrt{f}(1-z)}\sigma^3
\pm\frac{iG_{int}\phi}{\sqrt{f}(1-z)}\sigma^3+\frac{\lambda_n}{\sqrt{f}}\sigma^1\right)\psi_{1;2}=0
\end{equation}
We write the bilinears in short as
\begin{eqnarray}
I_{\pm} &=& \psi_1^{\dagger}\sigma^1\psi_1 \pm \psi_2^{\dagger}\sigma^1\psi_2,
\nonumber\\
J_{\pm} &=& \psi_1^{\dagger}\psi_1 \pm \psi_2^{\dagger}\psi_2,
\nonumber\\  
K_{\pm} &=& \psi_1^{\dagger}\sigma^3\psi_1 \pm \psi_2^{\dagger}\sigma^3\psi_2,
\end{eqnarray}
with $\bar{\psi}_1\equiv \psi_1^{\dagger}i\sigma^1$. Therefore
$I_{-}=(-i)\bar{\psi}\Gamma\psi$ because $\bar{\psi}=\psi^{\dagger}\Gamma^t$.
We rewrite the Dirac equation for the bilinears
\begin{eqnarray}
&& (\partial_z+2{\mathcal A})J_{\pm}+\frac{2m}{\sqrt{f}(1-z)}K_{\pm}+\frac{2\lambda_n}{\sqrt{f}}I_{\pm}+
\frac{2iG_{int}}{\sqrt{f}(1-z)}\phi K_{\mp}=0,  \nonumber\\
&& (\partial_z+2{\mathcal A})I_{\pm}+ \frac{2qA_0}{f}K_{\pm}+\frac{2\lambda_n}{\sqrt{f}}J_{\pm} =0, \nonumber\\
&& (\partial_z+2{\mathcal A})K_{\pm}-\frac{2qA_0}{f}I_{\pm}+\frac{2m}{\sqrt{f}(1-z)}J_{\pm}
+\frac{2iG_{int}}{\sqrt{f}(1-z)}\phi J_{\mp}=0.
\label{spinor}
\end{eqnarray}
The pairing parameter is obtained by averaging the current
$I_{-}$
\begin{eqnarray}
\label{deltagrel}\Delta=iG_{int}\langle I_{-}\rangle.
\end{eqnarray}
This system should be accompanied by the equation of motion for the neutral scalar field.
In the limit of $\omega=0$ and $k_i=0$ it is given by
\begin{equation}
-\frac{1}{\sqrt{-g}}\partial_z\left(\sqrt{-g}\frac{1}{g_{zz}}\partial_z\phi\right)+\frac{1}{2}V^{\prime}(|\phi|)
-\frac{1}{2}G_{int}\bar{\psi}\Gamma\psi=0
\end{equation}
where $g\equiv {\rm det}g_{MN}$.
In the dyonic black hole background, the equation of motion for the scalar is
\begin{equation}
\partial^2_z\phi+{\mathcal B}\partial_z\phi-\frac{m_{\phi}^2}{f(1-z)^2}\phi +\frac{iG_{int}}{2f(1-z)^2}I_{-}=0
\label{scalar}
\end{equation}
where
\begin{equation}
{\mathcal B}= \frac{2}{(1-z)}+\frac{f^{\prime}}{f}
\end{equation}
The system of equations eq.(\ref{spinor}) and eq.(\ref{scalar}) is solved, at the lowest Landau level, for the unknown 
$I_{\pm},J_{\pm},K_{\pm}$ and $\phi$. We do not consider the backreaction of the spinor and scalar fields to the gauge field, and therefore we omit
the Maxwell equation eq.(\ref{Maxwell}).

Since the magnetic field is encapsulated in the parameter mapping
eq.(\ref{mapping}), we may put $\lambda_n=0$ and use the rescaled fermion charge; furthermore, the terms
proportional to off-shell (discrete) momentum cancel out due to
symmetry reasons, as explained in Ref.\cite{czs2010}. Another key
property of the magnetic systems is that, at high magnetic fields,
the ratio $\mu_{eff}/T$ can approach zero at arbitrarily small
temperatures (including $T\to 0$). 

Next we set up boundary conditions at the IR and UV for the system of equations eq.(\ref{spinor}).
It is enough to establish the boundary conditions for the fermion components.
At the horizon we choose the incoming wave into the black hole. However, as we consider
static solutions $\omega=0$, it is enough to take a regular solution, not growing to infinity as we approach horizon. 
We write the Dirac equation at the horizon $z\sim 0$ for the upper component $\psi_1=(y_1,y_2)$,  
\begin{eqnarray}
&& \left(\partial_z+{\mathcal A}-\frac{i\mu qz}{f}\sigma^2+\frac{m+G_{int}\phi}{\sqrt{f}(1-z)}\sigma^3
+\frac{\lambda_n}{\sqrt{f}}\sigma^1 \right)\left(\begin{array}{c}
y_1\\y_2
\end{array}\right)=0
\nonumber\\
&& {\mathcal A} = \frac{1}{2}\left(\frac{3}{1-z}+\frac{f^{\prime}}{2f}\right)
\end{eqnarray}
where at $T=0$ the metric factor is $f=z(3-3z+z^2-3(1-z)^3)$. Near the horizon it becomes 
\begin{eqnarray}
&& \left(\partial_z+\frac{1}{2z}-\frac{i\mu_q}{6z}\sigma^2+\frac{m+G\phi}{z\sqrt{6}}\sigma^3
+\frac{\lambda_n}{z\sqrt{6}}\sigma^1 \right)\left(\begin{array}{c}
y_1\\y_2
\end{array}\right)=0.
\end{eqnarray}
Explicitly, the system is written as
\begin{eqnarray}
&& \partial_z y_1 +\frac{1}{z}\left(\frac{1}{2}+\frac{m+G_{int}\phi}{\sqrt{6}}\right)y_1
+\frac{1}{z}\left(\frac{\lambda_n}{\sqrt{6}}-\frac{\mu_q}{6}\right)y_2=0
\nonumber\\
&& \partial_z y_2 +\frac{1}{z}\left(\frac{1}{2}-\frac{m+G_{int}\phi}{\sqrt{6}}\right)y_2
+\frac{1}{z}\left(\frac{\lambda_n}{\sqrt{6}}+\frac{\mu_q}{6}\right)y_1=0.
\end{eqnarray}
The solution reads
\begin{eqnarray}
y_1 &=& C_1z^{-\frac{1}{2}-\nu}+C_2z^{-\frac{1}{2}+\nu}\nonumber\\
y_2 &=& \frac{1}{\frac{\mu_q}{6}-\frac{\lambda_n}{\sqrt{6}}}
\left(C_1(\frac{m+G_{int}\phi}{\sqrt{6}}-\nu)z^{-\frac{1}{2}-\nu}
+C_2(\frac{m+G_{int}\phi}{\sqrt{6}}+\nu)z^{-\frac{1}{2}+\nu}\right)
\end{eqnarray}
where $C_1,C_2$ are constants and 
\begin{equation}
\nu=\frac{1}{6}\sqrt{6(m+G_{int}\phi)^2+6\lambda_n^2-\mu_q^2}
\end{equation}
We choose the solution with the regular behavior $y\sim z^{-\frac{1}{2}+\nu}$.
The solution for $z_i$ in the lower component $\psi_2=(z_1,z_2)$ where $\psi=(\psi_1,\psi_2)$
is obtained from $y_i$ by a substitute $G_{int}\rightarrow -G_{int}$. 
We have for the bilinear combinations
\begin{eqnarray}
I_{\pm} &=& y_1^{\dagger}y_2+y_2^{\dagger}y_1 \pm (y\rightarrow z)\nonumber\\
J_{\pm} &=& y_1^{\dagger}y_1+y_2^{\dagger}y_2 \pm (y\rightarrow z)\nonumber\\
I_{\pm} &=& y_1^{\dagger}y_1-y_2^{\dagger}y_2 \pm (y\rightarrow z),
\end{eqnarray}
where $\psi_1=(y_1,y_2)$ and $\psi_2=(z_1,z_2)$.

We impose two boundary conditions for eq.(\ref{scalar}): at the horizon $\phi^{\prime}(z=0)=0$ and 
at the AdS boundary $\phi(z=1)=0$.

At the AdS boundary, the boundary conditions for the currents are known from
Ref.\cite{czs2010}: one should extract the normalizable components of
$J,I,K$ in order to read off the expectation values. However, a
normalizable solution is defined in terms of an absence of a source
for the fundamental Dirac field $\psi_{\alpha}$ rather than the
composite fields such as $J_{\pm}$. The solution is to put the
source of the Dirac field to zero and then to read off the desired
normalizable solution for $J_{\pm}$ directly. Under the assumption
that the electrostatic potential $A_0$ is regular, from
eq.(\ref{asymptotics}) the composite field densities behave near the
AdS boundary $z_0=1$  as
\begin{eqnarray}
&& \mathcal{J}_1 = \psi_1^\dagger\psi_1 \rightarrow a_1^2(1-z)^{3-2m} +b_1^2(1-z)^{3+2m},\nonumber\\
&& \mathcal{I}_1 = \psi_1^\dagger\sigma^1\psi_1 \rightarrow a_1b_1(1-z)^3,\nonumber\\
&& \mathcal{K}_1 = \psi_1^\dagger\sigma^3\psi_1 \rightarrow
a_1^2(1-z)^{3-2m}-b_1^2(1-z)^{3+2m}\label{asymptot1},
\end{eqnarray}
and
\begin{eqnarray}
&& \mathcal{J}_2 = \psi_2^\dagger\psi_2 \rightarrow a_2^2(1-z)^{5-2m} +b_2^2(1-z)^{5+2m},\nonumber\\
&& \mathcal{I}_2 = \psi_2^\dagger\sigma^1\psi_2 \rightarrow a_2b_2(1-z)^5,\nonumber\\
&& \mathcal{K}_2 = \psi_2^\dagger\sigma^3\psi_2
\rightarrow a_2^2(1-z)^{5-2m} -b_2^2(1-z)^{5+2m}\label{asymptot2}.
\end{eqnarray}
The currents we have defined in eq.(\ref{jdelta2}) are the averaged
densities, e.g. $J_1=\int d\omega d^2k \mathcal{J}_1$. A
normalizable solution in $\mathcal{J}_\pm=\mathcal{J}_1\pm
\mathcal{J}_2$ is thus defined by the vanishing of both the leading and
the subleading term.

In what follows the AdS evolution equations eq.(\ref{spinor}) and eq.(\ref{scalar}) with
appropriate boundary conditions are solved numerically with a
shooting method from the horizon. Unlike the recent study in Ref.\cite{Stefano-Bolognesi} where only in
the presence of the four-Fermi bulk coupling $G_{int}$
one finds a nontrivial solution for the averaged current $\langle I_-\rangle$
with the IR boundary taken at $z=0$, we will generically have a nonzero expectation value even for $G_{int}=0$. In Ref.\cite{Stefano-Bolognesi}, one needed to introduce an IR cutoff, such as the hard wall, positioned at a radial slice $z=z_\star$. In our setup, the choice of the boundary conditions in the UV guarantees that the condensate will form irrespectively of the IR geometry, as it specifically picks the quasinormal mode of the fermion.

We repeat the same calculations for the $x$-component order parameter 
\begin{equation}
\label{deltanewmat2}\tilde{\Gamma}\equiv i\Gamma^x\Gamma^5=\left(\begin{array}{cc}
0&1\\
1&0
\end{array}\right).
\end{equation}
The pairing current defined as
\begin{equation}
\label{jdelta22}
\tilde{I}=(-i)\int d\omega\int d^2k\bar{\psi}(\omega,k,z)\tilde{\Gamma}\psi(E-\omega,p-k,z)
\end{equation}
requires us to introduce the following currents 
\begin{eqnarray}
\nonumber \tilde{J}_\pm(E,p,z) &=& (-i)\int d\omega \int d^2 k
\left(\bar{\psi}_1(\omega,k,z)\sigma^1\psi_2(E-\omega,p-k,z)\pm
\bar{\psi}_2(\omega,k,z)\sigma^1\psi_1(E-\omega,p-k,z)\right)\nonumber\\
& \equiv & \tilde{J}_1(E,p,z)\pm \tilde{J}_2(E,p,z),\nonumber\\
\tilde{I}_\pm(E,p,z) &=& (-i)\int d\omega \int d^2 k
\left(\bar{\psi}_1(\omega,k,z)\psi_2(E-\omega,p-k,z)\pm
\bar{\psi}_2(\omega,k,z)\psi_1(E-\omega,p-k,z)\right)\nonumber\\
& \equiv & \tilde{I}_1(E,p,z)\pm \tilde{I}_2(E,p,z), \nonumber\\
\tilde{K}_\pm(E,p,z) &=& -\int d\omega \int d^2 k
\left(\bar{\psi}_1(\omega,k,z)\sigma^2\psi_2(E-\omega,p-k,z)\pm
\bar{\psi}_2(\omega,k,z)\sigma^2\psi_1(E-\omega,p-k,z)\right)\nonumber\\
\label{jdelta33} & \equiv & \tilde{K}_1(E,p,z)\pm \tilde{K}_2(E,p,z),
\end{eqnarray}
A tilde is used to distinguish the two cases of pairings involving $x$ and $y$ components.
Using the Dirac equation at $\omega=0$
\begin{eqnarray}
\left(\partial_z+{\mathcal A}-\frac{iqA_0}{f}\sigma^2+\frac{m}{\sqrt{f}(1-z)}\sigma^3+\frac{\lambda_n}{\sqrt{f}}\sigma^1\right)\psi_{1;2}
+\frac{iG_{int}\phi}{\sqrt{f}(1-z)}\sigma^3\psi_{2;1}=0
\end{eqnarray} 
where the pairing parameter is obtained by averaging the current
$\tilde{I}_{+}$
\begin{eqnarray}
\label{deltagrel}\tilde{\Delta}=iG_{int}\langle \tilde{I}_{+}\rangle
\end{eqnarray}
we get the following set of coupled equations
for the bilinears defined in eq.(\ref{jdelta33})
\begin{eqnarray}
&& (\partial_z+2{\mathcal A})\tilde{J}_\pm+\frac{2m}{\sqrt{f}(1-z)}\tilde{K}_\pm
+\frac{2\lambda_n}{\sqrt{f}}\tilde{I}_\pm+\frac{2iG_{int}}{\sqrt{f}(1-z)}\phi K_+=0,\nonumber\\
&& (\partial_z+2{\mathcal A})\tilde{I}_\pm+\frac{2q\Phi}{f}\tilde{K}_\pm
+\frac{2\lambda_n}{\sqrt{f}}\tilde{J}_\pm=0,\nonumber\\
&& (\partial_z+2{\mathcal A})\tilde{K}_\pm-\frac{2q\Phi}{f}\tilde{I}_\pm
+\frac{2m}{\sqrt{f}(1-z)}\tilde{J}_\pm+\frac{2iG_{int}}{\sqrt{f}(1-z)}\phi J_+=0.
\label{spinor2}
\end{eqnarray}
There are no minus components for the $G_{int}\phi$ term in the first and third equations of eq.(\ref{spinor2}), and these terms contain currents without tildes defined in eq.(\ref{jdelta2}).
The equation of motion for the scalar is
\begin{equation}
\partial^2_z\phi+{\mathcal B}\partial_z\phi-\frac{m_{\phi}^2}{f(1-z)^2}\phi +\frac{iG_{int}}{2f(1-z)^2}\tilde{I}_{+}=0.
\label{scalar2}
\end{equation}
The system of equations (\ref{spinor}), (\ref{scalar}) and (\ref{spinor2}), (\ref{scalar2}) differ only in the $G_{int}$ term: they are identical without it, though currents are defined differently. Therefore, provided there is no "source" in the equations of motion, i.e. 
there is no Yukawa interaction $G_{int}=0$,
the $x$ and $y$ components of gamma matrices produce the vacuum expectation values (VEVs) 
\begin{equation}
\langle I_{\pm}\rangle = \langle \tilde{I}_{\pm}\rangle
\end{equation}
and according to the definitions of the pairing parameters
\begin{equation}
\Delta\to \langle I_{-}\rangle,\;\;\; \tilde{\Delta}\to \langle I_{+}\rangle
\end{equation}
where $I_\pm$ is found from eq.(\ref{spinor}). However, the equations for the plus and minus components in eq.(\ref{spinor}) are identical. In particular,
\begin{equation}
\langle I_+\rangle = \langle I_-\rangle
\label{symmetry}
\end{equation}
which proves that $x$-$y$ rotational symmetry is intact and
\begin{equation}
\Delta=\tilde{\Delta}.
\end{equation}   
Further we consider only the $y$ component for simplicity.

\subsection{Quantum criticality in the electron-hole channel}

\subsubsection{Thermodynamic behavior}

\begin{figure}[ht!]
\begin{center}
\includegraphics[width=8cm]{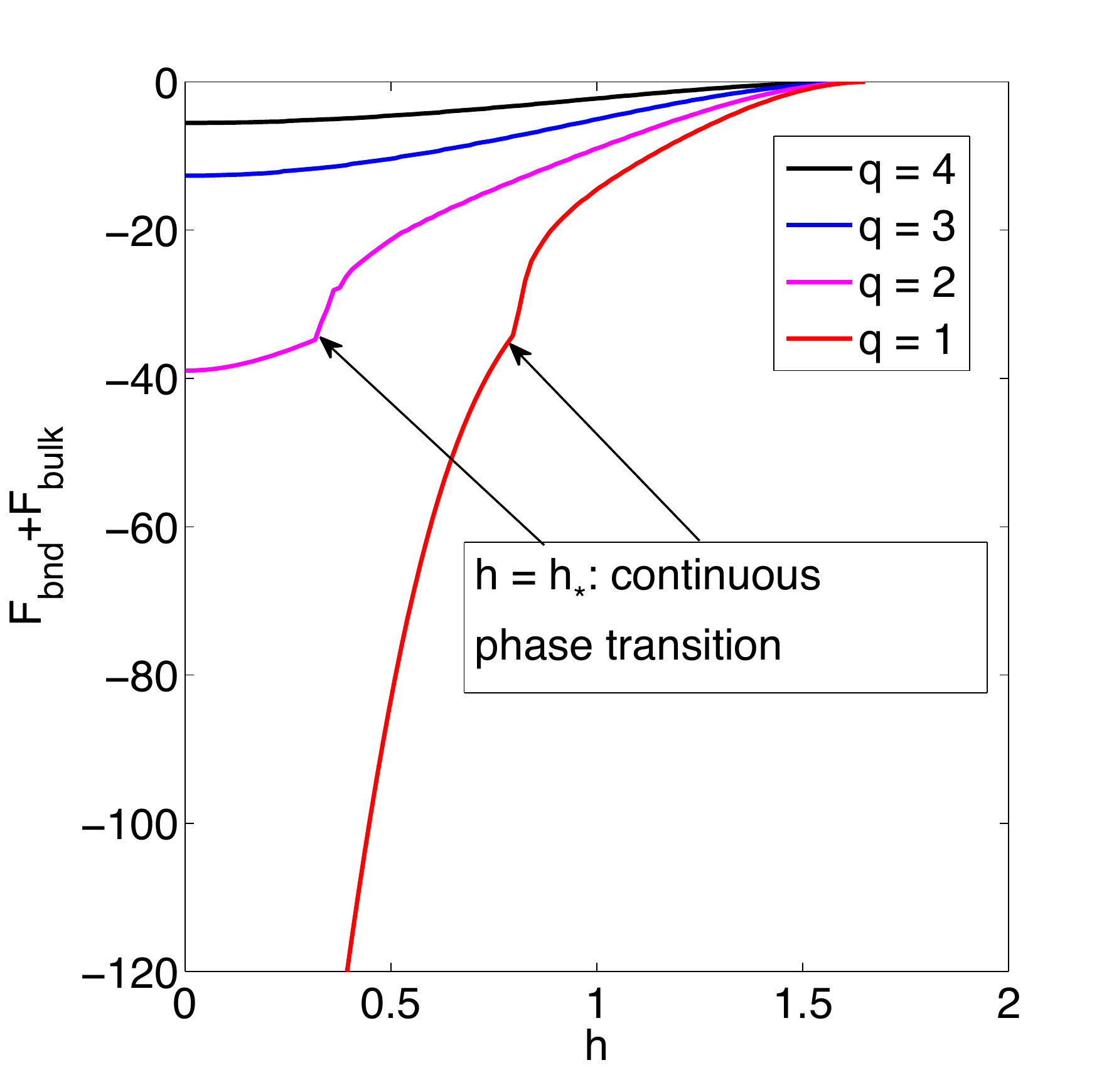}
\caption{Total (bulk plus boundary) free energy of the system
$\mathcal{F}(h)$ for increasing values of the charge $q$. An
explicit pairing term $G_{int}=2$ has been chosen in order to
suppress the stable Fermi surfaces and emphasize the phase
transition at $h=h_\star$. Still, for higher $q$ values, the
$\nu<1/2$ quasiparticles become subdominant compared to $\nu>1/2$
ones and the transition is lost. The bulk mass is $m=0.10$.}
\label{plot-fren}
\end{center}
\end{figure}

We will first use the bilinear formalism to inspect the
thermodynamics, in particular the phase transition that happens at
high magnetic fields and the behavior of the pair density after the
phase transition has occurred. To detect the transition, we can
simply plot the free energy eq.(\ref{free-act}) at a fixed temperature
as a function of the magnetic field. The action can be rewritten in
terms of the gauge field and currents as
\begin{equation}
\label{curr-act}S=\int
dzd^3x\left(\frac{1}{2}\Phi\partial_{zz}\Phi+\frac{1}{2}H^2-
I_-\Delta\right)
\end{equation}
and we need to include also the boundary term that fixes the
boundary values of the gauge field
\begin{equation}
\label{bnd-act} S_{bnd}=\int d^3x \sqrt{-h}A_\mu n_\nu
F^{\mu\nu}=\int d^3x \Phi\partial_z\Phi
\end{equation}
where $n_\mu=(0,0,0,1)$ is the unit normal to the AdS${}_4$
boundary, and $h$ is the induced metric for which
$\sqrt{-h}=z_0^{-3}$. Identifying $\Phi(z_0)=\mu$ and
$\partial_z\Phi(z_0)=\rho$ and using the Maxwell equation
(\ref{Maxwell}), we arrive at the final expression
\begin{equation}
\label{free-en}\mathcal{F}=\mathcal{F}_{bulk}+\mathcal{F}_{bnd}=\int
d^3x\left(\frac{(1-z_0)^{1+2m}}{2\sqrt{f}}J_+\Phi+\frac{1}{2}h^2
-I_-\Delta\right) +\int
d^3x\left(\frac{\mu}{2m+1}J^0_1(1-z_0)^{1+2m}+\frac{1}{2}\mu\rho\right)
\end{equation}
In particular, we see that $I_-$ is indeed the response to the bulk order parameter $\Delta$. When the coupling $G_{int}$ is set to zero, the
$I_-$ term in the bulk part of eq.(\ref{free-en}) will be absent. Let us
first see what happens in that case. The free energy is then
unaffected by the pairing, and we can only follow the dependence
on the magnetic field Fig.(\ref{plot-fren}). We see the
nonanalyticity in the free energy at the point $h=h_c$. The
underlying mechanism can be understood from the mapping
eq.(\ref{mapping}): it is the disappearance of the coherent
quasiparticle due to the lowering of the effective chemical potential
$\mu_{eff}$. The pairing arises as a byproduct of the interaction
with the boundary CFT and does not influence the transition.

With the contact interaction, corresponding to electron-hole
attraction in the infrared, we can further rewrite
eq.(\ref{free-en}) observing that generically
\begin{equation}
\label{kgrel}J_1(z,\omega)=\frac{1}{2}G_R(z,\omega)I_-(z,\omega),
\end{equation}
which gives the following result for the fermionic free energy
\begin{equation}
\label{free-delta}
\mathcal{F}_f=(1-z_0)^{1+2m}J_1(z_0,\omega=0)\left(\frac{3\mu}{4m+2}-2\Delta
G_R^{-1}(\omega=0)\right).
\end{equation}
The minus sign already makes it obvious that the derivative of the
free energy can change sign, signifying a new critical point. To
probe the transition point itself, however, we need to rewrite the
relation eq.(\ref{kgrel}) for on-shell values. Then the denominator
of $G_R$ vanishes, the current $J_1$ exactly captures the jump of
the particle number on the Fermi surface Ref.\cite{czs2010} and
eq.(\ref{noj}) becomes $J_1=3\mu/(2m+1)\times Z$, so we need to
replace $G_R\mapsto Z$, which gives the equation for the critical
point
\begin{equation}
\label{gintcrit} \mathcal{F}_f=(1-z_0)^{-2m}J_1(h)
\left[\frac{3^{3/2}q}{4(\Delta_\Psi-1)}\sqrt{1-\frac{h^2}{3}}-\frac{2G_{int}I_-\left(h\right)}{Z(h)}\right].
\end{equation}
We have also used $\Delta_\Psi=3/2+m$ in order to write the equation
purely in terms of the boundary quantities, and emphasized that $Z$
and $J_1$ are also complicated functions of $h$, since $h$
determines the effective chemical potential. Notice that only
$\mathcal{F}_{bnd}$ contributes to the fermionic term, while both
$\mathcal{F}_{bulk}$ and $\mathcal{F}_{bnd}$ contribute to the gauge
field term. For $G_{int}=0$, the second term vanishes and the free
energy can only have a nonanalyticity when $J_1(h_c)$ has it. It is
a first order transition already identified in the magnetic case in
Ref.\cite{Leiden:2010} and studied from a more general viewpoint in
Ref.\cite{czs2010}: the magnetic field depletes the Landau levels of
their quasiparticles and the Fermi surface vanishes. This
first-order jump happens at some critical $\mu_{eff}$ and we will
denote the corresponding value of the magnetic field by $h_c$. If,
however, $G_{int}$ becomes finite, we can see that the first term
decreases with $h$ while the second increases, since $Z(h)$
decreases. Thus, the overall free energy
$\mathcal{F}=\mathcal{F}_f+\mathcal{F}_{gauge}$ will have a saddle
point ($\mathcal{F}_{gauge}$ always decreases with $h$). We can now
conclude that the following behavior with respect to $G_{int}$ can
take place
\begin{itemize}
\item For $0\leq G_{int}<G_{int}^0$, the second term in
eq.(\ref{gintcrit}) is always negligible and the system only has
the first-order transition at $h=h_c$.

\item For $G_{int}^0<G_{int}<G_{int}^1$, the interplay of the
first and the second term in eq.(\ref{gintcrit}) gives rise to a
local stationary point (but not an extremum) at some $h=h_\star$. This
can potentially be a new critical point. In order to understand it
better we will later perform a detailed analysis of the infrared
behavior of the currents. It will turn out that it can be either a
second order transition or an infinite order Berezinskii-Kosterlitz-Thouless (BKT)-type transition.

\item For $G_{int}>G_{int}^1$, the Dirac hair cannot be formed
and we have $J_1=0$ for any magnetic field, including zero. Since
in this regime the pairing cannot occur even though $G_{int}$ is
large, this means we are in fact outside the applicability of the
mean field approach.
\end{itemize}
In Fig.(\ref{plot-fren}) we show the second, arguably most
interesting case. A second-order nonanalyticity in the free energy
is obvious, as long as the stable quasiparticles with $\nu>1/2$ do
not overpower the unstable quasiparticles that govern the
transition at $h=h_\star$.

The conclusion we wish to emphasize is that order parameter physics
is able to stabilize the non-Fermi liquids, while it is known
Ref.\cite{Hartnoll:es,czs2010} that in the absence of additional degrees of
freedom a consistent backreaction treatment tends to leave only the
stable, Fermi liquid surfaces. The physical nature of the point
$h_\star$ will be the object of further analysis. The next section
will reveal more on the actual pairing phenomenology, showing the
new phase to be characterized by an anomalous, growing dependence
$\Delta(h)$.

\begin{figure}[ht!]
\begin{center}
\includegraphics[width=16cm]{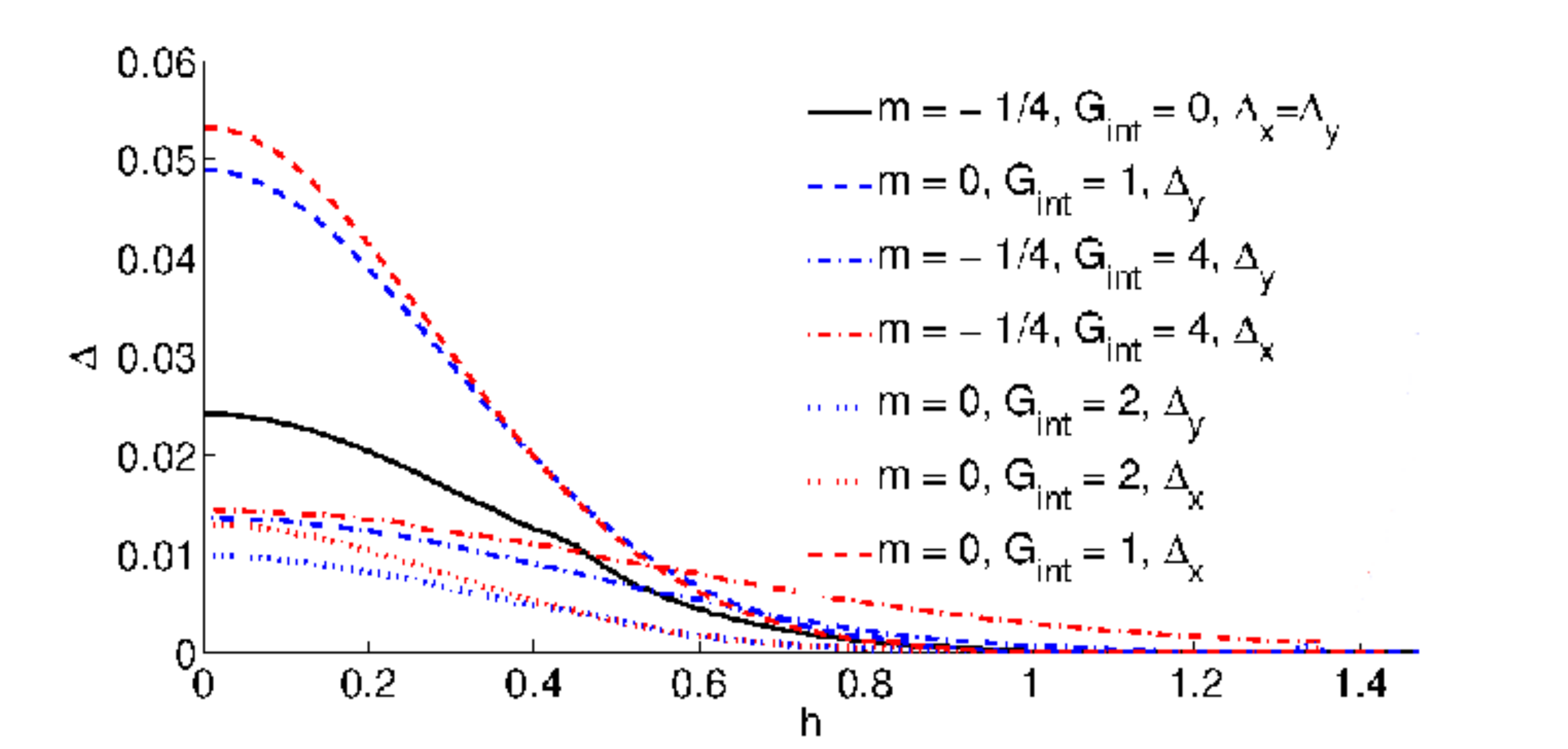}
\caption{Dependence for the $x$ and $y$ components of the pairing order $\Delta_x(h)$ and $\Delta_y(h)$
for $G_{int}=0$ (coinciding solid line) and for $G_{int}=1,2,4$ (dashed, dotted and dashed-dotted lines).
The coinciding solid line $\Delta_x=\Delta_y$ demonstrates the $x-y$ rotational invariance. For $G_{int}>0$,
increasing the bare coupling decreases $\Delta$ (and lowers $T_c$) which provides a way to tune
to the quantum critical point. Lowering the mass of the bulk fermion enhances pairing and increases $\Delta$ as seen for
$m=0$ and $m=-1/4$.}  
\label{hdep2}
\end{center}
\end{figure}

\subsubsection{Analysis of critical points}

Having analyzed the thermodynamics and found the existence of
critical points, we will now study the behavior of the order
parameter $\Delta$ in the most interesting regime, for
$G_{int}^0<G_{int}<G_{int}^1$, where the critical points are
expected to appear.

In a nutshell, we will find that the region between $G_{int}^0$ and
$G_{int}^1$ can be further subdivided into three regions, delimited
by the values $G_c^\star$, $G_c^{\star\star}$ and $G_c$,
characterized by one or two second order transitions or a BKT
transition. We will also show that the pairing is favored for high
effective chemical potentials when the density is high enough for
the gravitational interaction to produce bound states. Finally, at
small $h$ values the pairs vanish as
$\Delta\propto\exp\left(\left(T_c-T\right)^\beta\right)$ with
$0<\beta<1$ (presumably $\beta=1/2$) and finally reach zero density
$\Delta=0$ for $T\leq T_c$, while for higher magnetic fields the
trend is reversed and the order parameter starts growing with $h$.

\begin{figure}[ht!]
\begin{center}
(A)\includegraphics[width=7cm]{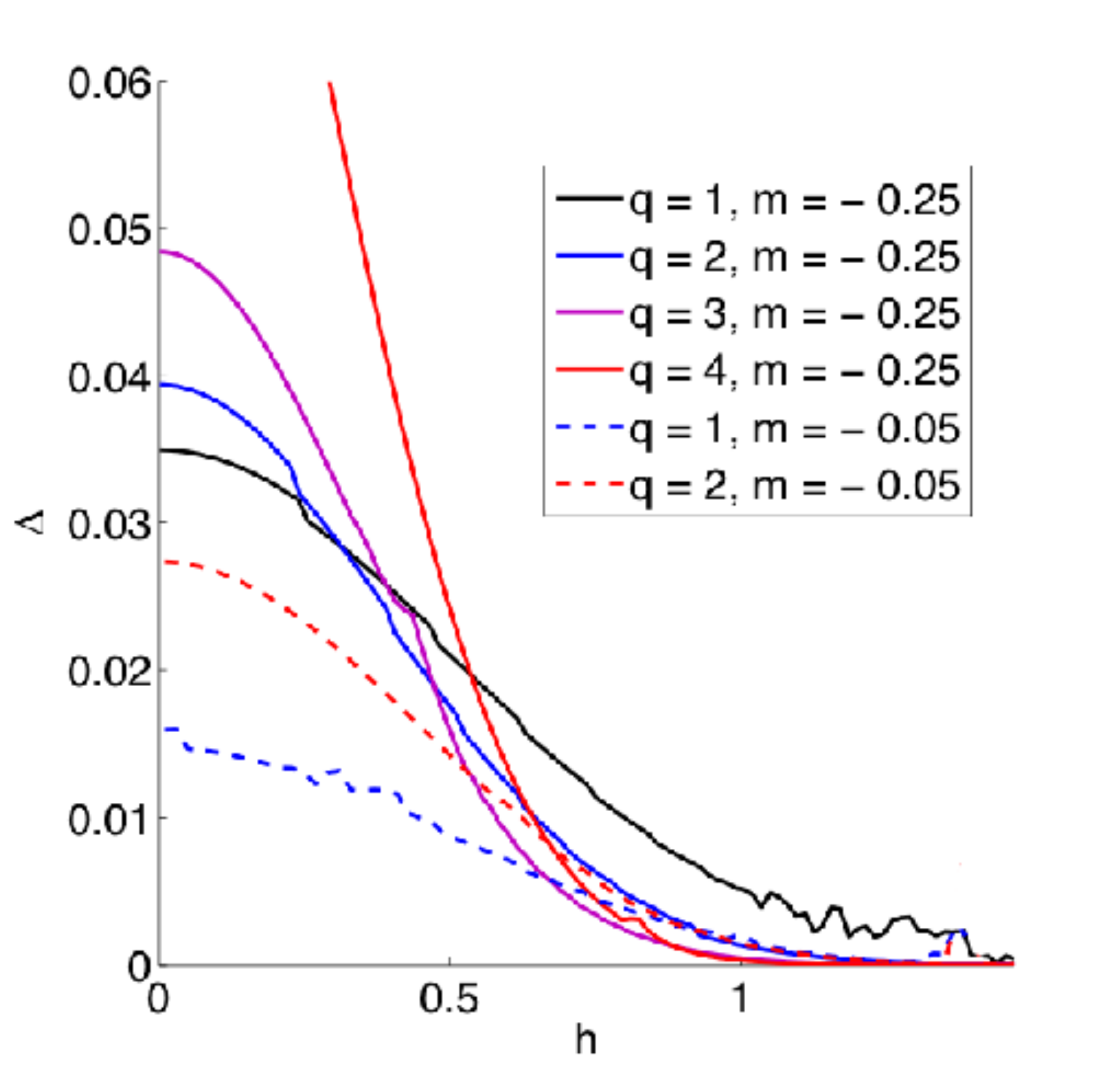}
(B)\includegraphics[width=9cm]{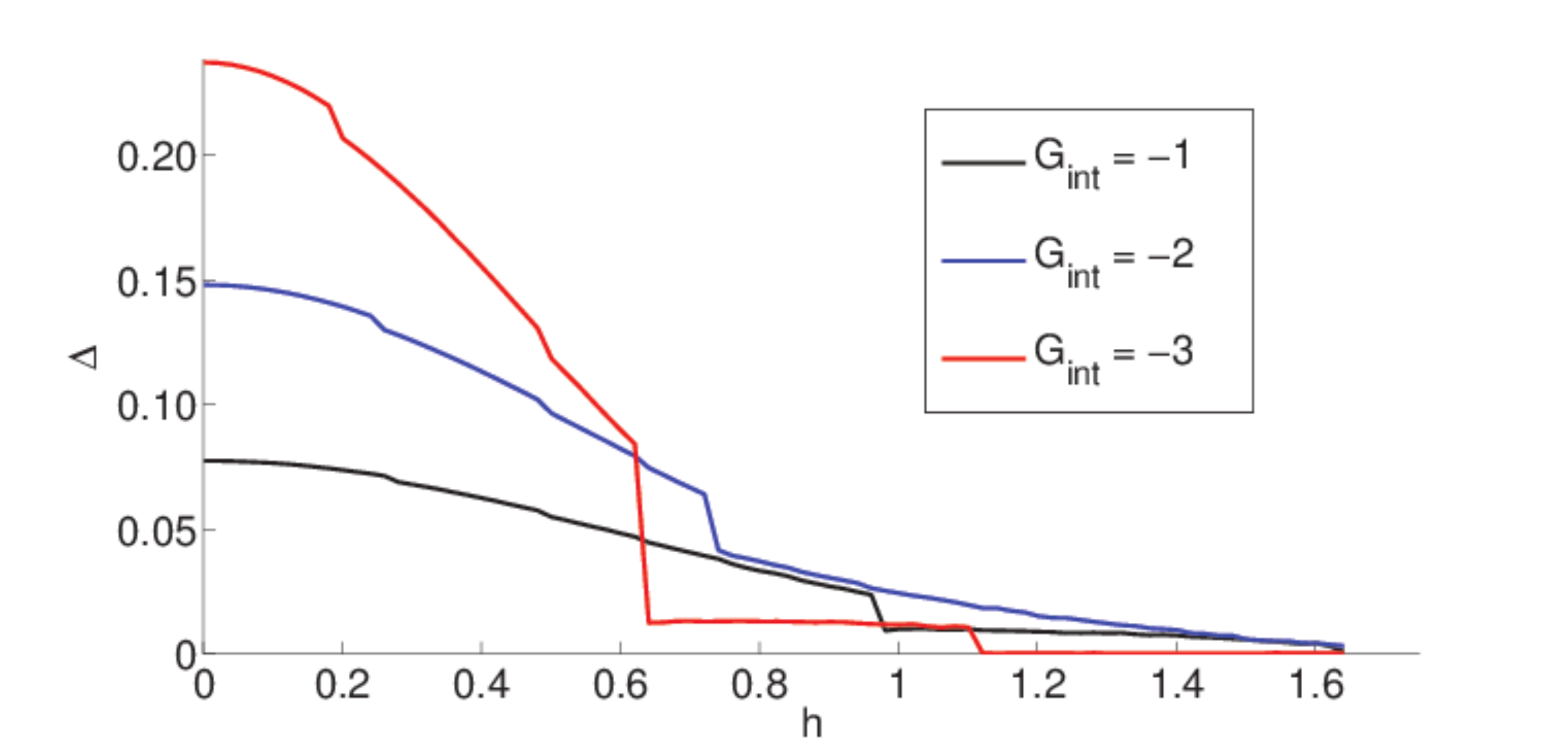} \caption{(A)
Dependence $\Delta(h)$ (in dimensionless units) for $m=-1/4$
(smooth lines) and $m=-1/20$ and for increasing values of the
fermion charge. At fixed fermion charge in the Fermi liquid regime
($\nu_{k_F}<1/2$), the magnetic field reduces the pair density,
while small charges reduce the number of pairing particles, thus
also reducing $\langle\Delta\rangle$. In the non-Fermi liquid
regime ($\nu_{k_F}>1/2$) for $h>h_c$ we observe an anomalous,
power-law growing behavior of the gap. (B) Dependence $\Delta(h)$
for $m=-1/4$ and the negative bulk coupling $G_{int}$. For increasing absolute values of the bulk coupling
$G_{int}$ the pairing order $\Delta$ is enhanced. A new value $h_\star$ arises where the order parameter
drops to zero due to competition between the channel $K_1$ and the
quasiparticle density channel $I_1$. For large absolute value of $G_{int}$
eventually $h_\star=0$ and we are out of the mean field regime.
The temperature is $T=5.6\times 10^{-4}$.} \label{hdep1}
\end{center}
\end{figure}

In order to construct the phase diagram, we will first study
$\Delta(h)$ at fixed temperature Fig.(\ref{hdep1}A). We see that
for $m=-1/4$ (smooth curves) the gap vanishes following a function
which is smoother than a power law. Indeed, it turns out that for
$h<h_c$ we have the infinite order BKT scaling behavior
\begin{equation}
\label{bkt} \Delta\sim\mu{\rm
exp}\left(-\frac{C}{2\sqrt{q(h_c-h)}}\right).
\end{equation}
The scaling eq.(\ref{bkt}) 
will be proven in section IV. Similar behavior
has been obtained in Ref.\cite{Iqbal:2010} where the scalar mass has
been tuned to the quantum phase transition: $\Delta\sim \mu{\rm
exp}\left(-\frac{C^{\prime}}{2\sqrt{m_c^2-m^2}}\right)$. Notice also
that the value $h_c$ is very high, corresponding to the magnetic
length of the order $\sqrt{h\mu_{eff}^2}\sim 10^2$ (we use
$1/\mu_{eff}$ as the natural unit of length).

The above behavior is characteristic of the normal metal parent
materials, i.e. $\nu_{k_F}>1/2$. At small values of $\nu_{k_F}$
(i.~e. $\Delta_\Psi$ close to $3/2$ or small $\mu_q$), the anomalous
growing dependence $\Delta(h)$ appears (found also in the previous
section at strong enough magnetic fields) as shown by the dashed
curves in Fig.(\ref{hdep1}A). The nature of the dependence
$\Delta(h)$ is rooted in the unstable Fermi surfaces with
$\nu_{k_F}\to 0$ and can be understood from the analysis of the
bilinear equations in the AdS${}_2$ region, which we postpone until the
next section.

We study the relation $\Delta(h)$ at different values of the pairing coupling
$G_{int}$. For $G_{int}>0$, $\Delta$ decreases as we increase $G_{int}$: repulsive interaction
destructs the pairing, as given in Fig.(\ref{hdep2}). For $G_{int}<0$,
$\Delta$ increases as absolute value of $G_{int}$ is increased: attractive interaction
triggers and enhances the pairing, as given in Fig.(\ref{hdep1}B). 
Combining the two cases, when the sign of $G_{int}$ is taken into account, the dependence $\Delta$
versus $G_{int}$ is decaying. Lowering the mass of the bulk fermion enhances pairing as can be seen
by comparing cases $m=0$ and $m=-1/4$ in Fig.(\ref{hdep2}).
As shown in Fig.(\ref{hdep2}), pairing parameters with the $x$ and $y$ component
are identical for $G_{int}=0$, which proves that the $x-y$ plane rotational symmetry is intact. As $G_{int}$
is switched on, it disrupts pairing in both channels in a slightly different way causing $\Delta_x$ and $\Delta_y$ 
to deviate from each other. An important novel feature distinguishing $G_{int}>0$ and $G_{int}<0$
is the appearance of the second anomalous branch for $G_{int}<0$ as seen in Fig.(\ref{hdep1}) where the magnetic field
enhances pairing: the rising $\Delta(h)$ manifests magnetic catalysis (MC). 

The motivation to consider $G_{int}>0$ was the ability to reduce the critical temperature to zero
and to tune to the quantum critical point. On the other hand, adding $G_{int}<0$ increases the critical temperature;
however we can tune to vanishing $T_c$ by adjusting other parameters such as the magnetic field.
Figures (\ref{hdep2}) and(\ref{hdep1}B) 
can be used to extract the quantum critical point
(QCP) $h=h_c$ when $\Delta=0$ (or $T_c=0$) at fixed $G_{int}$.
Upon varying the coupling $G_{int}$, the QCP becomes the quantum
critical line (QCL) $h_c(G_c)$ or $G_c(h_c)$. In Fig.(\ref{hdep1}B),
for growing $G_c$,
$h_c$ decreases in the normal branch and $h_c$ increases in the anomalous branch.
In the normal branch, $h$ depletes the particles from the Fermi surface
decreasing the pairing density. Therefore $h$ destroys condensate.
In the anomalous branch though, $h$ enhances the condensation
(magnetic catalysis).

\begin{figure}[ht!]
\begin{center}
(A)\includegraphics[width=8cm]{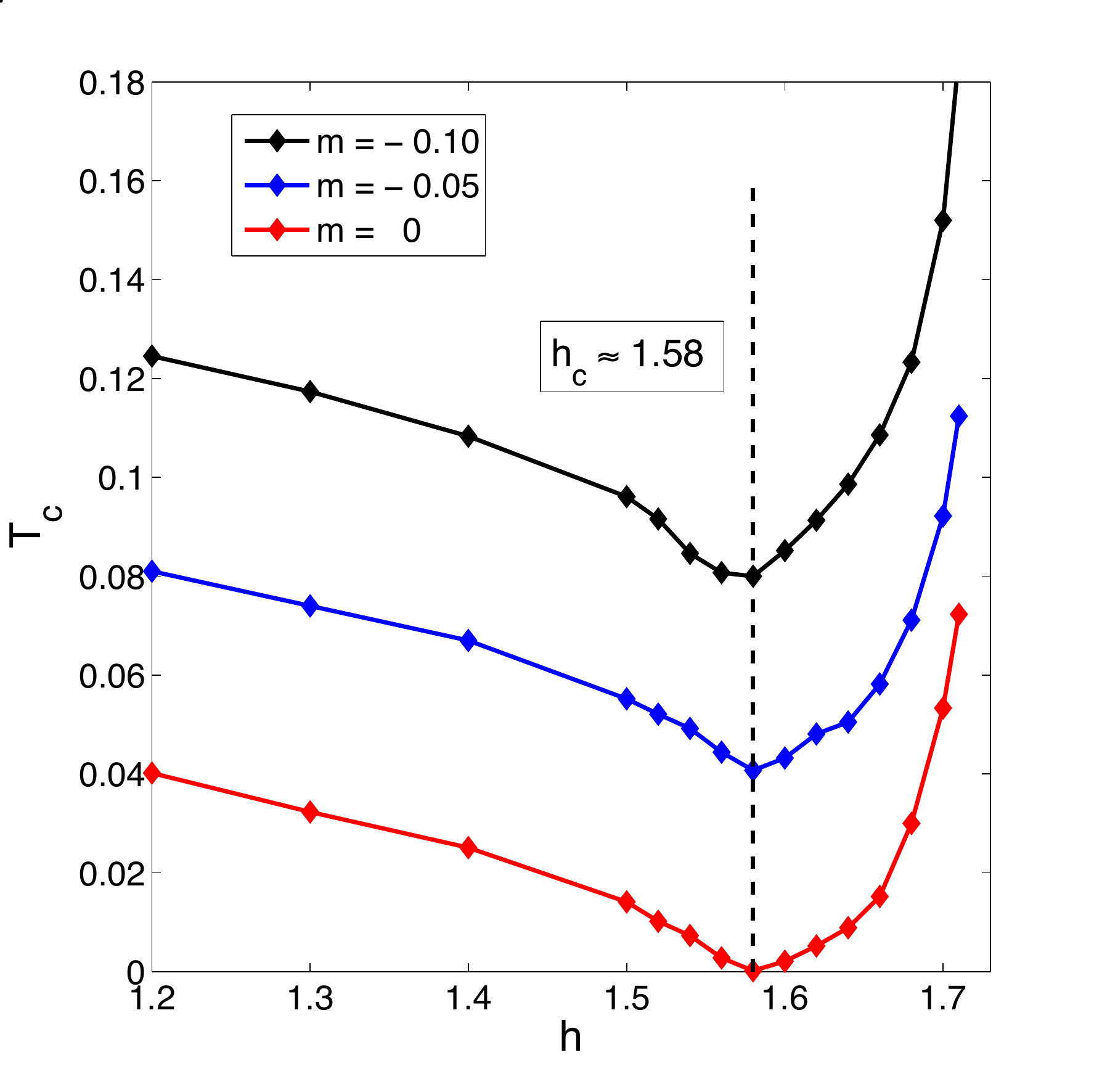}
(B)\includegraphics[width=8.5cm]{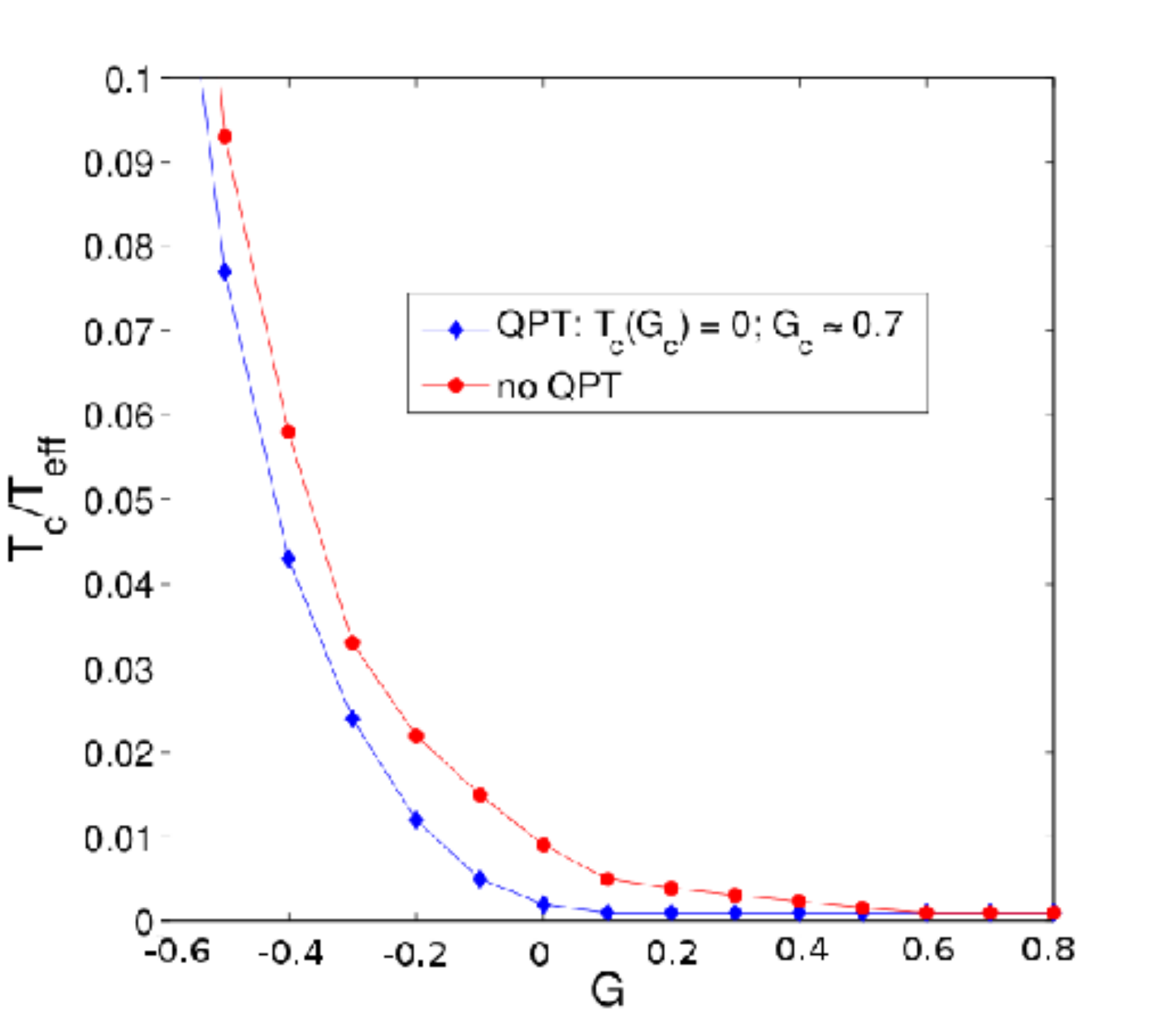}

\caption{Critical temperature $T_c$ vs the magnetic field (A)
and the coupling $G_{int}$ (B), for $q=1$. In (A), we again see
the anomalous branch starting at $h_c\approx 1.58$, that signifies
the exit from the Landau-Ginzburg regime and the mean field
scaling into a new phase. Higher curves do not possess the QCP and
arise when the system is always in the condensed phase. The
coupling is $G=0.7$. In (B), we see $T_c$ vanish at the QCP
$G_c\approx 1.1$, corresponding to the quantum phase transition
(QPT) with a non-mean field exponent $T_c\sim |G_c-G|^{\beta}$,
$\beta>1$. In the higher curve, $T_c$ remains nonzero for all
$G_{int}$, with no QPT for this set of parameters. The bulk mass
is $m=-0.10$.} \label{plot-tcrit}
\end{center}
\end{figure}

The next step toward the phase diagram is the dependence of the
critical temperature on the external magnetic field $T_c(h)$.
A typical situation is given in Fig.(\ref{plot-tcrit}A). We have
captured both branches so we see the expected twofold behavior,
with the decrease of $T_c$ up to $h=h_c$ and subsequent increase.
A precise tuning of the mass toward zero is necessary to enter the
quantum critical regime where $T_c(h_c)=0$. For reference, we have
also shown the cases $m=-0.10$ and $m=-0.05$, where the approach
of the critical point is seen but $T_c(h_c)$ is still a finite
minimum.

Fig.(\ref{plot-tcrit}B) shows the decreasing dependence of the
critical temperature $T_c$ vs. the coupling strength $G_{int}$.
For the blue curve $T_c$ vanishes at the QCP $G_c\approx 1.1$. It
corresponds to the quantum phase transition (QPT) of the second
order with a non-mean field exponent $T_c\sim |G_c-G|^{\beta}$,
$\beta>1$. For the red curve, $T_c$ remains nonzero for all
couplings $G_{int}$. It happens when the system is always in the
condensed phase (an extreme RN AdS black hole is unstable)
Ref.\cite{Faulkner:2010}. As seen from Fig.(\ref{plot-tcrit}B),
$G_{int}$ is a sensitive "knob" to adjust the critical temperature
$T_c$.

\begin{figure}[ht!]
\begin{center}
\includegraphics[width=8cm]{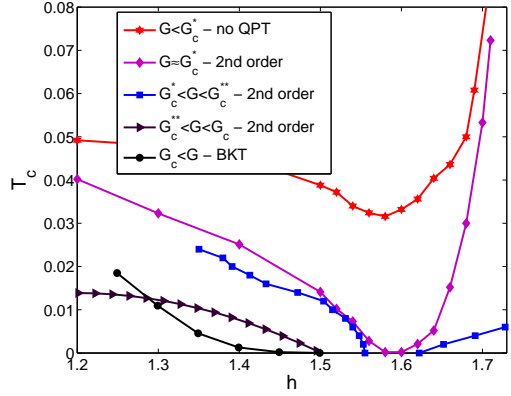}
\caption{Critical temperature $T_c$ vs. the magnetic field $h$ for
different couplings $G_{int}$. Depending on the coupling $G_{int}$,
there are the BKT and second order phase transitions. At $T_c=0$,
the QCP becomes QCL $(h_c,G_c)$ with a decreasing/increasing
dependence on $h_c$ as $G_c$ is increased which corresponds to
the normal/anomalous branch.} \label{plot-tcvsh}
\end{center}
\end{figure}

Finally, after studying the influence of the fermion charge $q$ and
the bulk mass $m$ on the relation $T_c(h)$, we conclude with the
Fig.(\ref{plot-tcvsh}), showing the critical temperature versus the
magnetic field for different couplings $G_{int}$. We find \emph{four
distinct regimes} located in the interval $G^0<G<G^1$ (we omit the "$int$"
subscript in $G_{int}$ for now). The delimiting points are denoted
by $G_c^\star$, $G_c^{\star\star}$ and $G_c$, with
$G^0<G_c^\star<G_c^{\star\star}<G_c<G^1$.

\begin{itemize}
\item For $G<G_c^\star$ the critical temperature is nonzero, as
demonstrated in Fig.(\ref{plot-tcrit}B) and also by the red curve
in Fig.(\ref{plot-tcvsh}). There is thus no QCP and the normal and
anomalous regimes are separated by a crossover.

\item For  $G_c^\star< G< G_c^{\star\star}$, there are two second
order phase transitions, one for the normal and one for the
anomalous branch. This case is represented by the blue curve in
Fig.(\ref{plot-tcvsh}), and can also be seen in Fig.(\ref{hdep1}A).
The quantum phase transition corresponding to the anomalous branch
scales with the non-mean field exponent $T_c\sim
|h_c-h|^{\delta^{\prime}}$, $\delta^{\prime}>1$. The limiting case
of $G = G_c^\star$ is given by the magenta curve, where the
two critical points coincide.

\item For $G_c^{\star\star}< G< G_c$, there is
the second order phase transition with the non-mean field exponent
$T_c\sim (h_c-h)^{\delta}$, $\delta<1$, which describes the normal
branch. This is the dark violet curve in Fig.(\ref{plot-tcvsh}), similar
to the regime  in Fig.(\ref{hdep1}B).\\

\item For $G>G_c$, there is an infinite order phase transition of
the Berezinsky-Kosterliz-Touless (BKT) type with the characteristic
exponential scaling $T_c\sim {\rm exp}\left(-\frac{C}{\sqrt{h_c-h}}\right)$.
This is the black curve in the figure.
\end{itemize}

\begin{figure}[ht!]
\begin{center}
\includegraphics[width=9cm]{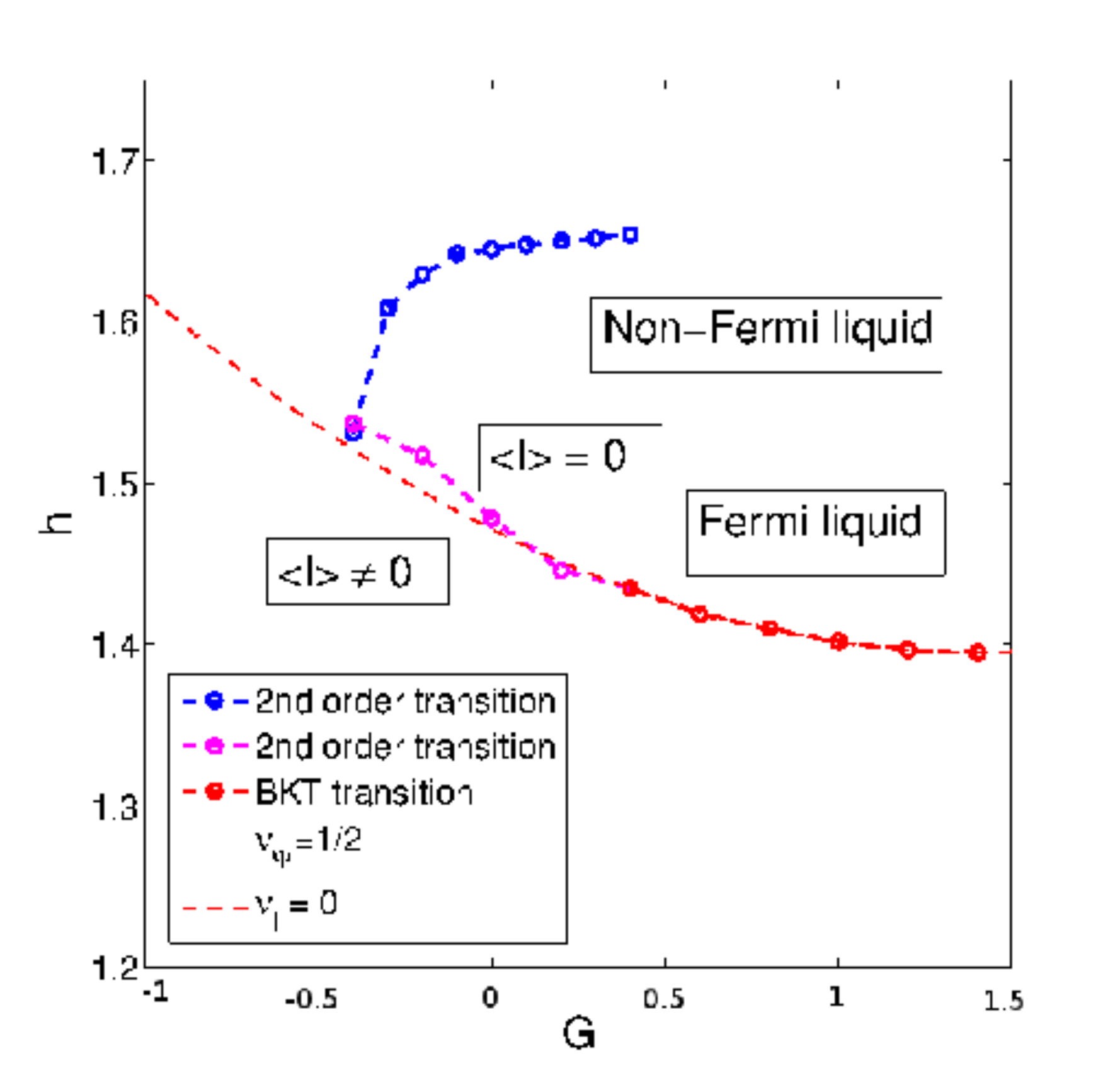}
\caption{Phase diagram $h$ vs. $G$ for the condensed/normal
(non-)Fermi liquids. $G$ and $h$ destroy the condensate except for
the non-Fermi liquid. The ordered non-Fermi liquid is enhanced and
stabilized by the strong magnetic field, which is also seen
experimentally in pyrolytic graphite.} \label{plot-phase-diagram}
\end{center}
\end{figure}

Finally, based on the data from Fig.(\ref{plot-tcvsh}) and some
additional calculations, we can draw the phase diagram in terms of
the magnetic field $h$ and the coupling $G_{int}$, given in Fig.(\ref{plot-phase-diagram}). 
The QCL (solid line) separates the condensed
(ordered) from uncondensed (disordered) phases. The position of the
QCL is extracted from the phase transition curve of the critical
temperature vs the magnetic field: the QCL where the critical
temperature vanishes is given by the relation $G_c(h_c)$. From the
dependence $G_c(h)$, one can translate the scaling exponents $T_c$
vs. $G$ into $T_c$ vs. $h$: $T_c\sim |G_c-G|^{\beta}\to
|G_c(h_c)-G|^{\beta}\to |h_c-h|^{\delta}$.

In Fig.(\ref{plot-phase-diagram}), increasing the coupling $G$
and the magnetic field $h$ destroy the pairing condensate except
in the non-Fermi liquid regime. This twofold behavior manifests
itself through a double-valued function $h_c(G_c)$ in some
parameter range. Indeed, the region with a condensed non-Fermi
liquid is enhanced by the magnetic field, which is a consequence
of the magnetic catalysis and the Callan-Rubakov effect discussed
in the next section.

A deeper understanding of the phase diagram can be reached by
considering the scaling dimensions of the condensate and the fermion
field. With some foresight from the next subsection, we note that
the IR conformal dimension of the operator which condenses $\delta_{\tilde{I}}=1/2+\nu_{\tilde{I}}$, where the bulk 
pairing current $\tilde{I}=\sqrt{\zeta}I$ is the gravity dual of the excitonic condensate, is given by eq.(\ref{equation_conformal_dimension})
\begin{eqnarray}
&&\nu_{\tilde{I}}=\sqrt{\frac{2}{3}}\sqrt{(m+\Delta)^2+2qh-\frac{\mu_q^2}{6}}.
\label{IRnu-I}
\end{eqnarray}
On the other hand, the IR conformal dimension of the fermion
operator $\delta_{\psi}=1/2+\nu_{\psi}$, where the bulk fermion field
$\psi$ is dual to the boundary fermion $\Psi$, is given by 
\begin{eqnarray}
&&
\nu_{\psi}=\frac{1}{6}\sqrt{m^2+k_F^2(h)-\frac{\mu_{q,eff}^2}{6}},\;\;
\mu_{q,eff}=\sqrt{3}q\sqrt{1-h^2/3}. \label{IRnu-psi}
\end{eqnarray}
Importantly, the ratio $\nu_{\tilde{I}}/\nu_{\psi}$ is first a decreasing and then an increasing
function of the magnetic field $h$ (see
left panel of Fig.(8) in Ref.\cite{Leiden:2010} for $\nu_\psi$). At the dashed line the IR
dimension $\nu_I$ of the operator with gravity dual pairing current
becomes imaginary, signaling the pairing instability.
This is analogous to the instability of a scalar operator, when the
Breitenlochner-Freedman (BF) bound in the AdS${}_2$ is violated but
the BF bound in the AdS${}_4$ remains unbroken. The dash-dotted
line corresponds to the locus of points in the phase diagram where
$\nu_\psi=1/2$, separating the Fermi liquid from the non-Fermi
liquid behavior as discussed in Ref.\cite{Faulkner:2009}. Since
$\nu_\psi(h)$ is a monotonically decreasing function, coherent
quasiparticles disappear at large magnetic field resulting in the
non-Fermi liquid regime at $\nu_\psi\leq\frac{1}{2}$ (upper part of
the phase diagram). Notably, there is a similarity between our phase
diagram Fig.(\ref{plot-phase-diagram}) and the phase diagram
obtained for a scalar field Fig.(14) in Ref.\cite{Faulkner:2011}, which
uses the double-trace deformation as the control parameter. This may
provide an insight of a mechanism of suppression/enhancement of
the ordered phase at small/large magnetic fields.

\begin{figure}[ht!]
\begin{center}
\includegraphics[width=9cm]{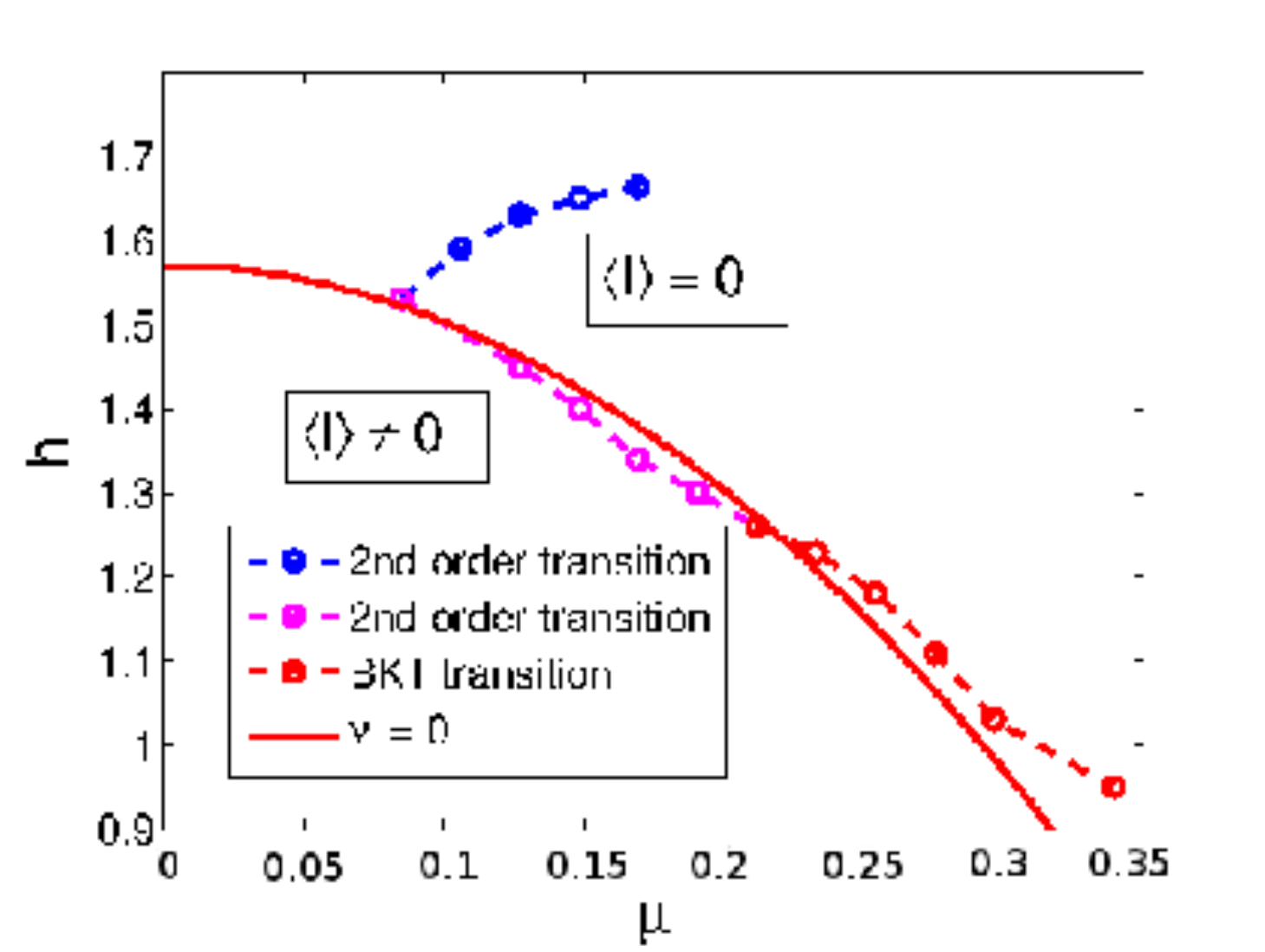}
\caption{Phase diagram $h$ vs. $q$ for the condensed/normal
(non-) Fermi liquids. Increasing the fermion charge at zero
temperature is equivalent to increasing the chemical potential. It
stabilizes the condensate in the anomalous regime and then
destabilizes it in the normal branch. We can thus qualitatively
relate $q$ to $G$, the coupling constant from the previous
figure.} \label{plot-phase-d2}
\end{center}
\end{figure}

We can redraw our phase diagram in terms of the magnetic
field $h$ vs. the chemical potential $\mu$, Fig.(\ref{plot-phase-d2}), to be able to compare our result
with the literature Ref.\cite{Schmitt:2010}. 

It is worth noting that our phase diagram
exhibits the same main features as the analogous phase diagram
obtained using the Sakai-Suggimoto model (Fig.(8) in Ref.\cite{Schmitt:2010}).
Primarily, it also has two regions of weak magnetic field where
the condensate is destroyed by the magnetic field ("inverse" magnetic
catalysis) and a regime of strong magnetic field which enhances the
condensate (magnetic catalysis). Likewise, Fig.(\ref{plot-tcvsh})
shows the same structure as the analogous Fig.(9b) in
Ref.\cite{Schmitt:2010}. Thus there are the two regimes with opposite dependence $\Delta(h)$.
We will discuss
the reasons for it in the next section.

\subsubsection{Pairing, double-trace deformations and conformal field theory}

We will conclude our study of the phase diagram by offering an
alternative viewpoint of the observed critical phenomena. Dialing
the pairing coupling to drive the system toward QPT can also be
understood as dialing the double-trace deformation in the boundary
theory Ref.\cite{Faulkner:2011}. For example, in the Gross-Neveu model
with vector $SU(N_f)$ symmetry, the four-fermion coupling operator
is relevant at the UV fixed point. Hence, as a relevant deformation
in UV, it can drive the RG flow of the system to a new IR fixed
point with spontaneous symmetry breaking. In holography, the
multitrace deformations which are introduced on the boundary and
correspond to the multiparticle states in gravity are a powerful
knob that can drive the theory either to a free CFT at the IR fixed
point or to a CFT with the spontaneously broken symmetry. An RG flow
of this kind has been considered in Ref.\cite{Gubser:2002}, where the
relevant double-trace deformation at the UV fixed point drives the
theory toward the asymptotically free IR fixed point. In the gravity
dual theory, it corresponds to different boundary conditions imposed
at the AdS${}_4$ boundary (alternative/standard quantization), and the
UV and IR CFTs are related by a Legendre transform
Ref.\cite{Gubser:2002}.

As an illustration, consider a scalar theory in the bulk as in
Ref.\cite{Iqbal:2010}. One can hope that this case 
captures the behavior of our system at least
qualitatively as a bilinear fermion combination bosonizes into a
scalar field. Fig.(8) shows
schematically the two loop beta function for the double-trace
coupling for decreasing magnetic field value. At strong magnetic
fields, Fig.(8) top, the theory exhibits the usual RG flow from the
strongly coupled UV fixed point (with a Landau pole at the QCP:
$g_c\to\infty$) to a free fermion (a noninteracting theory at $g\to
0$) at the IR fixed point, with no expectation value for the scalar
operator $O$. At the QCP i.e. $h=h_c$, Fig.(8) middle, the UV and IR fixed
points merge and annihilate, leading to the BKT scaling
Ref.\cite{Kaplan:2009}
\begin{eqnarray}
\Lambda_{IR}\sim \mu{\rm exp}\left(-\frac{C}{\sqrt{h_c-h}}\right)
\sim \mu{\rm exp}\left(-\frac{C^{\prime}}{\sqrt{g-g_c}}\right),
\label{eq-BKT}
\end{eqnarray}
which can be interpreted as a distance along the RG trajectory to
get to the nontrivial IR fixed point with broken symmetry. In this
case, the QPT is of infinite order and where the critical
temperature $T_c$ and the order parameter $\langle O\rangle$ are
governed by the exponential BKT scaling of eq.(\ref{eq-BKT}) as
$T_c\sim \langle O\rangle\sim \Lambda_{IR}$. When the magnetic field
$h$ is further decreased, Fig.(8) bottom, the theory becomes gapped leading
to an apparent conformality loss Ref.\cite{Kaplan:2009} and the QPT is
now of second order.

\begin{figure}[ht!]
\begin{center}
\includegraphics[width=10cm]{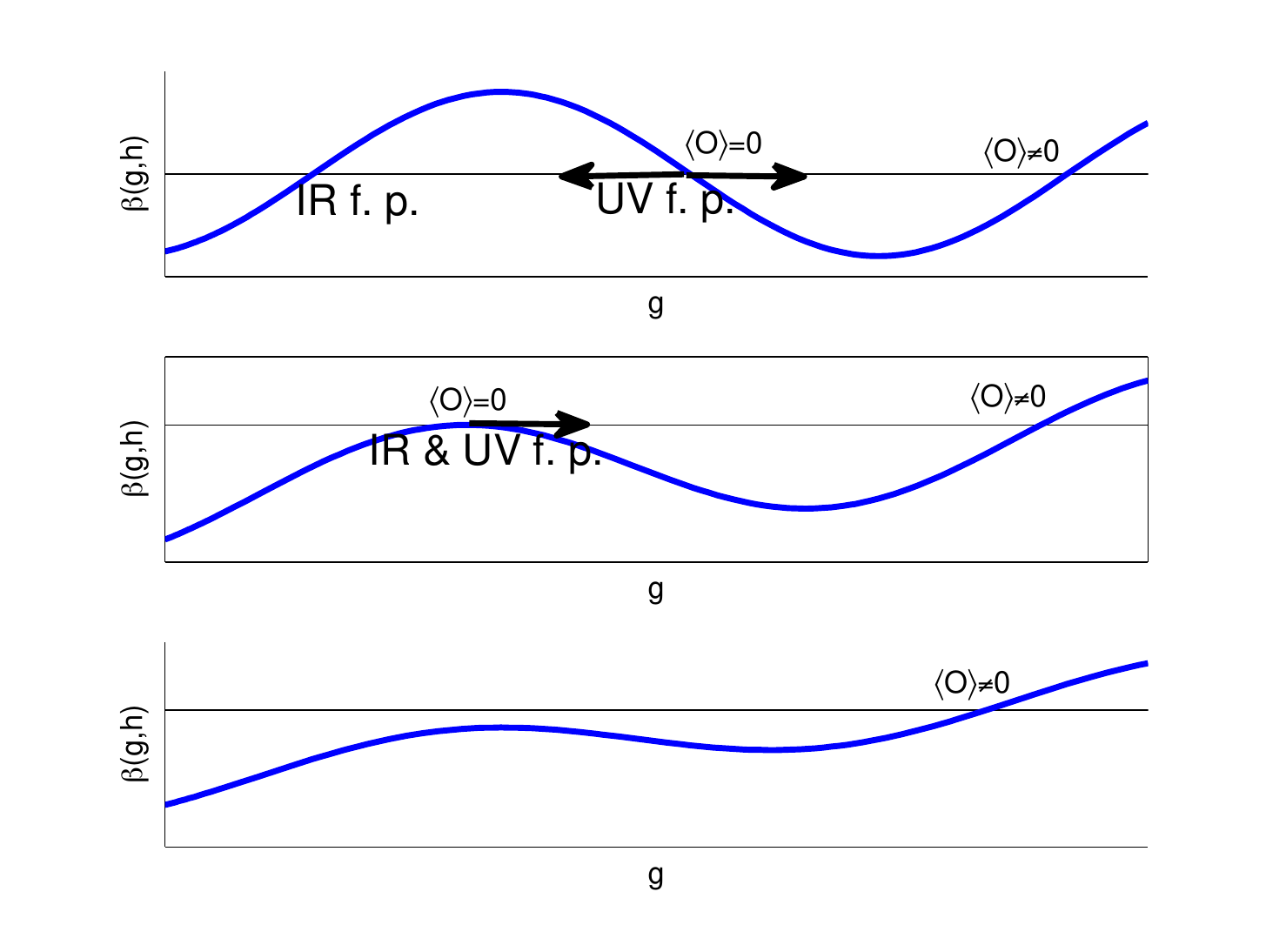}
\caption{Two loop beta-function for the double trace coupling for
decreasing magnetic field values. The disappearance of the original UV and
IR fixed points at the critical magnetic field $h=h_c$ leads to
conformality loss and the BKT scaling behavior, middle panel. At $h<h_c$,
bottom panel, the system flows to a new IR fixed point with spontaneous
symmetry breaking and nonzero condensate of a scalar field $\langle
O\rangle\neq 0$.} \label{plot-conformality-loss}
\end{center}
\end{figure}

In this paper we use the Yukawa coupling (or four-fermion coupling) in the bulk.
However, the results we obtain are in line with the theory having
a double-trace deformation on the boundary as described by Fig.(\ref{plot-conformality-loss}): 
we have observed the rise of a new
critical point. Fig.(\ref{hdep1}B) in particular conveys the
message: at some $h_\star<h_c$ we observe a transition from the
quasiparticle regime to an electron-hole condensate. Formally, it
comes from the competition between the pairing channel and the
particle-photon interaction, encoded by the bilinears $K_1$ and
$I_1$. Physically, it corresponds to the competition between the
Fermi surface "order" and the pairing order. At $h=h_c$, it is the
entrance into the non-Fermi liquid region ($\nu<1/2$) that drives
the transition. At very high $G_{int}$ values, the pairing is
again suppressed which we interpret as the consequence of the
Fermi surface depletion. The number density near the Fermi
momentum is given by the current $J_0$. In eq.(\ref{spinor}),
it is clear that the gauge field term, encoding for the chemical
potential (and implicitly density), is competing with the term
containing $\Delta(r)$, i.e. the term proportional to the coupling
$G_{int}$. When the latter is dominant, the pairing is highly
enhanced but only up to the point that all electrons are "used
up", and their total number density is small. Notice also how
$\Delta$ drastically increases at nonzero $G_{int}$, growing by
about an order of magnitude.

\subsection{AdS${}_2$ analysis of the critical exponents}\label{section:D}

Most of our conclusions so far were driven by numerical results,
with some qualitative analytical insight. A somewhat more detailed
analytical understanding of the model can be gained by considering
the far IR region, corresponding to the AdS${_2}$ throat of the RN
black hole.

We will follow the arguments of Ref.\cite{Iqbal:2010}, where it was shown
by analyzing the AdS${}_2$ region that a new IR scale $\Lambda_{IR}$
is generated which leads to the scaling behavior for the critical
temperature $T_c$ and the condensate $\Delta$ vs a tuning
parameter (the magnetic field in our case). The key point of this
analysis is to show that an instability for a scalar field develops
in a certain parameter range. In particular, for a neutral scalar
field the mass should be lower than the AdS${}_2$ BF bound,
$m^2R^2<-\frac{3}{2}$ (where $R$ is the AdS${}_4$ radius), which
corresponds to a point where the IR conformal dimension becomes
imaginary. For a charged scalar, the mass value can be slightly
higher if the product of the charge and the chemical potential,
$\mu_q$ is sufficiently large. We therefore consider a composite
bosonic field, which can be constructed as a bilinear combination of
$\psi$'s and in our case it is given by a bulk current.

Let us start by recalling that at $T=0$, the redshift factor develops a double zero near the horizon: $f\approx
6z^2$. Adopting the rescaled coordinates $\zeta,\tau$ instead of the dimensionless coordinates $z,t$:
\begin{eqnarray}
\label{ads2coor}
\frac{1}{\frac{1}{z}-1}=\frac{\omega}{6\zeta},\;\;
t=\frac{\tau}{\omega},
\end{eqnarray}
with $\omega\to 0$ and $\zeta,\tau$ finite, the metric eq.(\ref{geometry}) becomes near the horizon
\begin{equation}
\label{ads2metric}
ds^2=\frac{1}{6\zeta^2}\left(-d\tau^2 +d\zeta^2\right) + dx^2+dy^2,
\end{equation}
where the gauge field is
\begin{equation}
A_\tau = \frac{\mu}{6\zeta}.
\end{equation} In
this metric, the currents defined in eq.(\ref{jdelta2}) become
\begin{eqnarray}
\label{jdelta22} J(E,p,z) &=& (-i)\int d\omega \int d^2 k
\bar{\psi}(\omega,k,z)\sigma^1\psi(E-\omega,p-k,z),\nonumber\\
I(E,p,z) &=& (-i)\int d\omega \int d^2 k
\bar{\psi}(\omega,k,z)\psi(E-\omega,p-k,z), \nonumber\\
K(E,p,z) &=& -\int d\omega \int d^2 k
\bar{\psi}(\omega,k,z)\sigma^2\psi(E-\omega,p-k,z),
\end{eqnarray}
with $\bar{\psi}=i\psi^{\dagger}\sigma^1$.
The Dirac equation at $\omega=k=0$ assumes the form
\begin{equation}
\left(\partial_\zeta-i\frac{\mu_q}{\sqrt{6}e_{\hat{\zeta}}}\sigma^2
+\frac{(m+\Delta)}{e_{\hat{\zeta}}}\sigma^3+\frac{\lambda}{e_{\hat{\zeta}}}\sigma^1,
\right)\psi=0, \label{ads2-equation}
\end{equation}
giving the following equations of motion for the currents
\begin{eqnarray}
&& \partial_\zeta J+\frac{2(m+\Delta)}{e_{\hat{\zeta}}}K+\frac{2\lambda}{e_{\hat{\zeta}}}I=0\\
&& \partial_\zeta I+2\frac{\mu_q}{\sqrt{6}e_{\hat{\zeta}}}K+\frac{2\lambda}{e_{\hat{\zeta}}}J=0\\
&&
\partial_\zeta K-2\frac{\mu_q}{\sqrt{6}e_{\hat{\zeta}}}I+\frac{2(m+\Delta)}{e_{\hat{\zeta}}}J=0
\end{eqnarray}
where $e_{\hat{\zeta}}=\sqrt{6}\zeta$, $\mu_q=\mu q$, $h_q=h q$,
$\lambda=2|h_q|l$, $l=1,2,\dots$ and $\Delta=-\langle I\rangle$.
Differentiating the second equation for $I$ with respect to $\zeta$
and eliminating the derivatives of $J$ and $K$ currents from the
other two equations, we obtain the zero energy Schr\"{o}dinger
equation
\begin{eqnarray}
&& \partial_{\zeta}^2\tilde{I}-\frac{\nu_{\tilde{I}}^2-1/4}{\zeta^2}\tilde{I}=0,\label{I}\label{equationI0}\\
&& \nu_{\tilde{I}}=\sqrt{\frac{2\lambda^2}{3}-\frac{\mu_q^2}{9}},
\label{equation_conformal_dimension}
\end{eqnarray}
where $\tilde{I}=I\sqrt{\zeta}$. We assume that condensation occurs
for the lowest (first) Landau level ($l=1$) and it is caused by an
instability when $\nu_{\tilde{I}}$ becomes imaginary. Therefore we
can represent the conformal dimension as
\begin{eqnarray}
\tilde{\nu}_{\tilde{I}}=\sqrt{\frac{4}{3}(h_q^c-h_q)},\;\;
h_q^c=\frac{\mu_q^2}{12}
\end{eqnarray}
where $\nu_{\tilde{I}}\equiv i\tilde{\nu}_{\tilde{I}}$, and $h_q^c$
is found from the condition $\nu_{\tilde{I}}=0$. Generalizing for
$m\neq 0$ we get
\begin{eqnarray}
&& \nu_{\tilde{I}}=\sqrt{\frac{2}{3}(\lambda^2+m^2)-\frac{\mu_q^2}{9}},\label{nu-I}\\
&& h_q^c=-\frac{m^2}{2}+\frac{\mu_q^2}{12}, \label{nu}
\end{eqnarray}
in dimensionless units.

Now consider the scaling behavior near the quantum critical point,
$h\approx h_c$ or $G\approx G_c$ (solid red line in the phase
diagram Fig.(\ref{plot-phase-diagram}). As in Ref.\cite{Iqbal:2010},
imposing the Dirichlet boundary condition
$\tilde{I}(\zeta=\zeta_{IR})=0$ gives an oscillatory solution
of eq.(\ref{equationI0})
\begin{eqnarray}
I(\zeta)=\sin\left(\tilde{\nu}\log\frac{\zeta}{\zeta_{UV}}\right),
\end{eqnarray}
where $\zeta_{UV}$ is the location of the boundary of the AdS${}_2$ throat. In order to satisfy the boundary condition we should have
\begin{equation}
\tilde{\nu}\log\frac{\zeta_{IR}}{\zeta_{UV}}=\pi.
\end{equation}
According to the discussion in section IV of Ref.\cite{Iqbal:2010}, this
means that a new IR scale is generated
\begin{equation}
\Lambda_{IR}\sim \frac{1}{\zeta_h}\sim\mu
\rm{exp}\left(-\frac{\pi}{\tilde{\nu}}\right),
\end{equation}
where $\mu$ is the UV scale, that leads to the infinite order BKT
scaling behavior
\begin{eqnarray}
T_c\sim \mu \rm{exp} \left(-\frac{C}{\sqrt{h_q^c-h_q}}\right),\;\;
\Delta \sim \mu \rm{exp} \left(-\frac{C}{2\sqrt{h_q^c-h_q}}\right),
\label{scaling}
\end{eqnarray}
with $C=\frac{\pi}{\sqrt{4/3}}$ and $h_q^c$ given by eq.(\ref{nu}).
The factor of $2$ in the exponent comes from the difference in
operator dimensions in the intermediate conformal regime: the
current $I$ scales as a dimension $1/2$ operator and the temperature
scales with dimension $1$. Eq.(\ref{scaling}) describes the behavior
below the critical magnetic field $h<h_c$, which can be seen in
Fig.(\ref{plot-tcvsh}). Since $h_q=hq$, increasing the charge $q$
would produce higher curves.

Choosing the mass $m$ as a tuning parameter, we obtain the infinite
order BKT scaling behavior from the condition $\nu_{\tilde{I}}=0$ in
eq.(\ref{nu})
\begin{eqnarray}
T_c\sim \mu
\rm{exp}\left(-\frac{C^{\prime}}{\sqrt{m_c^2-m^2}}\right),\;\;
\Delta \sim \mu
\rm{exp}\left(-\frac{C^{\prime}}{2\sqrt{m_c^2-m^2}}\right),
\label{scaling2}
\end{eqnarray}
with $C^\prime=\frac{\pi}{\sqrt{2/3}}$ and
$m_c^2=-2h_q+\mu_q^2/6$. The scaling behavior from
eqs.(\ref{scaling}-\ref{scaling2}) describes the BKT regime found
also for the condensation of a scalar field in Ref.\cite{Iqbal:2010},
with the condensed phase for $h<h_c$ (or at $m^2<m_c^2$) and the
normal state with zero condensate at $h>h_c$ (or at $m^2>m_c^2$).

While the above analysis fits well into the results we have found
for the normal branch, the anomalous branch, where at high $h>h_c$
the magnetic field catalyzes and enhances the condensate is still to
be explained. The scaling behavior in this region is given by
\begin{eqnarray}
T_c\sim \Delta\sim |h-h_c|^\delta, \label{scaling3}
\end{eqnarray}
where $\delta>1$. In Figs.(\ref{hdep1}A,\ref{plot-tcrit}A), a
sharp increase with $h$ is found, which is in agreement with field
theory calculations of magnetic catalysis Ref.\cite{Shovkovy:2d} and
experiments on graphite in strong magnetic fields
Ref.\cite{Checkelsky:2009}. We leave the explanation of this regime
within the AdS${}_2$ analysis for further work.

For $m=0$, the equation of motion for $I$ can be reduced to a
Schr\"{o}dinger-like equation also in the general AdS${}_4$ case.
This is what we will do in the next subsection.

\subsection{The $m=0$ formalism}

As elucidated before in a slightly different context
Ref.\cite{Faulkner:2009}, nonzero contributions to the current
(corresponding to the quasiparticles at the boundary) are quantified
by counting the bound states at zero energy for the formal
wavefunction $I_-$ of the above equation. An important novel feature
in our setup is that the momentum is quantized due to the magnetic
field, and thus we cannot use the usual quasiclassical (WKB) formalism. Still, in the
massless limit we will be able to gain some more insight by
constructing an effective Schr\"{o}dinger equation with a formal WKB
momentum, that can be studied analytically.

Notice first that the RN geometry allows the spin connection term from eq.(\ref{spin-connection})
to be absorbed in the definition of the currents as it is a total
derivative Ref.\cite{Faulkner:2009}
\begin{equation}
\label{atotal}\mathcal{A}=\partial_z\left(-gg^{zz}\right)^{1/4}
\end{equation}
Upon implementing eq.(\ref{atotal}), the system of eqs.(\ref{spinor}) for
$m=0$ and in the static limit $\omega\to 0$ is simplified to
\begin{subequations}
\label{cursys}
\begin{eqnarray}
\label{cursys1}
e_{\hat{z}}\partial_zJ_\pm+2\Delta K_\mp+2e_{\hat{i}}\lambda I_\pm=0\\
\label{cursys2}
e_{\hat{z}}\partial_zI_\pm+2e_{\hat{t}}\Phi K_\pm+2\lambda e_{\hat{i}}J_\pm=0\\
\label{cursys3}
e_{\hat{z}}\partial_zK_\pm-2e_{\hat{t}}\Phi I_\pm+2\Delta J_\mp=0,
\end{eqnarray}
\end{subequations}
where the vierbeine of the metric eq.(\ref{geometry}) are
$e_{\hat{z}}=(1-z)\sqrt{f}$, $e_{\hat{t}}=(1-z)/\sqrt{f}$,
$e_{\hat{i}}=(1-z)$, and the scalar potential is rescaled as
$q\Phi\to \Phi$ to absorb $q$. As before, the magnetic field is
implemented by rescaling the chemical potential and the fermion
charge as given by eq.(\ref{mapping}), meaning that we can put
$\lambda=0$. The expectation values are given by the minus
component, with only three coupled equations for $J_-,K_-,I_-$
remaining to be solved. In order to understand the phenomenology of
the bulk pair current, it is useful to eliminate $J_-$ from
eq.(\ref{cursys}). Rescaling $I_-$ as
\begin{equation}
\label{jmresc} I_-\mapsto\tilde{I}_-\equiv
I_-\frac{e_{\hat{t}}\Phi}{e_{\hat{z}}}\equiv I_-\tilde{\Phi}
\end{equation}
we first easily eliminate $J_-$ and differentiate
eq.(\ref{cursys2}) with respect to $z$. The derivative
$\partial_zK_-$ can be expressed from eq.(\ref{cursys3}) and $K_-$
from eq.(\ref{cursys2}). In this way we arrive at the second order
equation involving $I_-$ only and having the form of the
Schr\"{o}dinger equation for $\tilde{I}_-$
\begin{equation}
\label{jscheq}\partial_{zz}\tilde{I}_--\left[\frac{2\partial_z\tilde{\Phi}}{\tilde{\Phi}}+4\tilde{\Phi}^2
-\Delta\partial_{zz}\log\tilde{\Phi}\right]\tilde{I}_-=0.
\end{equation}
Notice that the term containing the first derivative vanishes
automatically due to the transform eq.(\ref{jmresc}).

\begin{figure}[ht!]
\begin{center}
(A)\includegraphics[width=5cm]{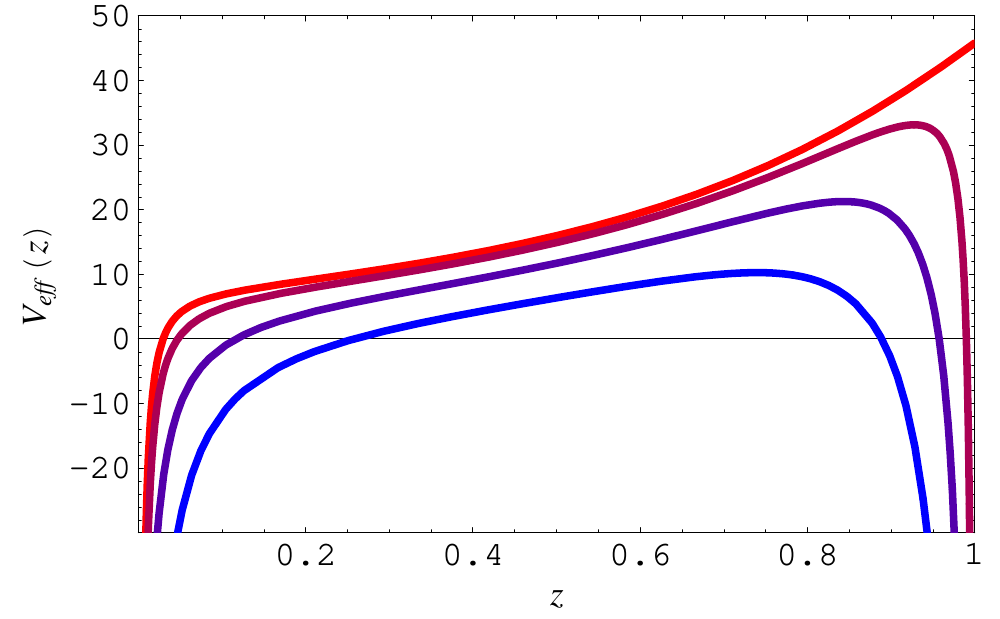}
(B)\includegraphics[width=5cm]{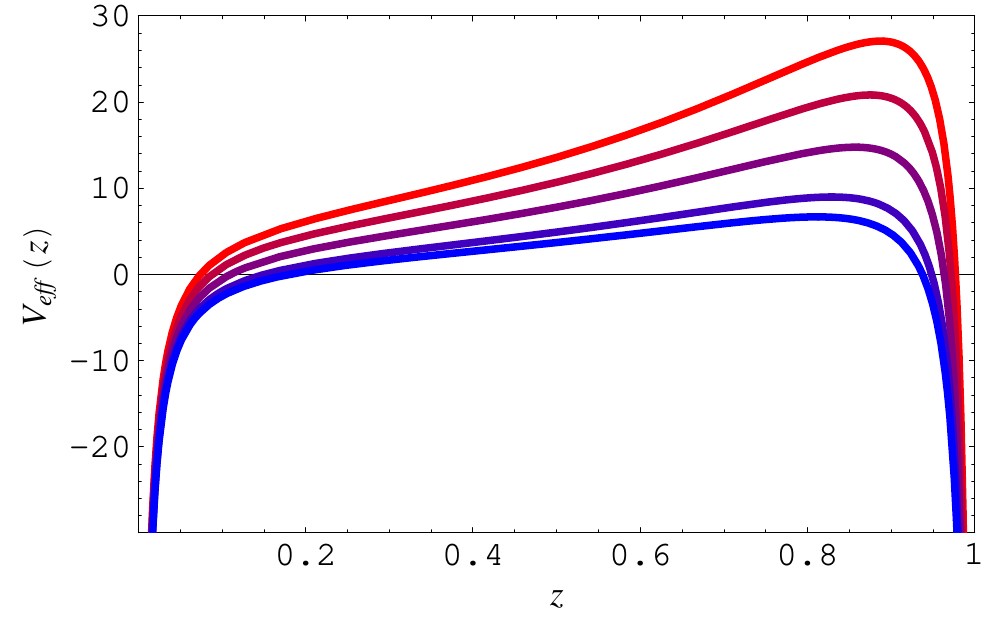} \caption{Effective
potential for the current $\tilde{I}_-$ for $m=0,q=2,T=0.001\times
10^{-3}$ and $h=0$, $G_{int}=0,1/3,2/3,1$ (red to blue, A)
and $G_{int}=0.2$, $h=0,0.50,1.00,1.50,1.71$ (red to blue, B). The
pairing interaction opens the near-boundary gap (A), which gets
wider but shallower as the magnetic field increases (B). The
competition between the broadening and the shallowing effect gives
rise to the transition between the normal and anomalous regime at
$h=h_c$.} \label{figpot}
\end{center}
\end{figure}

We are interested in the behavior of the current in the limit
$z\to z_0=1$. While the Schr\"{o}dinger formulation might in some
cases be more convenient also for computational reasons, the real
benefit is that we can use a formal WKB scheme to arrive at
surprisingly accurate solutions without solving the differential
equation. Eq.(\ref{jscheq}) has the form
$(\partial_{zz}-V_{eff}(z))\tilde{I}_-=0$, where the effective
potential obeys the inverse square law near the boundary (we also
use the relation eq.(\ref{deltagrel}))
\begin{equation}
\label{wkbpot}
V_{eff}(z\to 1)=\delta^2(1-z)-\frac{8\Delta}{(1-z)^2}+\mu^2(1-z)^2+O(1)
\end{equation}
where $\Delta\equiv\Delta(z\to z_0)\approx\mathrm{const.}$:
although, strictly speaking, one needs to compute $\Delta$
self-consistently given the value of $G_{int}$, for qualitative
considerations we may assume a constant $\Delta$ proportional to
$G_{int}$. The formal squared Dirac delta function is there to
enforce the condition $J_\pm(z_0)=I_\pm(z_0)=K_\pm(z_0)=0$.
The typical appearance of the potential is given in Fig.(\ref{figpot}).
The development of the electron-hole condensate can be seen as the
accumulation of bound states inside the potential well,
analogously to the similar logic for electron states in Fermi and
non-Fermi liquids, elucidated in Ref.\cite{Faulkner:2009} and applied
in Ref.\cite{Leiden:2011}. We can easily visualize our findings on the
transition points $h=h_\star$ and $h=h_c$ by looking at the
potential Fig.(\ref{figpot}). In the figures, we have left out the
Dirac-delta squared spike at the boundary, as it is completely
localized and only ensures that the currents reach zero at $z=1$,
exerting no influence on the behavior at small but finite $1-z$
values. Importantly, the near-boundary gap opens with $\Delta>0$,
supporting the electron-hole pair condensate near the boundary.
The influence of the magnetic field through the relation $q\mapsto
q\sqrt{1-H^2/(Q^2+H^2)}$ is subtler: it makes the potential well
both broader and shallower. The former generally facilitates the
formation of bound states, while the latter acts against it. It is
this competition that gives rise to the transition from the normal
toward the anomalous region at $h=h_c$.

\begin{figure}[ht!]
\begin{center}
\includegraphics[width=10cm]{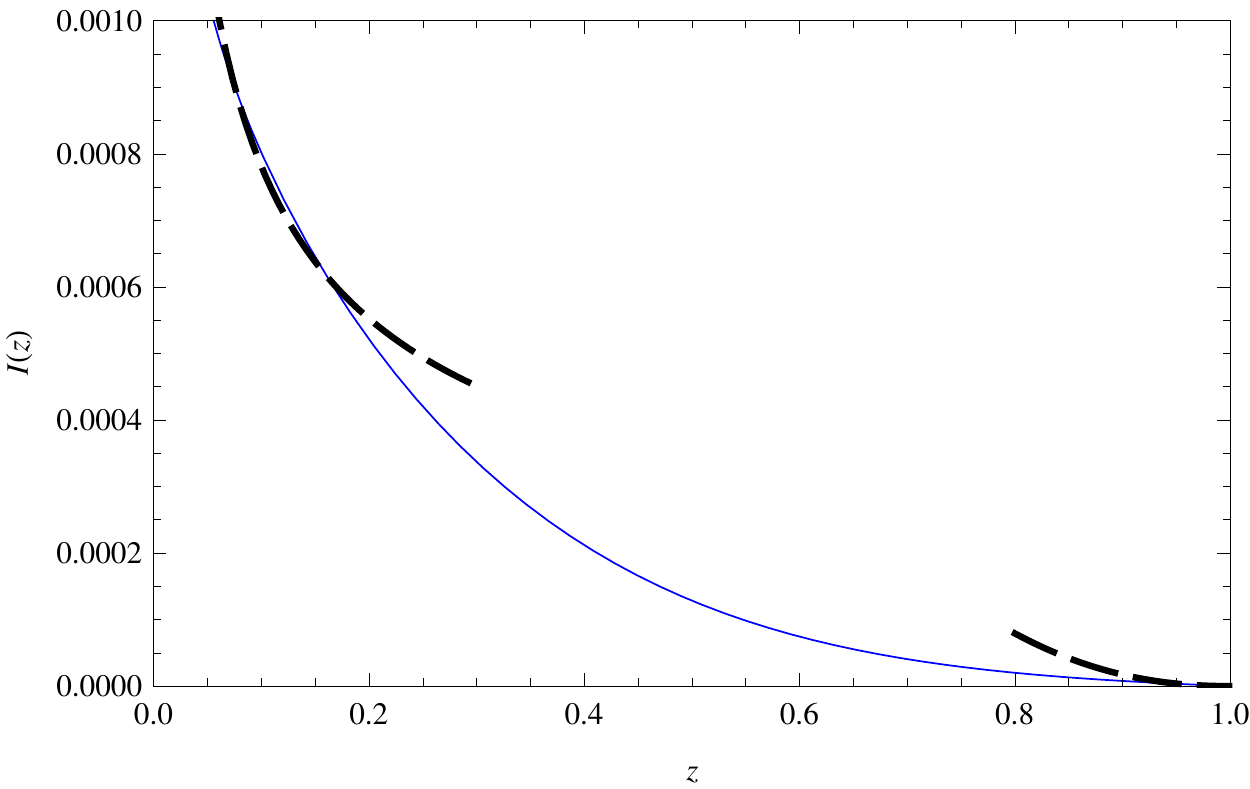}
\caption{Radial profile of the excitonic condensate (solid line)
from the numerical solution of the effective Schr\"odinger equation. The transformation law between radial coordinates
is $r=\frac{1}{1-z}$.
The
fits (dashed lines) are the asymptotic behaviors $(1-z)^3\to
\frac{1}{r^3}$ at $z\sim 1$ in the UV and $\frac{1}{z}\to \frac{1}{r-1}$ at $z\sim 0$ in the
IR.} \label{plot-radial-profile}
\end{center}
\end{figure}

Within the WKB approximation, the solution to eq.(\ref{jscheq}) can
be written as
\begin{equation}
\label{jscheqsol}\tilde{I}_-(z)=
\frac{(1-z)^2}{\sqrt{V_{eff}(z)}}\left(\exp\left(-\sqrt{-V_{eff}(z)}\right)+\exp\left(\frac{3\pi}{4}i
-\sqrt{V_{eff}(z)}\right)\right).
\end{equation}
We have constructed the solution by equating the WKB expansion with
the near-boundary expansion (eqs.(\ref{asymptot1},\ref{asymptot2})).
Notice that the phase shift is $3\pi/4$ instead of the usual
$\pi/4$, as the boundary itself provides an additional $\pi/2$ shift
due to the condition $I_{-}(z_0)\to 0$. The radial profile of the
condensate is depicted in Fig.(\ref{plot-radial-profile}). It can be
shown to have $\frac{1}{r^3}$ behavior at the UV boundary $r\to
\infty$, and it diverges as $\frac{1}{r-1}$ at the horizon in the IR
$r\to 1$.
We obtain the same asymptotic behavior when $\Delta=0$ in eq.(\ref{wkbpot}),
but we impose the hard wall near the horizon in the IR, which brings us in agreement with the results of Ref.\cite{Stefano-Bolognesi}.
The UV behavior follows from the boundary condition on the
fermion currents at the AdS boundary (putting the source term to
zero) and the appearance of a fermion mass gap,
to be discussed in more
detail later.

\begin{figure}[ht!]
\begin{center}
(A)\includegraphics[width=8cm]{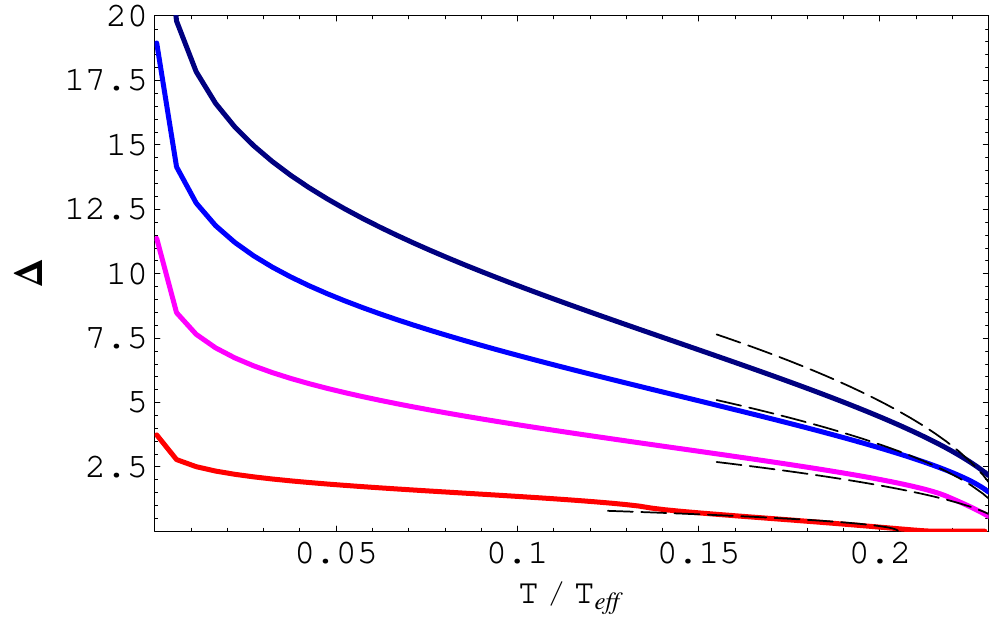}
(B)\includegraphics[width=8cm]{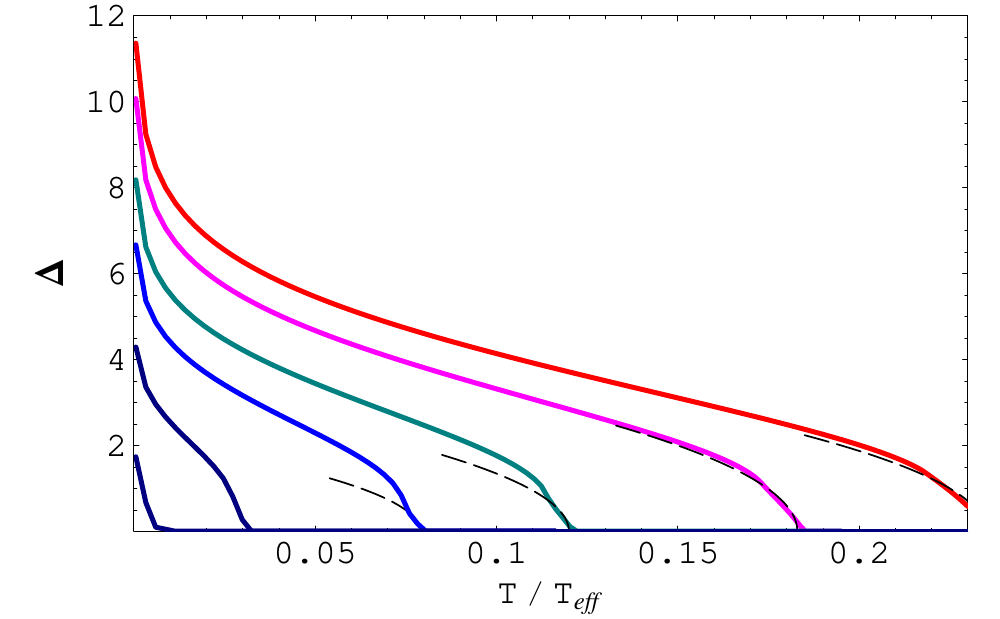} \caption{Order
parameter of the pair density $\Delta$ vs the temperature (all in
dimensionless units) (A) for $h=1$ and different values of the
coupling strength $q=1,3,5,7$ (red, magenta, blue, black) and (B)
for $q=3$ and different values of the magnetic field
$h=0,0.8,1.2,1.4,1.6,1.7$ (red to black). Pairing is favored in
the overdamped phase, with stable quasiparticles for $q\gg 1$ and
$\nu\sim 1$, and suppressed at very high magnetic fields when the
effective chemical potential is lowered and thus only a small
number of electrons is available for pairing.} \label{tempdepm0}
\end{center}
\end{figure}

\begin{figure}[ht!]
\begin{center}
(A)\includegraphics[width=8cm]{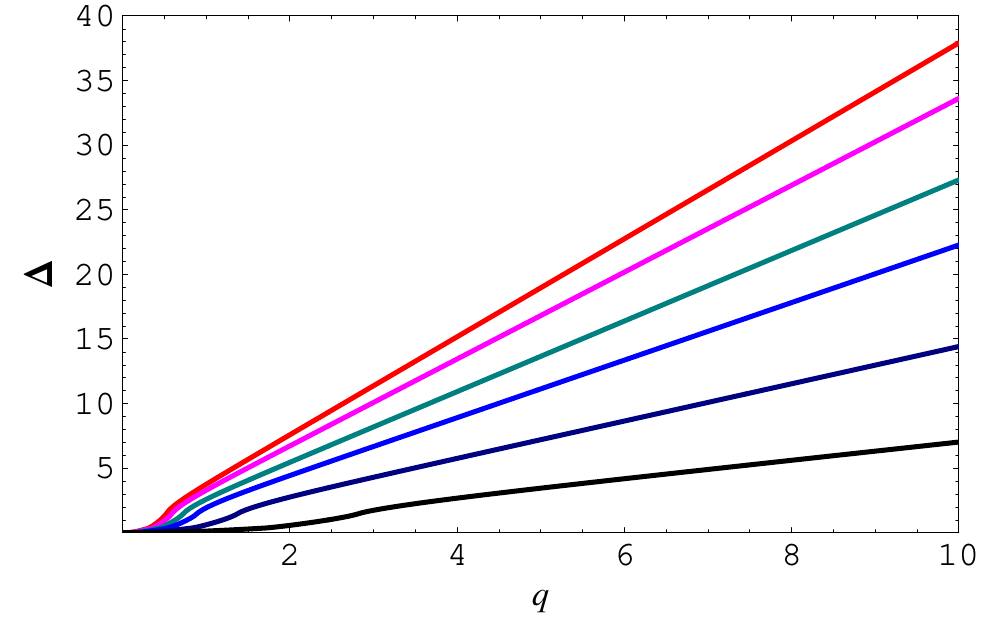}
(B)\includegraphics[width=8cm]{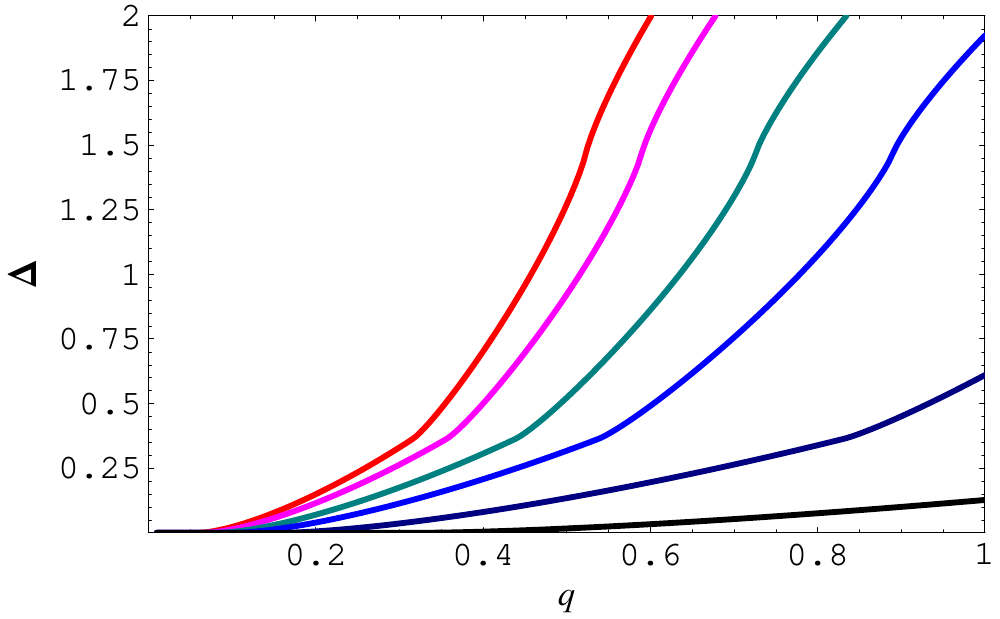} \caption{(A) Order
parameter of the pair density $\Delta$ vs. the fermion charge $q$,
for $T=5.6\times 10^{-4}$ and different values of the magnetic field
$h=0,0.8,1.2,1.4,1.6,1.7$ (red to black). The critical value $q_c$
is shifted again due to the shifting of the effective potential. (B)
Zoom-in near $q=0$ to better appreciate the transition.}
\label{gdephm0}
\end{center}
\end{figure}

Another advantage of the Schr\"{o}dinger approach is that solving
the Schr\"{o}dinger equation numerically is easier than solving
the current equations. In Figs.(\ref{tempdepm0}-\ref{gdephm0}) we
give the dependences $\Delta(h)$ and $\Delta(q)$, produced by
solving the equation (\ref{jscheq}). Qualitatively similar
behavior is seen in both cases. The WKB approach makes it
feasible to study also the dependence on the fermion charge $q$.
Fig.(\ref{gdephm0}) already shows that there is a critical value
$q=q_c$ below which no pairing can occur at all. We conjecture
that this value corresponds to $\nu<1/2$, i.~e. only stable
quasiparticles can pair up. While plausible, this is not easy to
see from the relations $\Delta(h)$ and $\Delta(T)$ that we
obtained in the $m\neq 0$ case.

\section{Spectra and the pseudogap}

In this section we will compute the spectra for the fermionic system with particle-hole
pairs. We invoke again eqs.(\ref{Dirac}) to derive the equations of
motion for the retarded propagator, which will directly give us the
spectral function as $A(\omega,k)=\rm{Im} G_R$.

Following Ref.\cite{Faulkner:2009}, we can write a single nonlinear
evolution equation for $G_R$. It will generically be a matrix
equation, due to the additional, pairing channel. Of course, we
can rewrite it as a system of four scalar equations for the four
components of the bispinor. We adopt the basis given in
eq.(\ref{matrices}) and the metric given by eq.(\ref{geometry}).
Introducing the notation $\psi=(\psi_1,\psi_2)^{T}$ with
$\psi_{\alpha}=(y_{\alpha},z_{\alpha})^T$ where $\alpha=1,2$, the resulting system reads
\begin{subequations}
\label{probeqs}
\begin{eqnarray}
\label{probeqs1}\left(\partial_z\mp m\sqrt{g_{zz}}\right)y_{1;2}=\mp i\sqrt{\frac{g_{zz}}{g_{ii}}}(\lambda-u)z_{2;1}-\Delta\sqrt{g_{zz}}y_{1;2}=0\\
\label{probeqs2}\left(\partial_z\mp
m\sqrt{g_{zz}}\right)z_{1;2}=\pm
i\sqrt{\frac{g_{zz}}{g_{ii}}}(\lambda+u)y_{2;1}+\Delta\sqrt{g_{zz}}z_{1;2}=0,
\end{eqnarray}
\end{subequations}
with
\begin{equation}
\label{probeu}
u=\sqrt{\frac{g_{ii}}{-g_{tt}}}\left(\omega+q\Phi(z)\right).
\end{equation}
Introducing $\xi_{\alpha}=iy_{\alpha}/z_{\alpha}$ as in
Ref.\cite{Faulkner:2009}, where the boundary Green's function is found
from the asymptotics of the solution at the boundary (eqs.(\ref{asymptotics}))
\begin{eqnarray}
G_{\alpha}=\lim_{z\to 1} \left(\frac{1}{1-z}\right)^{2m}\xi_\alpha(z)=
\lim_{\epsilon\to 0}\epsilon^{-2m}\xi_\alpha(1-z=\epsilon),
\end{eqnarray}
the equations of motion for $\xi_\alpha$ become
\begin{eqnarray}
\label{xieq}
\partial_z\xi_{1;2}=-2m\sqrt{g_{zz}}\xi_{1;2}
-\sqrt{\frac{g_{zz}}{g_{ii}}}(\lambda-u)
+\sqrt{\frac{g_{zz}}{g_{ii}}}(\lambda+u)\xi_{1;2}^2\mp
2\Delta\sqrt{g_{zz}}\xi_{1;2}
\end{eqnarray}
The infalling boundary conditions at the horizon are imposed
$\xi_\alpha= i$, while the amplitude of $y_\alpha$ remains
free (it cancels out in the propagator $G_R$) and can be chosen to be of
order unity for convenience in the numerical integration.

With no pairing channel, the morphology of the spectra is well known
and has been analyzed in detail in Refs.\cite{Faulkner:2009,Leiden:2010}:
near $k=k_F$, gapless quasiparticle excitations appear, belying a
Fermi surface. Let us now repeat the AdS${}_2$ analysis of
Ref.\cite{Faulkner:2009} for the equations with pairing. We will use the $(\zeta,\tau)$ coordinates introduced in the eq.(\ref{ads2coor}).
The near-horizon equation of motion now assumes the following form
\begin{equation}
\label{ads2eq}
\zeta\partial_\zeta\psi=\left(i\sigma^2\frac{\mu_q}{6}-\sigma^3\frac{(m+s\Delta)}{\sqrt{6}}
-\sigma^1\frac{k}{\sqrt{6}}\right)\psi,
\end{equation}
where $s=\pm 1$, and in the presence of magnetic field the role of
the momentum $k$ is taken over by Landau levels
$\lambda=\sqrt{2|qh|l}$. Near the AdS${}_2$ boundary ($\zeta\to 0$),
the equation can be solved analytically at the leading order
\begin{equation}
\label{ads2sol}
\psi=A \left(\begin{matrix}\frac{m+s\Delta}{\sqrt{6}}+\nu
\\ \frac{k}{\sqrt{6}}+\frac{\mu_q}{6}\end{matrix}\right)\zeta^{-\nu}
+B \left(\begin{matrix}\frac{m+s\Delta}{\sqrt{6}}-\nu
\\ \frac{k}{\sqrt{6}}+\frac{\mu_q}{6}\end{matrix}\right)\zeta^{\nu}
\end{equation}
with
\begin{equation}
\label{ads2nu}
\nu=\frac{1}{6}\sqrt{\mu_q^2-6\left(\left(m+s\Delta\right)^2+k^2\right)},
\end{equation}
and the self-energy scales as
\begin{equation}
\rm{Im}\Sigma\sim \omega^{2\nu}.
\end{equation}
As usual, the Fermi surface is stable for $\nu^2>1/4$, unstable for
$\nu^2<1/4$ and nonexistent for $\nu^2<0$.

In the bulk (and also as we move toward the boundary), the pairing
term acts by shifting the mass as $m\mapsto m\pm\Delta$, meaning
that the position of the quasiparticle pole is shifted, effectively
modifying the $k_F$ value, which removes the spectral weight from
the vicinity of $\omega=0$. It thus resembles a gap even though it
is, strictly speaking, not a gap since the poles in $\psi_{1}$ and
$\psi_{2}$ do not coincide (see also Ref.\cite{Faulkner_photoemission:2009}).
Nevertheless, we expect the size of the zero-weight region to be a
useful benchmark for the degree to which the pairing eats up the
(non-)Fermi liquid quasiparticles.

\begin{figure}[ht!]
\begin{center}
(A)\includegraphics[width=8cm]{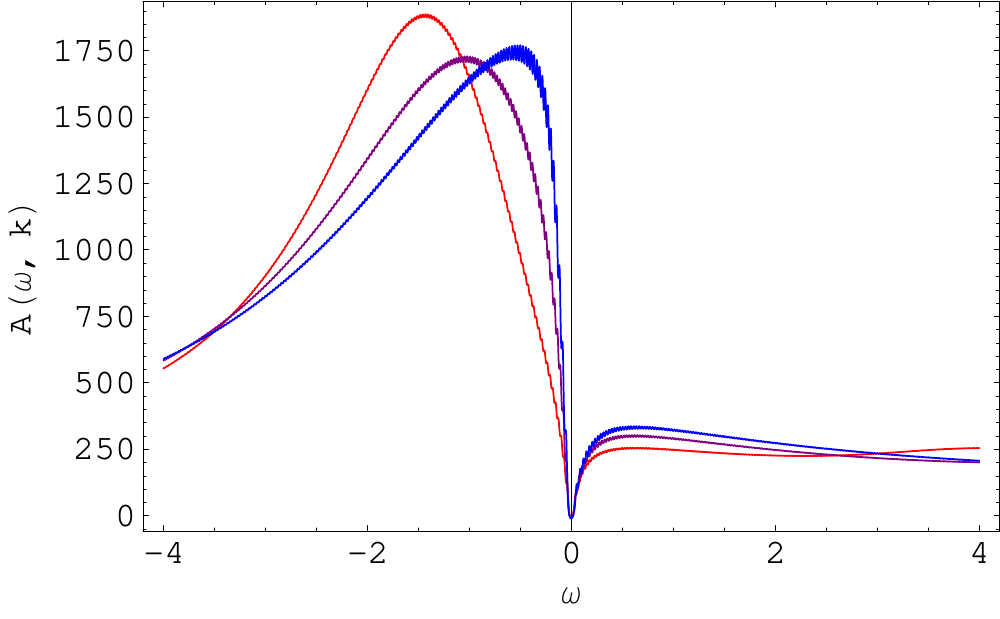}
(B)\includegraphics[width=8cm]{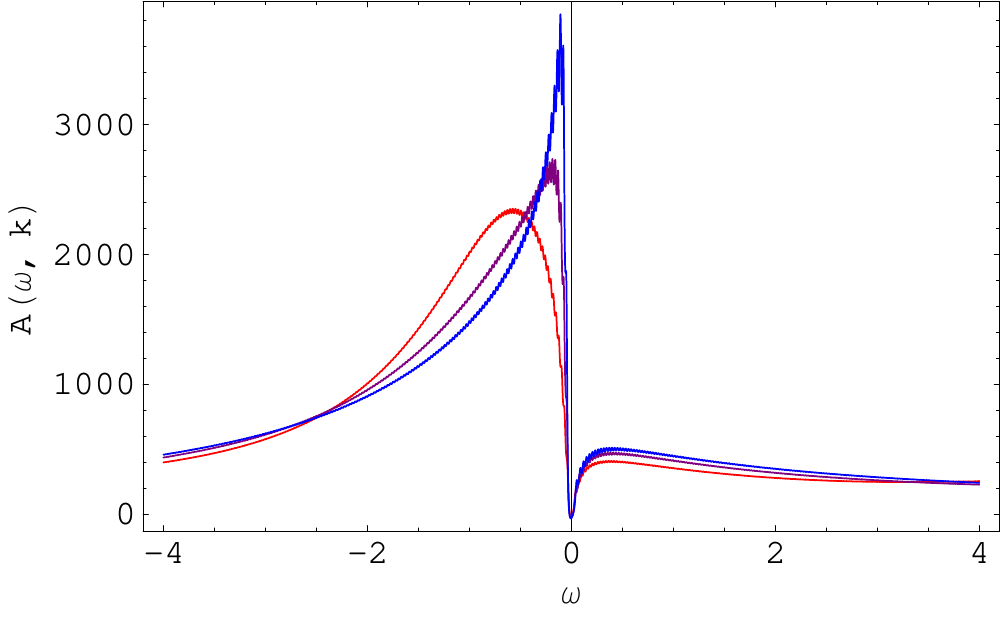}
(C)\includegraphics[width=8cm]{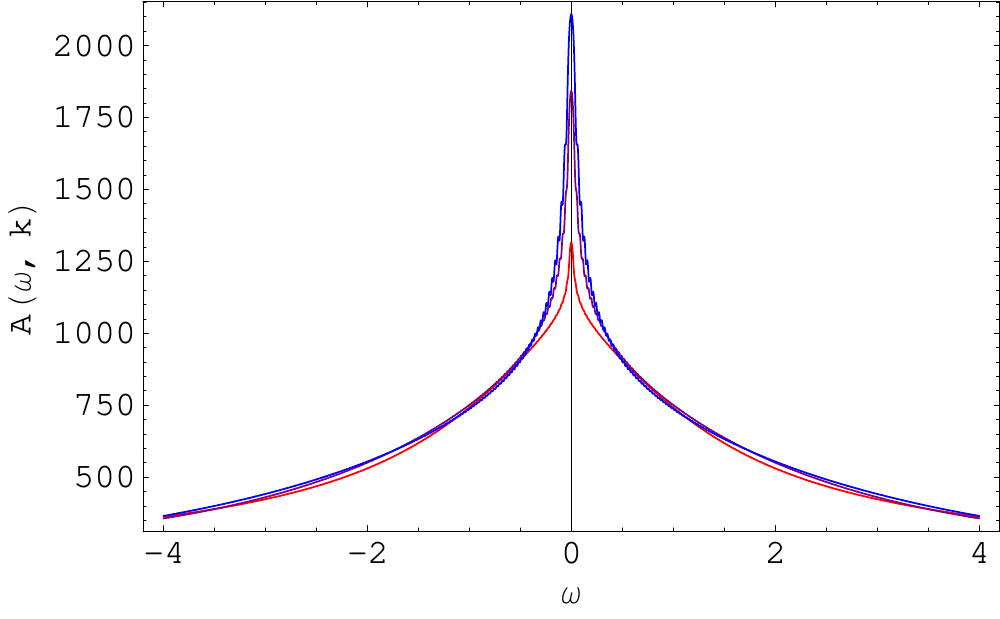}
(D)\includegraphics[width=8cm]{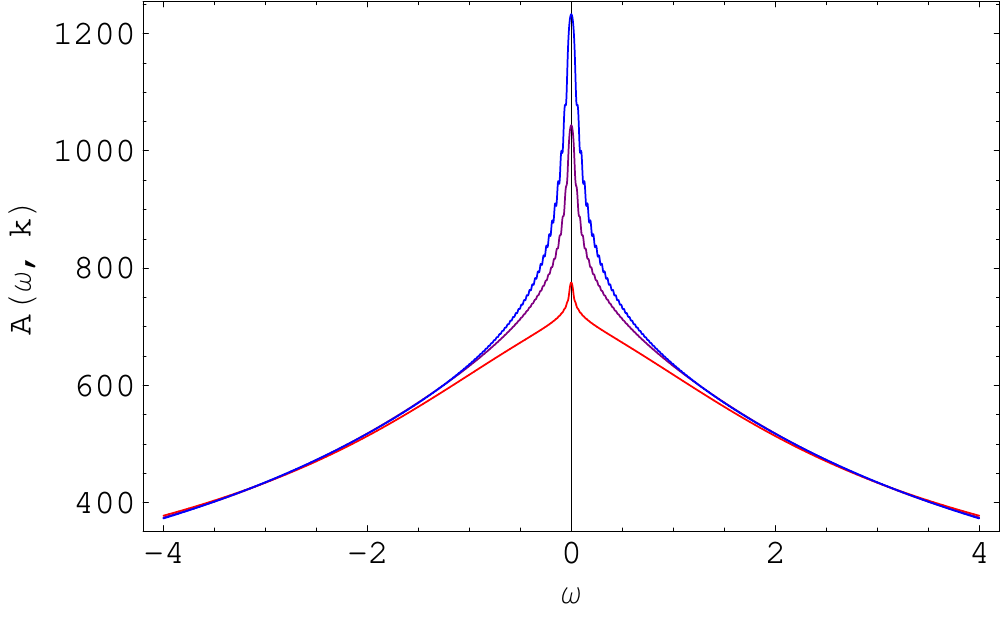}

\caption{The spectral function $A(\omega,\lambda)$ for $\Delta=0.2$,
$h=0.9,1.11,1.3,1.5$ (A,B,C,D) and three momentum values around
$k_F^\mathrm{eff}$. At $h<h_\star$ (A,B) we see that $\nu^2<0$,
corresponding to zero weight at $\omega\approx 0$, the phenomenon we
have dubbed the pseudogap. For $h>h_\star$ (C,D) we enter the
quasiconformal regime, with no Fermi surfaces left, the conformality
being only slightly broken by nonzero $\Delta$.} \label{spech}
\end{center}
\end{figure}

\begin{figure}[ht!]
\begin{center}
(A)\includegraphics[width=8cm]{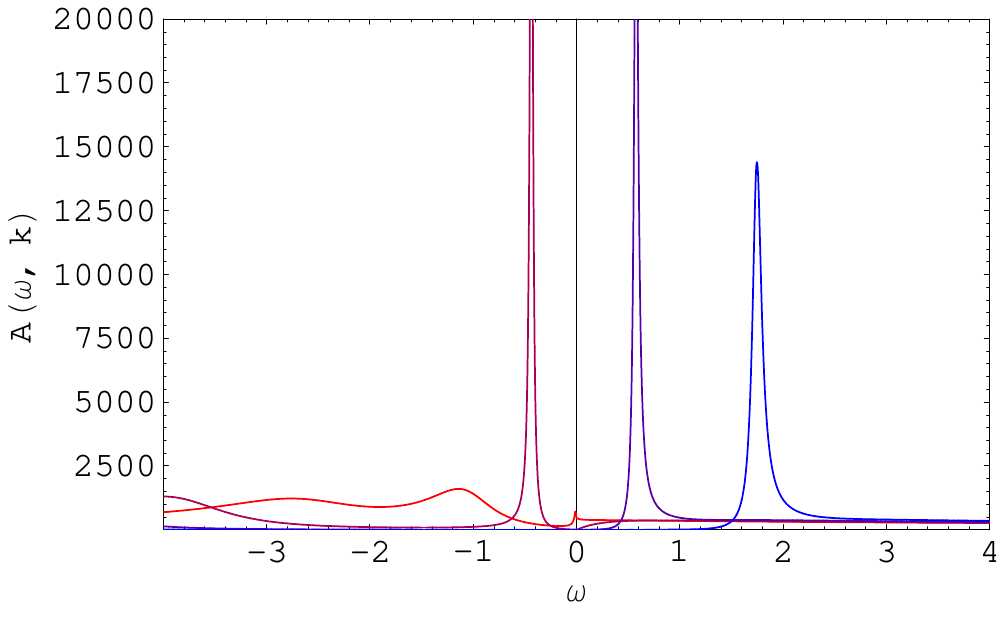}
(B)\includegraphics[width=8cm]{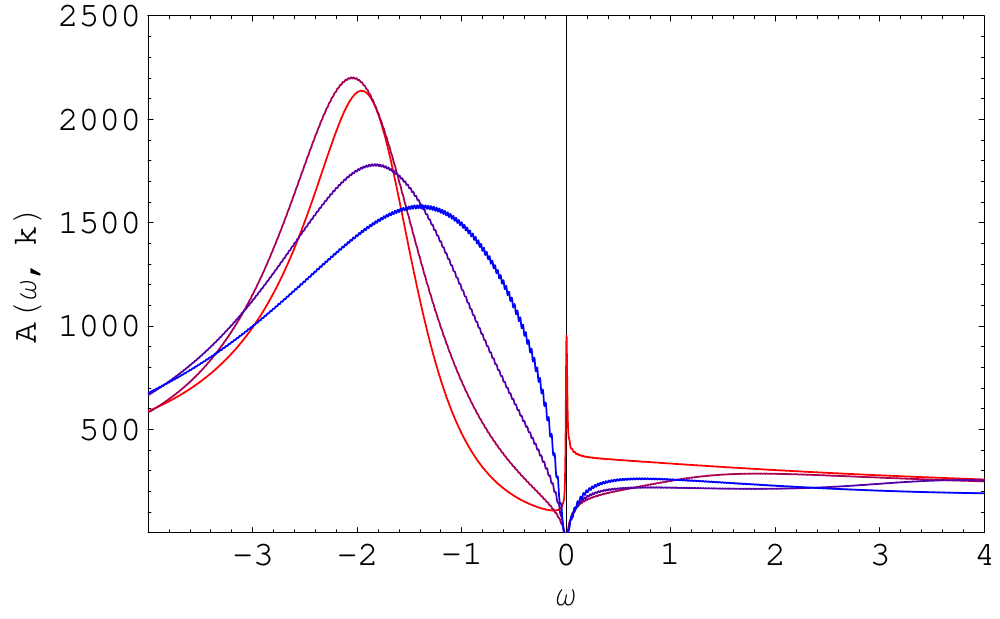}
(C)\includegraphics[width=8cm]{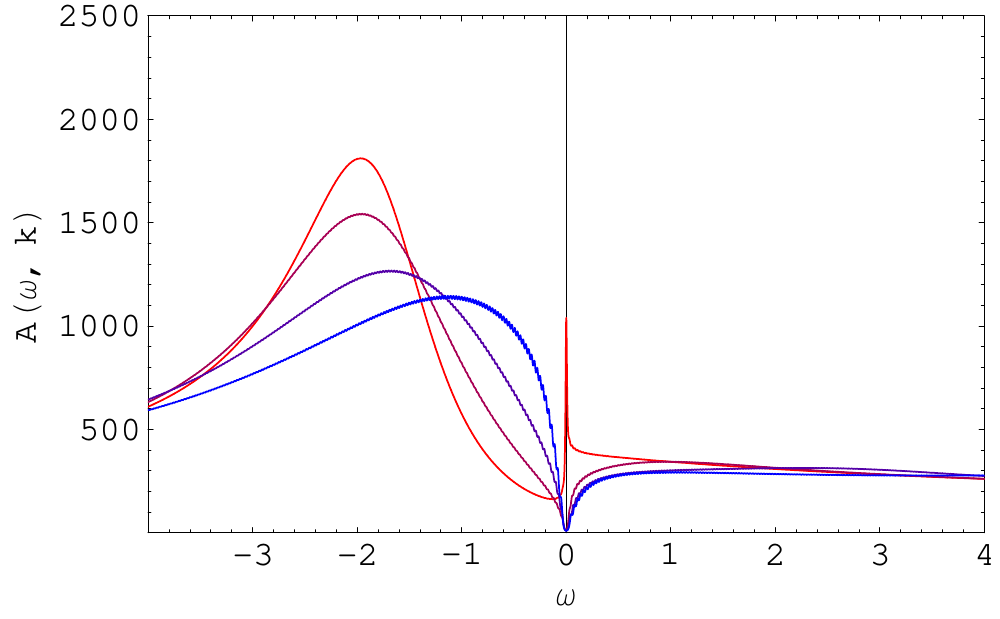}
(D)\includegraphics[width=8cm]{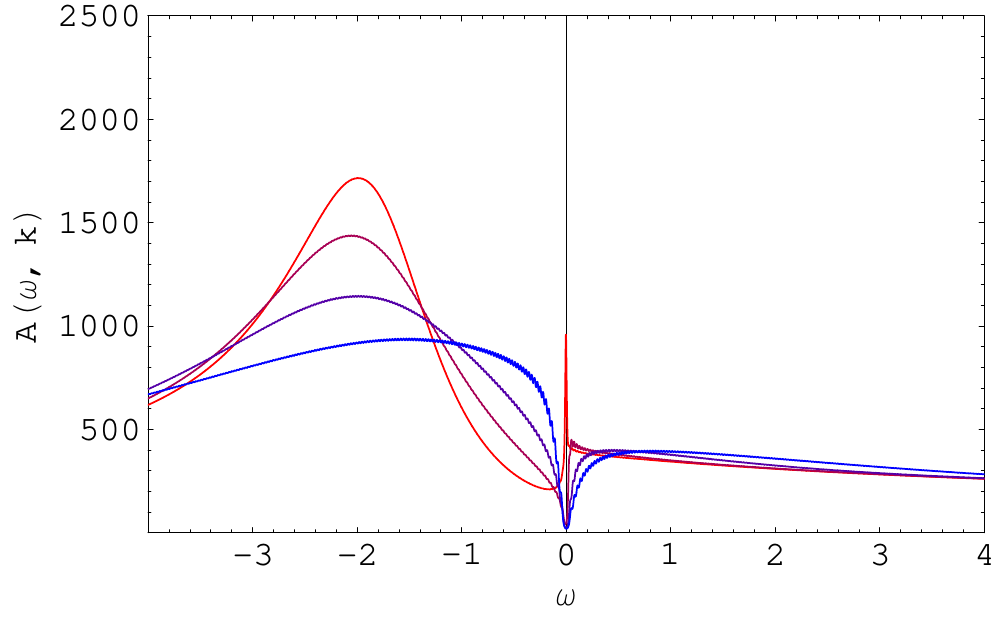}

\caption{The spectral function $A(\omega,\lambda)$ for $h=0.2$,
$\Delta=0.9,1.11,1.3,1.5$ (A,B,C,D) and four momentum values around
$k_F^\mathrm{eff}$. At $\Delta<\Delta_c\approx 0.2$ (A), the
quasiparticle peak survives; for higher $\Delta$ the influence of
the exact value of the pairing term is negligible, and the spectrum
always shows the pseudogap behavior.} \label{specdelta}
\end{center}
\end{figure}

The typical appearance of the spectrum is given in Fig.(\ref{spech}),
where we plot the spectra for $\Delta=0.2$ and for increasing
magnetic field values. Increasing the magnetic field leads to
destabilization of the quasiparticle (A,B), leading to a gap-like
behavior, destabilization of the quasiparticle as seen from the
asymmetry of the peak which loses its Fermi-liquid-like scaling.
Eventually (C,D) the effective chemical potential is so low that we
enter the "almost conformal" regime. Fig.(\ref{specdelta}) shows the
dependence on the pairing coupling: the peak at $\omega=0$ turns
into a dip, a "pseudogap" develops and we lose the quasiparticle.

\section{Discussion and conclusions}

Before concluding the paper, we will discuss possible universal
aspects of our findings, and show that the formation and enhancement
of the particle-hole condensate in a strong magnetic field is a
robust phenomenon seen in a number of distinct systems. We will
limit ourselves to short remarks only, as more detailed comparisons
with earlier work can be made by consulting the appropriate
references.

We found the exciton instability using a Dirac hair or bilinear approach. A Dirac hair method
uses bilinear combinations where a bilinear in a given channel develops an expectation value
at the UV boundary provided a source is switched off. 
Dirac hair is equivalent to a Tamm-Dancoff approximation (TDA),
planar diagrams of processes $2\rightarrow 2$ are included with no bulk fermion loops. 
In this sense, Dirac hair is a quantum mechanical treatment with one single classical wave function.
It is quite remarkable to see that the condensate develops on a "classical" level due to a nontrivial nature of the curved space-time with the help of the AdS/CFT dictionary, a phenomenon that was first obtained as a holographic superconductor Ref.\cite{Review:2010}.   

We have associated the rising critical temperature vs magnetic
field with the magnetic catalysis (MC), and the decreasing $T_c$ vs
$h$ with the inverse MC (anomalous and normal branches in
Fig.(\ref{plot-tcvsh}) for $G_c^{\star}<G<G_c^{\star\star}$,
respectively). We adopted the terminology from Ref.\cite{Schmitt:2010}.
It corresponds to a double-valued regime in the phase
diagram Fig.(\ref{plot-phase-diagram}). Similar behavior of
increasing $T_c$ vs. the scalar mass $m$ has been observed in
Ref.\cite{Iqbal:2010} under the action of a double-trace deformation,
for the alternative quantization starting at the critical mass
$m^2R^2\geq -\frac{27}{16}$. There it was associated with the formation
of a new condensed phase corresponding to the high temperature
regime. However, it was suggested that the high-T condensed phase is
thermodynamically unstable Ref.\cite{Iqbal:2010}. Likewise, in
Ref.\cite{Wang:2011}, exploring the phase diagram for a nonrelativistic
conformal field theory, the authors found the high temperature
condensate for $T\geq T_H$. The similarity of the dependences $\langle
O\rangle(T)$ at different chemical potentials $\mu$ and $T_c(h)$ at
different couplings $G$ to our Fig.(\ref{plot-tcvsh}) is obvious.
In that work, the high temperature condensate was related to the
high temperature instability predicted by Cremonesi et al.
Ref.\cite{Cremonesi}, and it was found to be thermodynamically
disfavored over the trivial vacuum by direct calculation of the
difference in the free energies Ref.\cite{Wang:2011}. However, the
particle-hole condensate found at high magnetic fields in our case
is crucially different from the unstable high temperature condensate
in Refs.\cite{Iqbal:2010,Wang:2011}. Though naively both the
magnetic field $h$ and the fermion mass $m$ destroy the condensate,
increasing $m^2$ (or $h$) drives the bulk system to the UV(or the IR). Indeed, from the
radial profile of the wave functions: at large $m$ the system
resides near the UV boundary and at strong $h$ it resides near the
RN black hole horizon in the IR, Fig.(5) in Ref.\cite{Leiden:2010}.
Therefore, from holographic viewpoint large magnetic fields can lead
to low-energy behavior and possible quantum critical phenomena,
involving different ordering in the system. The main argument in
favor of robustness and stability of our high-$h$ condensate is
provided by the magnetic catalysis effect. In strong magnetic
fields only the lowest Landau level contributes significantly to the
ground state. Therefore, the dynamics is effectively dimensionally
reduced as $d\to d-2$. In field theory this dimensional reduction
leads to an increase in the density of states Ref.\cite{Shovkovy:1994}
or in QCD to one-gluon exchange with a linear binding potential
Ref.\cite{Chernodub:2010}, with both effects working towards pairing and
enhancement of the condensate. In the AdS space, dimensional reduction
leads to a Schwinger model showing an instability which is very
similar to the Bardeen-Cooper-Schrieffer (BCS) pairing instability, where also the dynamics is
effectively one-dimensional at the Fermi surface. The exact mapping
between the magnetic catalysis at $h\neq 0$ and the BCS Cooper
pairing at $\mu\neq 0$ has been established in Ref.\cite{Shovkovy:1994}.

We obtained a nontrivial radial profile and a boundary VEV for the bulk excitonic condensate $\langle \bar{\psi}\Gamma \psi\rangle$ at vanishing source,
with the relation
\begin{eqnarray}
\langle\bar{\psi}_1\psi_1\rangle = \frac{1}{2}\langle\bar{\psi}\psi\rangle - \frac{1}{2}\langle\bar{\psi}\Gamma\psi\rangle,\;\;
\langle\bar{\psi}_2\psi_2\rangle = \frac{1}{2}\langle\bar{\psi}\psi\rangle + \frac{1}{2}\langle\bar{\psi}\Gamma\psi\rangle,
\end{eqnarray}
where $\Gamma=i\Gamma^2\Gamma^5$ and $\psi_{1,2}=\frac{1}{2}(1\mp \Gamma)\psi$ are the eigenvalues of the Dirac operator eq.(\ref{general-Dirac}) (the projectors
$\Pi_{1,2}=\frac{1}{2}(1\mp \Gamma)$ are constructed out of gamma matrices which enter the Dirac operator only Ref.\cite{Faulkner:2009}).
We need to find the boundary condensate whose gravity dual is $\langle \bar{\psi}\Gamma \psi\rangle$ where the bulk Dirac field $\psi$ corresponds to a fermionic operator $\Psi$,
$\psi\to\Psi$.
The AdS/CFT correspondence does not provide a straightforward
way to match a double-trace condensate to a boundary operator, though only gravity dual single-trace fields are easy to identify with the operators at the boundary.
For example, in holographic superconductors a superconducting condensate is modeled by
a charged scalar field $\langle\Phi\rangle$ (see e.g. Ref.\cite{Horowitz:2009}). As in Ref.\cite{Stefano-Bolognesi}, we find a boundary operator by matching discrete symmetries
on the gravity and field theory sides and considering the asymptotic behavior of the gravity dual condensate at the boundary. As a result
we associate a gravity dual excitonic order to some sort of a chiral condensate
\begin{equation}
\langle\bar{\psi}\Gamma\psi\rangle\leftrightarrow \bar{\Psi}\Psi
\label{chiral-condensate}
\end{equation}
or some combination of condensates which break chiral symmetry. In Ref.\cite{Stefano-Bolognesi}, this strategy provided the correspondence:
$\langle\bar{\psi}\Gamma^5\psi\rangle\leftrightarrow \bar{\Psi}\Psi$. There an explicit use of the chiral basis $\psi_{L,R}=\frac{1}{2}(1\mp \Gamma^5)\psi$
and the relation $\bar{\psi}_L\psi_R=\frac{1}{2}\langle\bar{\psi}\psi\rangle + \frac{1}{2}\langle\bar{\psi}\Gamma^5\psi\rangle$
made the correspondence evident. Specifically, by matching symmetries with respect to the discrete transformations eq.(\ref{discrete_symmetry}) we found that
$\langle\bar{\psi}\Gamma\psi\rangle$ and $\langle\bar{\Psi}\Psi\rangle$ are pseudoscalars under parity and are unaffected by the charge conjugation, therefore they both spontaneously
break the combination $\hat{C}\hat{P}$-symmetry. This finding is consistent with the existence of the parity odd mass in graphene associated with the excitonic order parameter
in the $2+1$-dimensional effective field theory of graphene Refs.\cite{Shovkovy:2d,Semenoff:2011}.
Also the asymptotic behavior of the bulk condensate at the boundary, which was found numerically Fig.(\ref{plot-radial-profile}) to be
$\langle\bar{\psi}\Gamma\psi\rangle\rightarrow C/r^3$ as $r\rightarrow\infty$,  allows us to use a standard AdS/CFT dictionary to identify $C$ as the response or VEV of the boundary operator.
The third power in the decay exponent indicates an extra mass scale. Indeed provided the response $\langle\Phi\rangle\sim 1/r^3$, the gauge-gravity duality
gives a strong coupling form of the magnetic catalysis in $2+1$-dimensional Ref.\cite{Stefano-Bolognesi}
\begin{equation}
\langle\bar{\Psi}\Psi\rangle\sim h\,M_F,
\label{MC}
\end{equation}
with the magnetic field $h$ and mass gap $M_F$ Ref.\cite{Stefano-Bolognesi}.
It can be compared to the weak coupling field theory result $\langle\bar{\Psi}\Psi\rangle\sim h$
(we absorbed dimensional electric charge in the definition of magnetic field $h$, i.e. in $2+1$-dimensions the operator dimension is given by
$[e]=\frac{1}{2}$ and $[h]=\frac{3}{2}$ with $[eh]=2$ and therefore we substitute
$|e|h\rightarrow h$) Ref.\cite{Shovkovy:2d}.
Strong coupling realization follows from the anomalous fermion dimension $[\Psi]=\frac{3}{2}$ compared to the weak coupling conformal dimension $[\Psi]=1$ (free value dimension)
in $2+1$-dimensional field theory.
An extra fermion mass gap $M_F$ appears as a consequence of the dimensional four-fermion coupling $G_{int}= 1/M_F$ in the bulk or introduction of the IR cutoff thought of as a hard wall
at the radial slice $z_{\star}=1/M$. The authors of Ref.\cite{Stefano-Bolognesi} have used the hard wall construction to obtain the strong coupling realization of the magnetic catalysis
eq.(\ref{MC}). It is remarkable that the chiral condensate is proportional to the magnetic field even at strong coupling, that manifests the essence of the magnetic catalysis.

Another aspect of the chiral condensate is related to the Callan-Rubakov effect. As found in
the field theory and also shown in the context of the gauge-gravity duality Ref.\cite{Stefano-Bolognesi2}, the
chiral condensate can be spontaneously created in the field of a
magnetic monopole. Due to the chiral anomaly $\partial j_5=F\tilde{F}$,
the chiral symmetry is spontaneously broken and the
chiral condensate $\langle\bar{\psi}\psi\rangle\sim {\rm e}^{i\Theta}/r^3$
is generated in the field of a monopole.
In AdS, a construction involving a
monopole wall (more precisely, a dyonic wall) and light fermions in
the bulk produces an analog of the Callan-Rubakov effect resulting in
the formation of the chiral-symmetry-breaking (CSB) condensate: $\langle\bar{\psi}\psi\rangle\neq 0$
Ref.\cite{Stefano-Bolognesi2}. The scaling behavior of the condensate
is, however, different in our setup:  as pointed out before, due to
the lowest Landau level (LLL), the dimensional reduction $3d\to 1d$ takes place in the bulk.
This reduces the equation of motion to an effective Schr\"odinger
equation for the condensate, given by eq.(\ref{jscheq}) with
the potential eq.(\ref{wkbpot}). Solving the equation, we have found the
IR behavior of the condensate near the horizon of the RN black hole
($1/r$) to be less divergent than the one near the monopole
$\frac{1}{r^3}$ or the monopole wall $\frac{1}{(r-r_w)^3}$
Ref.\cite{Stefano-Bolognesi2}.

\begin{figure}[ht!]
\begin{center}
\includegraphics[width=10cm]{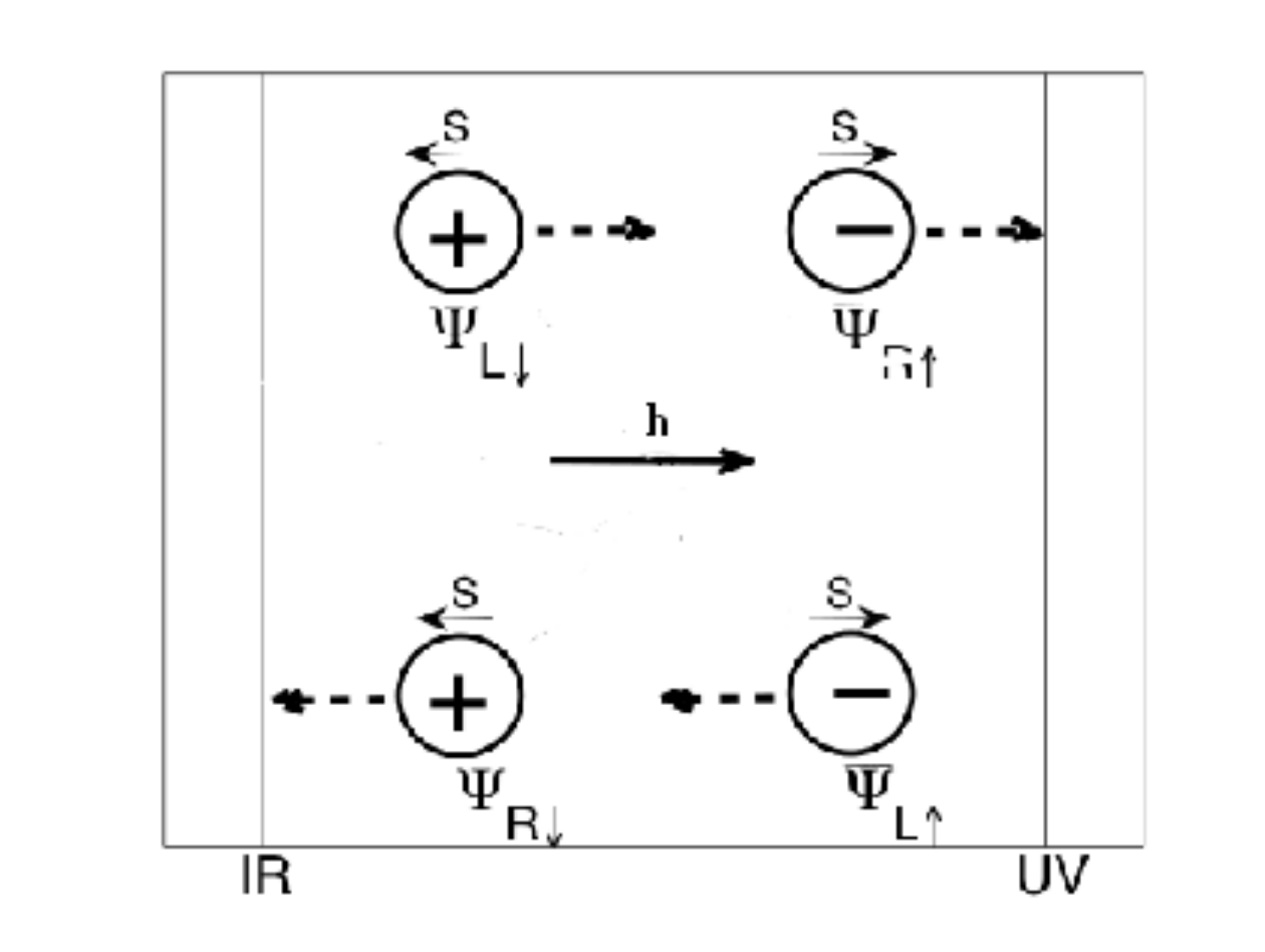}
\includegraphics[width=10cm]{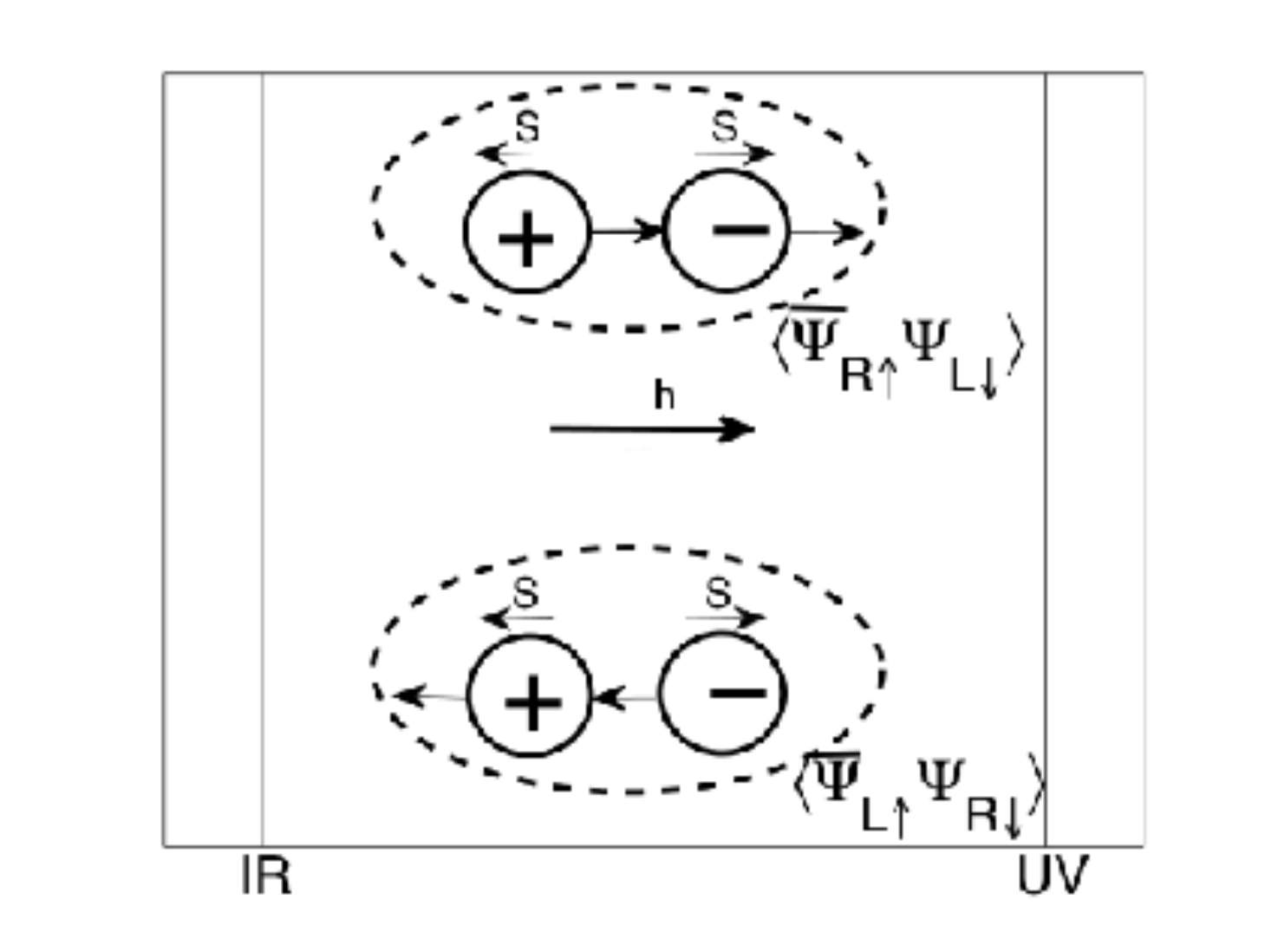}
\caption{Formation of the chiral symmetry breaking excitonic
condensate in the AdS space-time.
Individual "bouncing" events are shown schematically by the
dashed lines. In an individual event, helicity flips while spin and charge are conserved.} \label{plot-particle-hole}
\end{center}
\end{figure}

It turns out that the AdS space with two boundaries: the UV boundary and the IR hard wall, plays an additional role in
stabilizing the chiral condensate Refs.\cite{Stefano-Bolognesi,Stefano-Bolognesi2}.
It also provides an important hint for the interpretation of our current $\bar{\psi}\Gamma\psi$ in the boundary theory.
In particular, for the lightest states to condense,
we should take the LLL which only has one spin
state available (instead of two states available for the higher LLs). This means
that for a given charge, the spin direction is fixed. Therefore,
fixing the direction of motion and the charge fixes also the
helicity. Out of eight possibilities with a given charge, helicity
and direction, only four are available for the LLL, as depicted at
Fig.(\ref{plot-particle-hole}) (left). The charge $\pm$ denotes
$e^\pm$, positive/negative helicity is denoted by R/L, and
$S$ gives the spin orientation, lines with arrows show the momentum
direction and $h$ stands for the magnetic field. The following
bilinear combinations are possible when only LLL participate
\begin{itemize}
\item $\langle\bar{\psi}_{R\uparrow}\psi_{R\downarrow}\rangle$, $\langle\bar{\psi}_{L\uparrow}\psi_{L\downarrow}\rangle$
- (spin) scalar, charge neutral, momentum of the pair $\vec{P}=0$, chiral symmetry (CS) is not broken
\item $\langle\bar{\psi}_{R\uparrow}\psi_{L\downarrow}\rangle$, $\langle\bar{\psi}_{L\uparrow}\psi_{R\downarrow}\rangle$
- (spin) scalar, charge neutral, momentum of the pair $\vec{P}\neq 0$, CS is broken
\item $\langle\bar{\psi}_{R\uparrow}\bar{\psi}_{L\uparrow}\rangle$, $\langle\psi_{L\downarrow}\psi_{R\downarrow}\rangle$
- (spin) vector triplet, charged, momentum of the pair $\vec{P}=0$, CS is broken
\end{itemize}
We will not consider the first combination because it does not break the CS,
and in our case CS is broken otherwise there would be no preferred scale for the current $I_-$.
As for the third combination, it has been considered in the context of nonzero density QCD, where it describes the condensate of charged $\rho^\pm$ vector mesons Ref.\cite{Chernodub:2010}. It cannot be our order parameter either, since our current is a spin singlet.
It is tempting to regard the doublet $\Gamma^i\Gamma^5$, $i=1,2$ as a vector, and we leave it for a future work.

We are thus left with the second combination. One can think of this order parameter as a spin-density wave, or magnetization which precesses around the direction of the magnetic field.
The analog of the second combination has been considered within the Sakai-Sugimoto model as a holographic top-down
approach to QCD Ref.\cite{Schmitt:2010}. The setting of Ref.\cite{Schmitt:2010} is very different from ours: it identifies the embedding coordinate of a probe D-brane with a scalar field
dual to a single-trace fermion bilinear operator; the magnetic catalysis is modeled as bending of the probe brane under the influence of the magnetic field.
There the anisotropic spatially modulated CSB condensate in the form of a single plane wave of Larkin-Ovchinnikov-Fulde-Ferrell (LOFF) state has been found. To have a condensate in the form of the second combination,
we need to introduce the $SU(2)$ spin symmetry as in Ref.\cite{Iqbal:2010}.
We should note the difference with our case where the construction of the condensate is done in the bulk and there is a special effort involved to identify the boundary operator.
Provided the condensate of the second form is realized,
the AdS boundary and the IR hard wall play a stabilizing role in its formation Ref.\cite{Stefano-Bolognesi2}. As the pair $\langle\bar{\psi}_{R\uparrow}\psi_{L\downarrow}\rangle$ "bounces" from either of the boundaries it converts into the pair $\langle\bar{\psi}_{L\uparrow}\psi_{R\downarrow}\rangle$ conserving the total charge. This process can be decomposed into elementary "bouncing" events
\begin{itemize}
\item $\bar{\psi}_{R\uparrow}\to \psi_{R\downarrow}$, $\psi_{L\downarrow}\to \bar{\psi}_{L\uparrow}$ - helicity is conserved, spin flips, mixing of charge occurs
\item $\bar{\psi}_{R\uparrow}\to \psi_{L\uparrow}$, $\psi_{L\downarrow}\to \bar{\psi}_{R\downarrow}$ - helicity flips, spin and charge are unaffected
\end{itemize}
In the first case a particle deposits the charge at the boundary, which is picked up by the antiparticle, thus conserving the total charge
of the particle-hole pair. The main difference between the two cases is either the "bouncing" event involves spin flip or not, therefore either helicity is conserved or broken, respectively.
By imposing the AdS boundary condition which breaks CS, helicity gets inverted by the boundary and CS breaking propagates from the boundary into the bulk.
Then CS breaking occurs due to the boundary condition before the chiral condensate forms,
which affects the propagation of the fields in the bulk in accordance with the second case stabilizing the condensate Ref.\cite{Stefano-Bolognesi}.

\begin{figure}[ht!]
\begin{center}
\includegraphics[width=10cm]{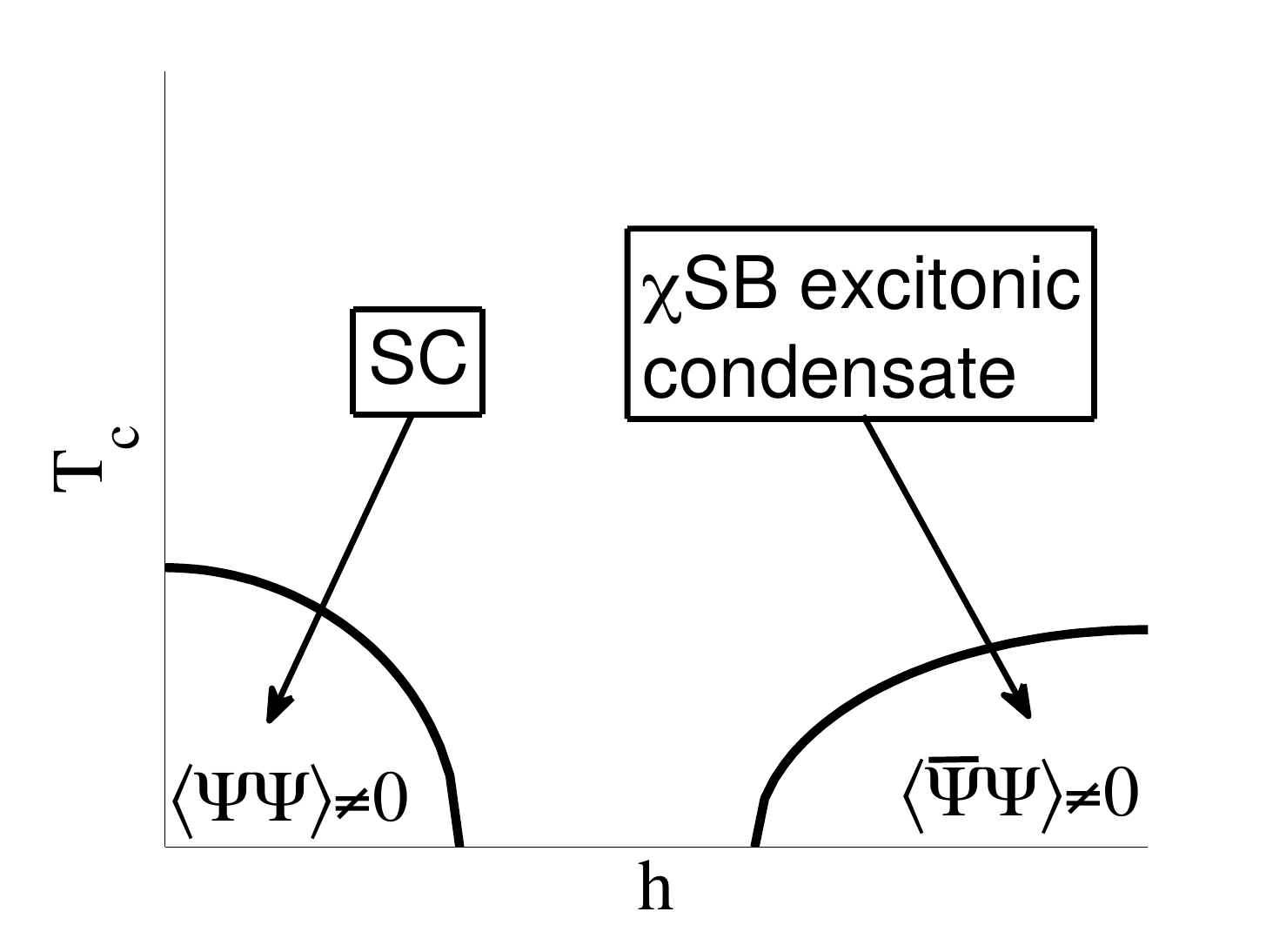}
\includegraphics[width=10cm]{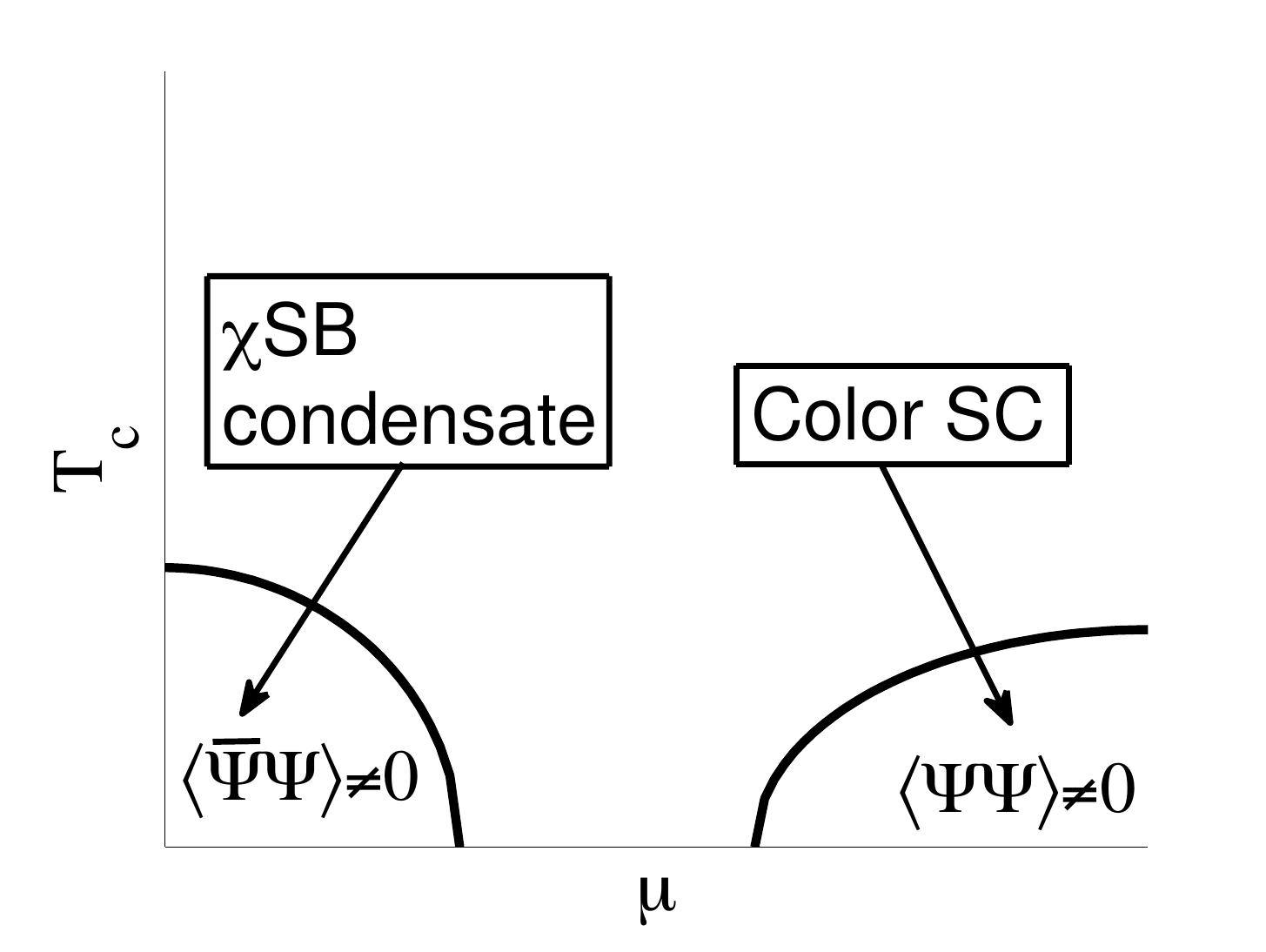}
\caption{Analogy between the phase diagram of a condensed matter
system at nonzero magnetic field and the QCD phase diagram at
nonzero density. In strong magnetic fields, the excitonic condensate
is mapped to the asymptotic regime of high chemical potential QCD
with the color superconductor phase.} \label{plot-diagram}
\end{center}
\end{figure}

Next we discuss an analogy between the MC and the BCS Cooper pairing, and mapping
between the Gross-Neveu model (or the Nambu-Jona-Lasinio (NJL)) in the presence of the magnetic field and the BCS model
at nonzero chemical potential. The reason this mapping works is that effectively the dynamics in both cases is one dimensional:
in the strong magnetic field the motion is constrained to Larmor orbits and
includes only states from the lowest Landau level, while in a high density system only states at the Fermi surface
contribute to the dynamics. We can draw the following analogy Ref.\cite{Schmitt-talk:2012}
\begin{equation}
\begin{array}{|c|c|}
\hline {\rm MC} & {\rm BCS}\\
\hline (3+1){\rm d}\to (1+1){\rm d} & (1+1){\rm d}\\ 
{\rm LLL\;\; and}\;\; \varepsilon=0\;\;{\rm surface} & {\rm Fermi\;\;surface}\;\; \varepsilon=\mu \\
\varepsilon=\sqrt{k_z^2+2|eh|n} & \varepsilon=k-k_F,\; k=\sqrt{\vec{k}^2} \\
{\rm excitonic:}\;\; \Delta\sim G\langle\bar{\psi}\psi\rangle & {\rm SC:}\;\;\Delta\sim G\langle\psi\psi\rangle \\
\Delta\sim \sqrt{eh}\;{\rm exp}(-\frac{const}{G\nu_0}) &
\Delta\sim \mu\;{\rm exp}(-\frac{const}{G\nu_F})\\
\nu_0\;\; {\rm is\;\; DOS\;\;at}\;\; \varepsilon=0 & \nu_F \;\;{\rm is \;\;DOS\;\; at}\;\;\varepsilon=\mu \\
h\;\;{\rm enhances},\;\;\mu\;\;{\rm destroys}\;\;\Delta& \mu\;\;{\rm enhances},\;\;h\;\;{\rm destroys}\;\;\Delta \\
\delta\Omega\sim h(\mu^2-\frac{\Delta^2}{2}) & \delta\Omega\sim \mu^2(\delta\mu^2 -\frac{\Delta^2}{2})\\
h\gg\mu,\Delta & \mu\gg\delta\mu,\Delta \\
{\rm it\;\;can\;\;have}\;\;\mu=0 & {\rm it\;\; can\;\;have}\;\;h=0 \\
T_c \;\; {\rm grows\;\; with}\;\;h\;\;({\rm MC}) & T_c\;\;{\rm decreases\;\;with}\;\; h\\
T_c\;\;{\rm decreases\;\;with}\;\;\mu & T_c\;\;{\rm grows\;\;with}\;\; \mu\;\;({\rm SC})\\
\hline 
\end{array}
\label{mapping_models}
\end{equation}
where the acronym DOS stands for the density of states.   
Effectively one dimensional dynamics in both cases leads to similarities in formulas for the pairing gap $\Delta$
and the gain in thermodynamic potential $\delta\Omega$ as compared to the normal unpaired state. In the BCS, the density of states
at the Fermi surface $\varepsilon=\mu$ defines the gap $\Delta$, and there is a an energy cost $\mu^2\delta\mu^2$ to bring two Fermi surfaces together
to pair in case of nonzero mismatch $\delta\mu$ between them. In MC, the density of states at the $\varepsilon=0$ surface separating electrons and holes 
contributes to the gap,
and a similar cost in energy $h\mu^2$ exists to involve both particles and holes to pair. The gain from the pairing is proportional to $\Delta^2$ in both cases,
and is linear in $\mu^2$ for the BCS and in $h$ for the MC, manifesting the essence of both phenomena. These simple formulas for $\delta\Omega$
can be obtained when there is a hierarchy of scales: the largest scale is $\mu$ in the BCS and it is $h$ in the MC. 

The comparison given in eq.(\ref{mapping_models}) provides the following mapping between parameters in the two systems
at a nonzero density and at a nonzero magnetic field Ref.\cite{Shovkovy:1994} 
\begin{eqnarray}
\begin{array}{ccc}
{\rm MC} & \longleftrightarrow & {\rm BCS}\\
\langle\bar{\psi}\psi\rangle\neq 0 & \longleftrightarrow &     \langle\psi\psi\rangle\neq 0\\
{\rm finite}\;\; h & \longleftrightarrow & {\rm finite}\;\; \mu \\
{\rm small}\;\; \mu & \longleftrightarrow & {\rm small}\;\; \delta\mu \\
h\gg\mu & \longleftrightarrow & \mu\gg\delta\mu
\end{array}
\label{mapping2qcd}
\end{eqnarray}
where the last line expresses the hierarchy of scales.
A similar mapping has been obtained in case of the Gross-Neveu and the BCS models, where the
magnetic field $h$ maps to the chemical potential mismatch
$\delta\mu$ and is relevant for the inhomogeneous superconductors in
the incommensurate phase Ref.\cite{Thies:2006}. Based on Fig.(\ref{plot-tcvsh}) and
using the above described mapping, we can speculate and
draw an analogy between the condensed matter phase diagram in $T_c$
vs. $h$ and the QCD phase diagram in $T_c$ vs. $\mu$, as depicted in
Fig.(\ref{plot-diagram}). The high magnetic field phase is mapped to
the color superconductor state at very large densities (for example color-flavor-locked
(CFL) phase) in QCD, while weak magnetic fields which do not destroy
superconductivity are mapped to the chirality-broken phase in QCD.
The robust feature of the phase diagram in Fig.(\ref{plot-tcvsh}) is the existence of two regions, with small and large-$h$ where the condensate is destroyed and enhanced,
respectively,
by the magnetic field. We found numerically that both branches are thermodynamically favored compared to the normal states, as can be seen in
Fig.(\ref{plot-fren}). In the Sakai-Sugimoto model Ref.\cite{Schmitt:2010}, analytical formulas
for the free energy difference between condensed and normal states have been obtained, proving the stability of both condensed states.
The strong-$h$ regime ("direct" magnetic catalyses) has a remarkably simple
form Ref.\cite{Schmitt:2010}
\begin{equation}
\delta\Omega\sim -h\left(\frac{\Delta(h)^2}{2}-\mu^2\right),
\label{Clogston-stability}
\end{equation}
which is exactly the result obtained in the field theory eq.(\ref{mapping_models}); compare also with Fig.(\ref{plot-fren}). 
The condition for a thermodynamically stable ordered phase with the excitonic condensate is given by
\begin{equation}
\mu\leq \frac{\Delta(h)}{\sqrt{2}},
\end{equation}
which according to the mapping eq.(\ref{mapping2qcd}) coincides with the familiar Clogston limit in the SC:
$\delta\mu\leq \frac{\Delta}{\sqrt{2}}$. However, there is an important difference between formation of the excitonic
and superconducting condensates. In MC the excitonic condensate $\Delta(h)$ is a growing function with $h$
Fig.(\ref{hdep1}A) which insures that eq.(\ref{Clogston-stability}) is always satisfied at high enough magnetic fields.
This finding is important, since it
demonstrates the robustness of the chiral condensate.

Though MC and BCS both have one dimensional dynamics, the mapping between the two models may come as a surprise. Indeed properties of both systems (one is a magnetic and the other is a dense medium) including the symmetry breaking pattern when a condensate forms are quite different. However, we speculate that these two systems
can be unified on the gravity side using the duality between electric and magnetic fields. In the gravity dual description the two phenomena can be represented as follows
\begin{equation}
\begin{array}{|c|c|}
\hline {\rm Holographic\;\;MC} & {\rm Holographic\;\;SC}\\
\hline {\rm dyonic\;\;AdS\;\;RN\;\; BH,} & {\rm AdS\;\;RN\;\;BH}\\
{\rm Schwarzschild\;\;BH} & \\
|H|> |Q| & |Q|>|H| \\
{\rm it\;\;can\;\;be}\;\;Q=0 &{\rm it\;\;can\;\;be}\;\;H=0 \\
Z_2 ({\rm chiral\;\;SB})\;\;{\rm broken}  & U(1)\;\;{\rm broken}\\
{\rm magnetic\;\;field\;\;enhances\;\;it} & {\rm magnetic\;\;field\;\;destroyes\;\;it}\\
{\rm electric\;\;field\;\;destroyes\;\;it} &  {\rm electric\;\;field\;\;enhances\;\;it}\\
\hline{\rm Callan}-{\rm Rubakov\;\;effect} & {\rm dual\;\;Callan}-{\rm Rubakov\;\;effect}\\
\hline 
\end{array}
\label{duality}
\end{equation}
which shows the electromagnetic duality: the invariance under an interchange of the electric and magnetic charges of the black-hole $(|Q|,|H|)\longrightarrow (|H|,|Q|)$.
The motivation for this duality is a similarity in expressions for the gap and the energy gain of the ordered phase between two systems as given by eq.(\ref{mapping_models}).
Probably the underlying reason for the duality is a symmetry of how both charges of the black hole enter in the redshift factor. They always enter in the combination $Q^2+H^2$, which defines also the Hawking temperature of the black hole or the temperature of the system $T\sim r_0(1-\frac{Q^2+H^2}{3r_0^4})$ where $r_0$ is the radius of the horizon of the BH.
Similarly, according to the Montonen-Olive conjecture, the spectrum in the Georgi-Glashow model is invariant under the electromagnetic
$Z_2$ duality $(q,g)\longrightarrow (g,-q)$ as a consequence of the fact that the Bogomol'nyi bound is invariant under electromagnetic duality
(Bogomol'nyi bound for the mass of the 't Hooft-Polyakov monopole is $M\geq a\sqrt{q^2+g^2}$), and the spectrum of the Georgi-Glashow model saturates
this bound Ref.\cite{Figueroa-O'Farrill:1998}. Notably, the mass of the black hole
\begin{equation}
M=r_0^3+\frac{Q^2+H^2}{r_0}
\end{equation}
is also invariant under the electromagnetic duality. The electromagnetic duality eq.(\ref{duality}) holds on a classical level and is destroyed by quantum corrections. It stays intact for the supersymmetric theories though.  

In this work a four-fermi interaction has been used as a control parameter to go from one regime mimicking the SC to the other one of MC. The robustness of both regimes can be
seen in a symmetric form of the dependence $T_c$ vs $h$, top of Fig.(\ref{plot-diagram}).  
In the application to nonzero density QCD, this means that at strong enough magnetic fields the chiral symmetry is spontaneously broken by a chiral condensate.
Moreover, due to a dimensional reduction QCD as well as plain QED are in the confined regime even on the perturbative level:
they can be reduced to a Schwinger model where one-gluon (one-photon) exchange in one dimension leads to a linear rising potential in the configuration space
(a similar argument provides confinement along the boost direction for theories in the light-front quantization). Evidence of the QED confinement in a strong magnetic field
can be provided by the existence of a $2{\rm e}$ bound state which contributes to the fractional quantum Hall effect (QHE) Ref.\cite{Kopelevich:2012}.  
Summarizing, quark gluon plasma (QGP) at strong magnetic fields
is probably confined and has a broken chiral symmetry, as opposed to QGP at zero magnetic fields which is in a deconfined and chiral symmetry invariant phase.
This finding might have some implications for the chiral magnetic effect in heavy-ion collisions at the RHIC Ref.\cite{Dima}.

A general note is that the low-energy behavior of the non-Fermi liquids is governed by a
nontrivial IR fixed point which arises from the near-horizon region
with AdS${}_2$ geometry Ref.\cite{Faulkner:2009}. This IR fixed point arises as a consequence of
the interplay between the emergent quantum critical bosonic modes
and the fermions at finite density. In other words, the class of systems studied
is both metallic and quantum critical at low energies. On the
gravity side, this is reflected by the instability of the background
(Reisner-Nordstrom black hole in the AdS space) unless order parameter fields are introduced to stabilize
it Ref.\cite{Hartnoll:es}.

We have explored the quantum critical aspects of the system by using the magnetic
field as a knob to tune the system to a quantum critical point.
Indeed, the magnetic field as an external parameter driving the system
to quantum criticality is used in experiments on heavy fermions
and graphene. We have shown that by increasing the magnetic field,
the system evolves from the normal metallic to a quantum critical phase,
where the stable quasiparticle is destroyed. The quantum critical
point is controlled by the IR fixed point with the scaling dimension
$\nu=\frac{1}{2}$, where the Fermi velocity vanishes $v_F=0$ (see Fig.(10) in Ref.\cite{Leiden:2010}) but the
Fermi momentum stays finite $k_F\neq 0$ (see Fig.(9) in Ref.\cite{Leiden:2010}). It is important that we are able to deduce
the position of the quantum critical point from our calculations.
The phase transition can be
understood as the formation of a semiclassical condensate on the
gravity side near the AdS${}_4$ boundary. Using the bilinear
formalism developed in Ref.\cite{czs2010}, we have also calculated the
thermodynamic parameters of both phases. We found that the
particle-hole pairing instability arises for both $\nu>\frac{1}{2}$
corresponding to $h<h_c$ and $\nu<\frac{1}{2}$ corresponding to
$h>h_c$. In a holographic superconductor, a superconducting instability
has been shown to exist only for $\nu>\frac{1}{2}$ Ref.\cite{Hartman:2010}. 
This shows the remarkable difference in nature between superconducting and excitonic instabilities:
the existence of an excitonic condensate
beyond the critical point $\nu=\frac{1}{2}$ is a quantum critical phenomenon.
The magnetic field acts as a catalyzer of the particle-hole
pairing because of the dimensional reduction
$d\rightarrow d-2$ in the magnetic field
Ref.\cite{Shovkovy:2d}. 

Other thermodynamic and transport quantities including
the heat capacity and DC conductivity are calculated for both normal- and anomalous- paired phases 
in Appendix B. The results support our findings obtained from the bilinear holographic approach
on the nature and scaling behavior of the two phases.

The critical temperature of the normal-paired phase transition
follows the expected behavior for $h<h_c$: the critical temperature
$T_c$ decreases with increasing $h$, with the scaling $T_c\propto
\mu {\rm exp}\left(-C/\sqrt{q(h_c-h)}\right)$. At $h>h_c$, however,
we find anomalous behavior: $T_c$ grows with increasing $h$. To the
best of our knowledge, this is the first example of non-mean field
scaling from an AdS${}_4$ holographic model. Mathematically, it
follows from the fact that, for $\nu<\frac{1}{2}$, we have the
scaling $T_c\sim \delta^{2\nu-1}$ with $\delta$ small and
decreasing. Physically, such behavior is consistent with the fact
that the system is driven through the quantum critical point at
$h_c$ where $T_c=0$, and beyond the quantum critical point at
$h>h_c$ it can be characterized as a quantum critical metal
possessing new properties. In the existing literature, a novel
antiferromagnetic vior has been predicted for heavy fermions
driven through the quantum critical point Ref.\cite{Schofield:2005}.
Such an anomalous behabehavior for $T_c$ vs. $h$ has been seen in
experiments on highly oriented pyrolytic graphite at strong magnetic
fields $h>h_c$ Ref.\cite{Kopelevich:1999}. Furthermore, the anomalous
branch matches the properties of excitons in bilayer interfaces and
cold atom realizations Ref.\cite{excitonrev}, and can further be related
to the behavior of chiral condensates in holographic QCD models,
signaling the universal significance of the twofold normal-anomalous
regime in the phase diagram.

\section*{Acknowledgments}
We thank Gerald Dunne, Tom Faulkner, Daniel Fernandez-Fraile, Tom Hartman, Sean Hartnoll, Nabil Iqbal, Dima Kharzeev,
Alex Kovner, Hong Liu, John McGreevy, Mark Mezei, Piero Nicolini, Andrei Parnachev, Rob Pisarski, Andreas Schmitt and Igor Shovkovy for
helpful inputs and discussions. We also thank Dirk Rischke,
Francesco Giacosa and Elina Seel for reading the manuscript and valuable
corrections and suggestions. As our paper was close to completion, we learned of a similar work
done by Stefano Bolognesi and David Tong, which is now published in Ref.\cite{Stefano-Bolognesi}: we are very grateful to the authors
for sharing their results and illuminating discussions.
The work was supported in part by the
Alliance program of the Helmholtz Association, contract HA216/EMMI
``Extremes of Density and Temperature: Cosmic Matter in the
Laboratory'' and by ITP of Goethe University, Frankfurt
(E.G.).

\clearpage
\appendix

\section{Bulk Green's function and zero modes}\label{section:bulk}

We express the bulk Green's
function through the boundary one as in Ref.\cite{Hartman:2010}. The
bulk Green's function is a solution of the free Dirac equation,
\begin{equation}
\hat{D}(\Omega,k_l){\mathcal
G}^R(z,z',\Omega,k_l)=\frac{1}{\sqrt{-g}}i\delta(z,z'),
\end{equation}
with the free radial Dirac operator
$\hat{D}(\Omega,k)=\Gamma^iD_i$, which includes the mass term,
chemical potential and the magnetic field but has zero gap,
$\Delta=0$, i.e. the equation \eqref{Dirac}. The bulk Green's function is constructed from the
modes $\psi(z)$ which are solutions of the free Dirac equation
\begin{equation}
\hat{D}(\Omega,k_l)\psi_{radial}(z)=0.
\end{equation}
Due to the choice of the Dirac matrices in eq.(\ref{matrices}),
$\psi$ decouples into two-component spinors,
$\psi_{radial}=(\psi_1,\psi_2)^{T}$.
Therefore the bulk retarded Green's
function has a block-diagonal form
\begin{equation}
\label{gcal} {\mathcal
G}^R(z,z',\Omega,k_l)=\left(\begin{array}{cc}
                      {\mathcal G}_1^R & 0 \\
                        0 & {\mathcal G}_2^R
                     \end{array}
\right),
\end{equation}
\def\alp{\alpha}
\def\ome{\omega}
\def\Ome{\Omega}
where the components ${\mathcal G}_\alpha$, $\alpha=1,2$ are
constructed from the solutions to the Dirac equation
Ref.\cite{Hartman:2010}
\begin{equation}
{\mathcal
G}^R_\alpha(z,z^\prime,\Omega,k_l)=\frac{i}{W(\psi^{in}_\alpha,\psi^{bdy}_\alpha)}\times\left\{
\begin{array}{c}
\psi_\alpha^{in}(z)\bar{\psi}_{\alpha}^{bdy}(z^{\prime})\;\; z<z^\prime\\[0.1in]
\psi_\alpha^{bdy}(z)\bar{\psi}_{\alpha}^{in}(z^{\prime})\;\;
z>z^\prime
\end{array} \right.,
\label{bulk-boundary}
\end{equation}
with $\bar{\psi}_\alpha=i\psi_\alpha^{\dagger}\sigma^1$ 
and
$W_\alpha$ are the components of the Wronskian
\begin{eqnarray}
\label{Wronskian}
W(\psi^{in}_\alpha,\psi^{bdy}_\alpha)=-\frac{\sqrt{-g}}{2\sqrt{g_{zz}}}
\left(\bar{\psi}^{bdy}_\alpha\sigma^3\psi^{in}_\alpha
-\bar{\psi}^{in}_\alpha\sigma^3\psi^{bdy}_\alpha
\right).
\end{eqnarray}
The retarded Green's function eq.(\ref{bulk-boundary}) must
satisfy the following two conditions. At the boundary
($z,z'\rightarrow 1$) where $\psi^{radial}_{\alpha}\sim
a_{\alpha}(1-z)^{3-\Delta_{\psi}}+b_{\alpha}(1-z)^{\Delta_{\psi}}+\ldots$ with $\Delta_{\psi}=\frac{3}{2}+m$
it must be the normalizable
solution, i.e. $\psi^{bdy}_{\alpha}=b_{\alpha}(1-z)^{\Delta_{\psi}}+\ldots$. At the horizon
($z,z'\rightarrow 0$) where $\psi_{radial} \sim {A_{\alpha}
z}^{-i\omega/4\pi T}+B_{\alpha}{ z}^{i\omega/4\pi T}$, the retarded propagator
corresponds to the ingoing solution $\psi^{in}_{\alpha}
={z}^{-i\omega/4\pi T}A_{\alpha}$. This infalling solution behaves near the boundary as
\begin{equation}
\label{bndpsir}
 \psi_\alpha^{in}\sim a_\alpha (1-z)^{3-\Delta_{\Psi}}
+b_\alpha (1-z)^{\Delta_{\Psi}}+\ldots.
\end{equation}
In principle, the coefficients in $\psi^{bdy}$ and $\psi^{in}$ are different, i.e. $b^{bdy}$ and $b^{in}$ (we omit the difference for simplicity).
This determines the $z$-independent Wronskian
$W_\alpha=-i\mathrm{Re}(b_\alpha^{bdy\dagger}\sigma_1\sigma_3 a^{in}_{\alpha})$
after substituting the asymptotic behavior near the AdS-boundary.
The Wronskian is directly proportional to the spectral function of the dual CFT.
The two spinor components of each spinor $a_{\alpha}$ and $b_{\alpha}$ are not independent, but are related by the Dirac equation Refs.\cite{Vegh:2009,Leiden:2009,Pomarol}. Defining up/down spin eigenstates with respect to $\gamma^{z}=-\sigma^3$,

\def\up{\uparrow}
\def\down{\downarrow}
\begin{eqnarray}
  \label{eq:1}
  a_{\alpha} \equiv \left(\begin{matrix}a_{\up} \\ a_{\down} \end{matrix}\right)=
  \left(\begin{matrix}1 \\ 0 \end{matrix}\right)\;,\;\;b_{\alpha} \equiv
 \left(\begin{matrix}b_{\up} \\ b_{\down} \end{matrix}\right)=
  \left(\begin{matrix}0 \\ 1 \end{matrix}\right)
\end{eqnarray}

Substituting this into the Wronskian
\begin{eqnarray}
  \label{eq:4}
  W = -ib_{\alpha\down}^{\dagger}a_{\alpha\up} 
\end{eqnarray}
and recalling the expression for the boundary propagator,
\begin{equation}
\label{gfinish}{G}_\alpha=\frac{b_{\alpha\down}}{a_{\alpha\up}}.
\end{equation}
one finds that
\begin{eqnarray}
  \label{eq:5}
  W(\psi^{in}_{\alpha},\psi^{bdy}_{\alpha}) = -i|b_{\alpha\down}|^2 G_{\alpha}^{-1} = - \frac{i}{G_\alpha}
\end{eqnarray}
The result is similar to the one for fermion transport in Ref.\cite{Faulkner:2010}.

This expression for the bulk propagator in terms of boundary spectral functions shows us that the contribution to the effective action  
is dominated by the poles of the boundary Green's function. These poles precisely correspond
to the values where $\psi^{in} \propto \psi^{bdy}\equiv \psi^0$ is the zero mode with $a_{\alpha}=0$.

\section{Thermodynamics and transport at zero magnetic field}\label{app.section:1}

Quantum critical behavior is associated, among other things, with
unusual scaling exponents of the heat capacity and the resistivity
with temperature. In this section, we obtain an equation of state
and find the scaling behavior of the specific heat and the DC
conductivity with temperature. Following a prescription worked out
in detail for conductivity Ref.\cite{Faulkner:2010}, we bypass the bulk
calculations and do our calculations directly in the boundary field
theory making use of the gravity-``dressed`` fermion propagators
Ref.\cite{Faulkner:2009}. Since the two-point Green's function is
obtained from the AdS/CFT correspondence, it is ``exact`` in terms of
gauge coupling corrections, and the lowest order diagrams on the
field theory side should suffice. However, we lack the knowledge of
the gravity-``dressed`` gauge-fermion vertex. Nevertheless, for the
quantities considered below, the scaling behavior should not change
when vertex corrections are taken into account.

\subsection{Single-particle spectral functions and dispersion relations}

Using AdS/CFT, one finds that, close to the Fermi surface
($\omega/\mu\ll 1$) and at low temperatures ($T/\omega\ll 1$), the
retarded fermion Green's function is given by
Ref.\cite{Faulkner:2009}
\begin{equation}
G_R(\omega,\vec{k})=\frac{h_1v_F}{v_Fk_{\perp}-\omega+v_Fh_2{\rm
e}^{i\theta-i\pi\nu_{k_F}}\omega^{2\nu_{k_F}}}+O\left(\frac{\omega}{\mu}\right).
\label{green-function}
\end{equation}
Here $k_\perp=k-k_F$, the last term in the denominator defines the
self-energy $\Sigma$, $h_1$ and $v_F$ are real constants obtained
from the UV (bulk) physics, $h_2$ is positive with contributions
from both the UV and IR regions, the phase $\theta$ is such that the poles
of eq.(\ref{green-function}) are in the bottom frequency half-plane
corresponding to stable quasiparticle poles and $\nu_{k_F}$ is the
IR conformal dimension at the Fermi momentum. At $T=0$, it is given
by (in dimensionless units)
\begin{equation}
\label{nu0}\nu_{k_F}=\frac{1}{6}\sqrt{6(m^2+k_F^2)-\mu_q^2},
\end{equation}
with $\mu_q=\mu q$. The IR conformal dimension $\nu_{k_F}$ defines
the quasiparticle dispersion. Writing the Green's function pole in
eq.(\ref{green-function}) as $\omega_c(k)=\omega_{*}(k)-i\Gamma(k)$,
at leading order $\omega\sim 0$ we get the following dispersion
relations
\begin{equation}
\omega_*\sim \left\{\begin{array}{cc}
                   v_Fk_{\perp}, & \nu_{k_F}>\frac{1}{2}\\
                   k_{\perp}/\ln k_{\perp}, & \nu_{k_F}=\frac{1}{2}\\
                   k_{\perp}^{1/2\nu_{k_F}}, & \nu_{k_F}<\frac{1}{2}\\
                  \end{array}\right..
\label{dispersion}
\end{equation}
For $\nu_{k_F}=1/2$, the leading order coefficients in front of
$\omega$ and $\omega^{2\nu_{k_F}}$ diverge and cancel exactly,
leaving the subleading logarithmic dependence
$\tilde{c}_1\omega\ln\omega$ where $\tilde{c}_1$ is a real
constant \footnote{The logarithmic dependence for the real part of the
self-energy defines the dispersion for $\nu_{k_F}=\frac{n}{2}$,
$n\in Z_{+}$. Therefore, the linear spectrum is valid for
$\nu_{k_F}\neq \frac{n}{2}$.}. As $\nu_{k_F}$ is decreased we move
from a metal (Fermi liquid) at $\nu>1/2$ to a marginal metal at
$\nu=1/2$ to a quantum critical metal (non-Fermi liquid) at
$\nu_{k_F}<1/2$, the dispersion eq.(\ref{dispersion}) becomes
softer. This has consequences for the behavior of thermodynamic
properties, e.~g. the heat capacity.

The imaginary part of the self-energy $\Sigma\sim
\omega^{2\nu_{k_F}}$ gives rise to the following width of the
quasiparticle dispersion
\begin{equation}
\Gamma\sim\left\{\begin{array}{cc}
                   k_{\perp}^{2\nu_{k_F}}, & \nu_{k_F}>\frac{1}{2}\\
                   k_{\perp}/\ln k_{\perp}, & \nu_{k_F}=\frac{1}{2}\\
                   k_{\perp}^{1/2\nu_{k_F}}, & \nu_{k_F}<\frac{1}{2}\\
                  \end{array}\right..
\label{dispersion2}
\end{equation}
Comparing eqs.(\ref{dispersion}) and (\ref{dispersion2}), we see
that the pole represents a stable quasiparticle only for
$\nu_{k_F}>1/2$ when the width is much smaller than the real part:
$\Gamma/\omega_*\ll 1$, while a coherent quasiparticle is replaced
by an unstable pole for $\nu_{k_F}\leq 1/2$ where
$\Gamma/\omega_*=const$. The imaginary part of the self-energy
becomes important for the behavior of transport coefficients, e.~g.
conductivity, where the dissipation processes play the key role.

We rewrite eq.(\ref{green-function}) as
\begin{eqnarray}
G_R(\omega,\vec{k})=\frac{h_1v_F}{v_Fk_{\perp}-\omega+\Sigma(\omega,k_F)},
\end{eqnarray}
with the self-energy $\Sigma=\Sigma_1+i\Sigma_2$. Therefore
the spectral function defined as
$A(\omega,\vec{k})=\frac{1}{\pi}{\rm Im}G_R(\omega,\vec{k})$ is
given by
\begin{eqnarray}
A(\omega,\vec{k})&=& \frac{1}{\pi}\frac{h_1v_F
\Sigma_2(\omega,k_F)}{(\omega-v_Fk_{\perp}+\Sigma_1(\omega,k_F))^2+
\Sigma_2(\omega,k_F)^2}. \label{spectral}
\end{eqnarray}
From the above form we can directly read off the structure: a
sharp quasiparticle near $k=k_F$ and $\omega=0$ goes through the
infrared scaling region for $\omega/T<1$ and eventually asymptotes
to the universal conformal scaling in the UV, i.~e. for
$\omega,k\gg 1$.

\subsection{Equation of state and specific heat}
\label{sec:equat-state-spec}

Having established the formal structure of the single-particle
propagator, we can use it to construct the Landau-Ginzburg action
for our system. An effective potential in the Cornwall-Jackiw-Tomboulis (CJT) formalism is given
by Ref.\cite{CJT}
\begin{eqnarray}
\Gamma_{eff} = \frac{1}{2}{\rm Tr}\ln S^{-1} +\frac{1}{2}{\rm
Tr}(S_0^{-1}S-1)+\Gamma_2[S],
\end{eqnarray}
where $S$ is a dressed fermion propagator, $\Gamma_2$ is the sum of
all two-particle irreducible (2PI) diagrams, and the trace ${\rm
Tr}$ involves also the summation over the Matsubara frequencies and
the integration $\int d^2x$. The last two terms can be simplified
with the help of the Dyson-Schwinger equation, to give
\begin{eqnarray}
\Gamma_{eff} =  \frac{1}{2}{\rm Tr}\ln S^{-1} -\frac{1}{4}{\rm
Tr}(\Sigma S),
\end{eqnarray}
where the self-energy is $\Sigma=S^{-1}-S_0^{-1}$.

The fact that we have a finite quasiparticle width, encoding for
inelastic/dissipative processes, allows us to calculate the
transport coefficients, which would otherwise be infinite.
However, the imaginary part of the self-energy gives rise to a
branch cut in the fermion propagator along ${\rm Im}\omega=0$ in a
complex $\omega$ plane
Refs.\cite{Denef1:2009,Denef2:2009,Basagoiti:2002,Tomoi}. In the
calculation of the Matsubara sum we should take into account the
contributions from poles and from the discontinuities along the
branch cuts Refs.\cite{Basagoiti:2002,Tomoi}
\begin{eqnarray}
T\sum_{odd\;m}F(i\omega_m) =
\sum_{poles}n(z_i)Res(F,z=z_i)-\sum_{cuts}\int_{-\infty}^{\infty}
\frac{d\zeta}{2\pi i}n(\zeta)Disc\;F, \label{contour-integral}
\end{eqnarray}
with the analytical continuation $i\omega_m\rightarrow z$, and the
Fermi distribution function $n(x)$. In the contour integral one
can use either $n(x)\equiv n(\frac{x}{T})$ or
$\tanh(\frac{x}{2T})$ functions with prefactors $(-\frac{1}{2\pi
i})$ and $(-\frac{1}{4\pi i})$ respectively, as both give the same
result for the observables. The calculation of Matsubara sums
using a perturbative expansion in the imaginary part of the
self-energy has been developed in Ref.\cite{Armen}.

For simplicity we will take $h_1v_F\rightarrow -1$ which will not
change our results qualitatively. Using the retarded fermion
propagator, an effective potential is found to be
\begin{eqnarray}
\Gamma_{eff} &\rightarrow & -\frac{1}{4\pi i}
\frac{V_2}{T}\int\frac{d^2k}{(2\pi)^2}
\int_C dz \tanh\frac{z}{2T} \times\nonumber\\
&&\hspace{1.5cm} T\left(\frac{1}{2}\ln\frac{z-v_F
k_{\perp}+\Sigma(z,k_F)}{T} -\frac{1}{4}\frac{\Sigma(z,k_F)}{z-v_F
k_{\perp}+\Sigma(z,k_F)}\right),
\end{eqnarray}
where we have substituted the Matsubara sum by the contour
integral. The original contour $C_0$ going around the poles along
the imaginary $z$ axis was deformed into the contour $C$ going
along the real $z$ axis and then along the arcs at infinity with
vanishing contribution, denoted by $\Gamma$ Ref.\cite{Basagoiti:2002}.
In the case of a pure real self-energy the result for the contour
integration is (see Ref.\cite{Denef1:2009})
\begin{eqnarray}
\Gamma_{eff} &\rightarrow &
\frac{V_2}{T}\int\frac{d^2k}{(2\pi)^2}\sum_{z_*}
\left(\frac{1}{2}T\ln\left( 2\cosh \frac{z_*}{2T}\right)
+\frac{1}{4} \Sigma(z_*)\tanh\frac{z_*}{2T}\right),
\label{pressure}
\end{eqnarray}
where $z_*$ are the poles of the retarded propagator, and the
sum over all allowed poles is taken. As was shown in
Ref.\cite{Denef1:2009}, when the self-energy and hence the poles
include an imaginary part, the following substitution of hyperbolic
functions with $\Gamma$ functions should be made
Ref.\cite{GradsteinRyzhik}
\begin{eqnarray}
 |\Gamma(\frac{1}{2}+iz)|^2 &=& \frac{\pi}{\cosh (\pi z)},\nonumber\\
 |\Gamma(iz)|^2 &=& \frac{\pi}{z \sinh(\pi z)}.
\label{generalization}
\end{eqnarray}
We can now use $\Gamma_{eff}$ to compute all thermodynamic
quantities, using the relations
\begin{eqnarray}
p=\frac{T}{V_2}\Gamma_{eff},\;\; s= \frac{\partial p}{\partial
T},\;\; c=T\frac{\partial s}{\partial T},\;\; n=\frac{\partial
p}{\partial \mu},
\end{eqnarray}
where the role of $\mu$ is played by $k_F$, and we get the equation
of state
\begin{eqnarray}
p = \int\frac{d^2k}{(2\pi)^2}\sum_{z_*}\left(-\frac{1}{2}
T\ln\left(\frac{1}{2\pi}|\Gamma(\frac{iz_*}{2\pi
T}+\frac{1}{2})|^2\right) +\frac{1}{4}
\frac{\Sigma(z_*)|\Gamma(\frac{iz_*}{2\pi
T}+\frac{1}{2})|^2}{\frac{|z_*|} {2\pi
T}|\Gamma(\frac{iz_*}{2\pi T})|^2}\right), \label{eq-of-state}
\end{eqnarray}
where the summation over complex poles $z_*$ is performed. We only
take into account the contribution of the pole closest to
$\omega=0$, with the imaginary part of the self-energy scaling as
$\Sigma(z)\sim z^{2\nu}$. Near the Fermi surface, the one-loop
contribution dominates over the self-energy term for Fermi liquids
$\nu>\frac{1}{2}$, while the self-energy becomes leading for
non-Fermi liquids $\nu<\frac{1}{2}$.

What we are truly interested in are the temperature scaling
relations for these quantities, in particular for the specific heat
$c$. The first term in eq.(\ref{eq-of-state}) gives the following
contributions to $c$
\begin{eqnarray}
&& \sim\frac{1}{T^2}\int\frac{d^2k}{(2\pi)^2}{\rm
Re}\left(z_*^2\Psi^{\prime}(\frac{iz_*}{2\pi T}+\frac{1}{2})
+z_*^{*\,2}\Psi^{\prime}(-\frac{iz^*}{2\pi T}+\frac{1}{2})
\right),\nonumber\\
&& \frac{1}{T^2}\int\frac{d^2k}{(2\pi)^2}{\rm Re}\left(\sim
z_*T\Psi(\frac{iz_*}{2\pi T}+\frac{1}{2});\; \sim
z_*^*T\Psi(\frac{-iz_*^*}{2\pi T}+\frac{1}{2})\right)
\end{eqnarray}
where
$\Psi^\prime(x)=\frac{d\Psi}{dx}=\frac{d^2\ln\Gamma}{dx^2}$. The
second term gives the following contribution
\begin{eqnarray}
\frac{1}{T^2}\int\frac{d^2k}{(2\pi)^2}{\rm Re} \left( \sim
T\Sigma(z_*)F[\Gamma];\;\sim z_*\Sigma(z_*)F[\Gamma];\; \sim
\frac{z_*^2\Sigma(z_*)}{T} F[\Gamma] \right),
\end{eqnarray}
where $F[\Gamma]$ denotes a combination of $\Gamma$ functions and
their first and second derivatives. Here, the momentum integration
is performed around the Fermi surface, $d^2 k\rightarrow k_F
dk_{\perp}$ with $k_{\perp}=k-k_F$, the poles $z_*=\omega_c-i\Gamma$
are given by eqs.(\ref{dispersion},\ref{dispersion2}) for the three
cases of interest, and $\Sigma(z)\sim z^{2\nu}$.

For a Fermi liquid, one has $\nu>\frac{1}{2}$ and $z_\perp\sim
k_\perp$ (the real part is dominant). The first term then gives
$\frac{1}{T^2}\int dk_\perp z_*^2\rightarrow  T$ and the same
behavior from the other combination, while in the second term we
have $\Sigma\sim k_{\perp}^{2\nu}$. Therefore, the second term gives
$\frac{1}{T^2}\int dk_\perp\Sigma(z_*) z_* \rightarrow T^{2\nu}$ and
the same behavior for the other two combinations \footnote{This is
related to the fact that in eq.(\ref{eq-of-state}) for the effective
action the one-loop term dominates over the self-energy for
$\nu>\frac{1}{2}$.}. Thus for a Fermi liquid at low temperatures we
have
\begin{eqnarray}
c\sim T. \label{capacity}
\end{eqnarray}
We thus reproduce the linear temperature dependence of the heat
capacity known for Fermi liquids.

For a non-Fermi liquid, we have instead $\nu<\frac{1}{2}$ and
$z_\perp\sim k_\perp^{\frac{1}{2\nu}}$ (for both real and imaginary
parts). The first term gives $\frac{1}{T^2}\int dk_{\perp}
k_{\perp}^{\frac{1}{\nu}}\rightarrow T^{\frac{1}{\nu}-1}$ and
$\frac{1}{T^2}\int dk_{\perp} k_{\perp}^{\frac{1}{2\nu}}T\rightarrow
T^{\frac{1}{2\nu}}$. The second term gives $\frac{1}{T^2}\int
dk_\perp\Sigma(z_*) T\rightarrow T^{2\nu}$ and subleading behavior
for the other two combinations. For $\nu<\frac{1}{2}$, the
self-energy dominates over the one-loop contributions in the
pressure and at low temperatures we have
\begin{eqnarray}
c\sim T^{2\nu}. \label{capacity2}
\end{eqnarray}
This result for the heat capacity reflects the scaling behavior of
the self-energy. Finally, for $\nu=\frac{1}{2}$, all the terms are
$\sim T$, so for the marginal liquids we have $c\sim T$. One can
understand it physically from the dispersion relation
eq.(\ref{dispersion}). As the dispersion becomes softer, the number
of states per energy interval increases, and thus the heat capacity
increases as well
\begin{equation}
c_{\rm qcm}>c_{\rm m},
\end{equation}
where $``{\rm m}``$ stands for the normal metal and $``{\rm qcm}``$
for the quantum critical metal.

It is illustrative to repeat the derivation of the equation of state
using the spectral function as given in eq.(\ref{spectral}). The density
of states can be written through a spectral function as follows
\begin{eqnarray}
n = T\sum_{m}\int\frac{d^2k}{(2\pi)^2}A(i\omega_m,\vec{k})
 \rightarrow  -\frac{1}{4\pi i} \int\frac{d^2k}{(2\pi)^2}\int_C dz A(z,\vec{k})f(z),
\end{eqnarray}
where $f(z)=\tanh(\frac{z}{2T})$. One can also use the Fermi
distribution function $f(z)=n(z)$ with a prefactor
$(-\frac{1}{2\pi i})$, which gives the same result for the
observables. The pressure is given by
\begin{eqnarray}
p=\int_{-\infty}^{\mu} d\mu' n,
\end{eqnarray}
where in our case $\mu\equiv k_F$. For simplicity we again take
$h_1v_F\rightarrow 1$. We expand the spectral function with respect
to the imaginary part of the self-energy, which we treat as a small
parameter in this calculation Ref.\cite{Armen}
\begin{eqnarray}
&& A(z,\vec{k}) \approx 2\pi\delta(z-z_*)-\Sigma_2(z,k_F){\cal P^{\prime}}\frac{1}{z-z_*}, \nonumber\\
&& {\cal P^\prime}\frac{1}{z-z_*} \equiv \frac{\partial}{\partial
z}\left({\cal P}\frac{1}{z-z_*}\right).
\end{eqnarray}
The pole of the propagator $z_*$ is a solution of the equation
$z-v_Fk_\perp-\Sigma_1(z,k_F)=0$ which does not contain the imaginary
part of the self-energy $\Sigma_2$. Substituting this representation
into the equation for the pressure, we have
\begin{eqnarray}
p = -\frac{1}{4\pi i} \int \frac{d^2k}{(2\pi)^2}\int_{-\infty}^{k_F}
dk_{F}^{\prime}\int_{-\infty}^{\infty} dz \left(2\pi
\delta(z-z_*)+\Sigma_2(z){\cal
P^{\prime}}\frac{1}{z_*-z}\right)f(z).
\end{eqnarray}
The frequency integral in the first term gives the familiar
expression for the number density
\begin{eqnarray}
n=\int\frac{d^2k}{(2\pi)^2} f(z_*),
\end{eqnarray}
where usually $f$ is a Fermi distribution function, and the
dispersion relation is given by $z_*$ (in standard notation $z_*
\rightarrow \varepsilon_k$). Here we have $f(x)=\tanh(\frac{x}{2})$,
and therefore integrating over $k_F$ gives $\int d k_{F}^\prime
\tanh\frac{z_*}{2}\rightarrow \ln(2\cosh\frac{z_*}{2})$ where, at
the leading order $z_*\sim (k-k_F)$. In the second term we exchange
the order of integrations in $z$ and $k_F$. Therefore,
$\int_{-\infty}^{k_F}dk_F^\prime{\cal
P^\prime}\frac{1}{z_*(k_F^\prime)-z}\rightarrow -
\frac{1}{z_*(k_F)-z}$, and there is no $k_F$ dependence in
$\Sigma_2(z)\sim z^{2\nu}$ at the leading order. The second integral
is $\frac{1}{2\pi i}\int_{-\infty}^\infty dz
\Sigma_2(z,k_F)f(z)\frac{1}{z_*-z}\rightarrow
\Sigma_2(z_*)f(z_*)$. Combining all terms together we have
\begin{eqnarray}
p & = & \int\frac{d^2k}{(2\pi)^2}\sum_{z_*}
\left(\frac{1}{2}T\ln\left( 2\cosh \frac{z_*}{2T}\right)
+\frac{1}{4} \Sigma_2(z_*)\tanh\frac{z_*}{2T}\right),
\end{eqnarray}
which is exactly eq.(\ref{pressure}). Here, $z_*$ is the pole of the
fermion propagator without the imaginary part $\Sigma_2$, and
summing over the poles is understood. If we take $z_*$ to be the
pole of the full propagator, $z_*$ becomes imaginary and a
generalization of hyperbolic functions to the $\Gamma$ functions is
necessary as in eq.(\ref{generalization}). Then we arrive at
eq.(\ref{eq-of-state}) for the pressure of the system.

\subsection{DC conductivity from the Kubo formula}\label{app.section:2c}

We calculate the DC conductivity in the boundary theory using the
gravity-``dressed'' retarded/advanced fermion propagators. Strictly
speaking, we need also the ``dressed'' vertex, to satisfy the Ward
identities. As argued in Ref.\cite{Faulkner:2010} however, the boundary
vertex which is obtained from the bulk one can be approximated by a
constant in the low temperature limit. Also, according to
Ref.\cite{Basagoiti:2002}, the vertex only carries the singularities of
the product of the Green's functions. Therefore, dressing the vertex
will not change the temperature dependence of the DC conductivity at
low $\omega$ Ref.\cite{Basagoiti:2002}.

We can start from the Kubo formula for conductivity
\begin{equation}
 \sigma = -\frac{\partial}{\partial\omega}{\rm
 Im}\Pi_{AA}(\omega,\vec{k}=0)|_{\omega=0}.
\label{conductivity}
\end{equation}
The polarization operator $\Pi_{AA}$ is given by
\begin{eqnarray}
\Pi_{AA}(i\nu_n,0)=\int\frac{d^2k}{(2\pi)^2}T\sum_{\omega_m}G(i\omega_m+i\nu_n,\vec{k})
\Lambda_A(i\omega_m+i\nu_n,i\omega_m,\vec{k})G(i\omega_m,\vec{k})\Lambda_A^{(0)}(\vec{k}),\nonumber\\
\end{eqnarray}
where the fermion frequency is $\omega_m=(2m+1)\pi T$, and the boson
frequency is $\nu_n=2n\pi T$, and in the low temperature limit
$\Lambda_A(i\omega_m+i\nu_n,i\omega_m,\vec{k})=\Lambda_A^{(0)}(\vec{k})$.
Usually the most difficult step is to take the Matsubara sum. Here
we can do it in two ways. The first way consists of analytically
continuing in the complex plane $i\omega_m\rightarrow z$ and
replacing the Matsubara sum by a contour integral with the Fermi
distribution function $n(x)=\frac{1}{{\rm e}^x+1}$ whose poles sit
at the Matsubara frequencies along the imaginary axis. The second
way is to use the spectral representation. In both cases we follow
Ref.\cite{Basagoiti:2002}, where transport coefficients were calculated
with propagators including their imaginary parts.

Taking the first way, we have for the fermion Matsubara sum
\begin{eqnarray}
H(i\nu_n,\vec{k})=T\sum_{\omega_m}G(i\omega_m+i\nu_n,\vec{k})G(i\omega_m,\vec{k})\rightarrow
-\frac{1}{2\pi i} \int_C dz G(z+i\nu_n,\vec{k})G(z,\vec{k})n(z),\nonumber\\
\label{contour-integral}
\end{eqnarray}
where the contour along the imaginary $z$ axis can be deformed to
the contour C which goes along two branch cuts, ${\rm Im}Z=0$ and
${\rm Im}z=-\nu_n$, and the large arcs $\Gamma$ with vanishing
contribution Ref.\cite{Basagoiti:2002}. The fermion propagator has a
branch cut along ${\rm Im}z=0$ Refs.\cite{Tomoi,Basagoiti:2002}.
Therefore we can rewrite
\begin{eqnarray}
H(i\nu_n) &=& -\frac{1}{2\pi i}\int_{-\infty}^{\infty}d\zeta
n(\zeta)G(i\nu_n +\zeta)(G_R(\zeta)-G_A(\zeta))
\nonumber\\
&-&\frac{1}{2\pi i}\int_{-\infty}^{\infty}d\zeta n(\zeta)G(-i\nu_n
+\zeta)(G_R(\zeta)-G_A(\zeta)),
\end{eqnarray}
where the difference of the retarded and advanced functions in the
first bracket is due to the discontinuity along ${\rm Im}z=0$ and in
the second bracket it is due to the discontinuity along ${\rm
Im}z=-\nu_n$. This contribution corresponds to the second term in
eq.(\ref{contour-integral}), and there are no pole contributions
Ref.\cite{Basagoiti:2002}. We use the usual prescription for retarded
and advanced Green's functions, $G_R=G(\omega+i0^+)$ and
$G_A=G(\omega-i0^+)$ and suppress the momentum indices. Taking
$i\nu_n\rightarrow \omega+i0^+$, we have
\begin{eqnarray}
H(\omega) &=& -\frac{1}{2\pi i}\int_{-\infty}^{\infty}d\zeta
n(\zeta)G_R(\omega+\zeta)(G_R(\zeta)-G_A(\zeta))
\nonumber\\
&-& \frac{1}{2\pi i}\int_{-\infty}^{\infty}d\zeta
n(\zeta+\omega)G_A(\omega+\zeta)(G_R(\zeta+\omega)-G_A(\zeta+\omega)),
\end{eqnarray}
where we changed the integration variable in the second integral
$\zeta-\omega\rightarrow \zeta$. In the limit $\omega\rightarrow
0$, the dominant contribution comes from the pair $G_RG_A$, and it
is inversely proportional to the distance between the poles given
by the imaginary part $\Sigma_2$. The combinations $G_RG_R$ and
$G_AG_A$ with the poles on one side of the real axis make a much
smaller contribution due to the cancellation between the residues at
the poles. Therefore, as $\omega\sim 0$, we have
\begin{eqnarray}
H(\omega,\vec{k})\rightarrow -\frac{1}{2\pi i}
\int_{-\infty}^{\infty} d\zeta
(n(\zeta+\omega)-n(\zeta))G_R(\zeta+\omega)G_A(\zeta),
\end{eqnarray}
and
\begin{eqnarray}
{\rm Im}\Pi_{AA}(\omega,0) &=&
\frac{1}{2\pi}\int\frac{d^2k}{(2\pi)^2}\Lambda_A^{(0)}(\vec{k})
\int_{-\infty}^{\infty}\frac{d\zeta}{2\pi}(n(\zeta+\omega)-n(\zeta))G_R(\zeta+\omega,\vec{k})\times\nonumber\\
&&
\Lambda_A(\zeta+\omega+i0^{+},\zeta-i0^{-},\vec{k})G_A(\zeta,\vec{k}).
\end{eqnarray}
In the small-$T$ limit the vertex is a constant. We integrate
around the Fermi surface, and therefore the momentum integral is
$\int\frac{d^2k}{(2\pi)^2}\rightarrow
\frac{k_Fdk_{\perp}}{(2\pi)^2}$ with $k_{\perp}=k-k_F$. We
exchange the order of integration and perform first the momentum
integration Refs.\cite{Hartman:2010,Faulkner:2010}. For
$\omega\sim 0$, we have
\begin{eqnarray}
&& \hspace{-2cm}\int_{-\infty}^{\infty} \frac{dk_{\perp}}{2\pi}
\frac{1}{(\frac{\zeta}{v_F}-k_{\perp}+\Sigma(\zeta,k_F)+i0^{+})
(\frac{\zeta}{v_F}-k_{\perp}+\Sigma^{*}(\zeta,k_F)-i0^{+})}=\nonumber\\
&&
\frac{i}{\Sigma(\zeta,k_F)-\Sigma^{*}(\zeta,k_F)}=\frac{1}{2{\rm
Im}\Sigma(\zeta,k_F)}.
\end{eqnarray}
Writing $n^{\prime}(\zeta)=-\beta n(\zeta)(1-n(\zeta))$, we have
for $\omega\sim 0$
\begin{eqnarray}
\sigma\rightarrow
\Lambda^{(0)\;2}k_Fh_1^2\int_{-\infty}^{\infty}\frac{\beta
d\zeta}{2\pi} \frac{n(\zeta)(1-n(\zeta))}{{\rm
Im}\Sigma(\zeta,k_F)}, \label{resultcond}
\end{eqnarray}
where we have dropped constant terms. Note that we get the same
result for the conductivity also if we use $\tanh\frac{x}{2}$ in the
contour integral eq.(\ref{contour-integral}) since
$n^\prime(x)=-2\tanh^{\prime}(\frac{x}{2})$. For the Landau Fermi
liquid $\Sigma(\omega) \sim \omega^2$ at small $T$
Refs.\cite{Landau,Faulkner:2010}. We get
\begin{eqnarray}
\sigma \sim T^{-2},
\end{eqnarray}
meaning that we recover the standard result for the resistivity of
the Fermi liquid: $\rho\sim T^2$.

In our case, $\Sigma (\omega)\sim \omega^{2\nu_{k_F}}$, which
produces
\begin{eqnarray}
\sigma \sim T^{-2\nu_{k_F}}, \label{sigma}.
\end{eqnarray}
This result agrees with the DC conductivity obtained in
Ref.\cite{Faulkner:2010}. For the marginal liquid,
$\nu_{k_F}=\frac{1}{2}$, we recover the resistivity $\rho\sim T$,
which is empirically found in the strange metal phase.

\begin{itemize}
\item  It is interesting that the scaling behavior of the DC
conductivity is the same as the single particle scattering rate.
On the gravity side it is explained by the fact that the
dissipative part of the current-current correlator is controlled
by the rate of the bulk fermion falling in the horizon, given by
the single-particle scattering rate. Comparing the resistivity in
the quantum critical metal $``{\rm qcm}``$ to the one in the
normal metal $``{\rm m}``$,
\begin{equation}
\rho_{\rm qcm}>\rho_{\rm m},
\end{equation}
which indicates that the quantum critical metal becomes
increasingly insulating as $\nu_{k_F}$ is decreased. This suggests
that there is some sort of ordering in the system, not necessarily
associated with a gap.
\end{itemize}

To check our calculation, we get the DC conductivity using the
spectral representation
\begin{eqnarray}
G(i\omega_m,\vec{k})=\int\frac{dk_0}{2\pi}\frac{A(k_0,\vec{k})}{k_0-i\omega_m},
\end{eqnarray}
where the spectral function $A(k_0,\vec{k})$ is given by
eq.(\ref{spectral}). For the product of the Green functions we use
the following formula
\begin{eqnarray}
T\sum_{m}\frac{1}{i\omega_m-\omega_1}\frac{1}{i\omega_m+i\nu_n-\omega_2}=
\frac{n(\omega_1)-n(\omega_2)}{i\nu_n+\omega_1-\omega_2}.
\end{eqnarray}
Taking $i\nu_n\rightarrow \omega+i0^{+}$, the polarization
operator is given by
\begin{eqnarray}
\Pi_{AA}(\omega,0) =
\int\frac{d^2k}{(2\pi)^2}\frac{d\omega_1}{2\pi}\frac{d\omega_2}{2\pi}
\frac{n(\omega_1)-n(\omega_2)}{\omega+\omega_1-\omega_2}\Lambda_A^{(0)2}
A(\omega_1,k_{\perp})A(\omega_2,k_{\perp}).
\end{eqnarray}
Performing the integration over $\omega_2$, we have
\begin{eqnarray}
{\rm Im} \Pi_{AA}(\omega,0) =
\int\frac{d^2k}{(2\pi)^2}\frac{d\omega_1}{2\pi}
(n(\omega_1)-n(\omega_2))\Lambda_A^{(0)2}
A(\omega_1,k_{\perp})A(\omega_1+\omega,k_{\perp}).
\end{eqnarray}
In the limit $\omega\sim 0$, the momentum integration proceeds as
\begin{eqnarray}
\int\frac{d^2k}{(2\pi)^2}A^2(\omega_1,k_{\perp})\rightarrow
k_F\int\frac{dk_{\perp}}{2\pi}A^2(\omega_1,k_{\perp})\rightarrow
\frac{k_Fh_1^2}{\Sigma_2(\omega_1,k_F)},
\end{eqnarray}
with $\Sigma_2={\rm Im}\Sigma$. Therefore, the DC conductivity
given by eq.(\ref{conductivity}) is
\begin{eqnarray}
\sigma \rightarrow \Lambda_A^{(0)2} k_F h_1^2 \int \frac{\beta
d\omega_1}{2\pi} \frac{n(\omega_1)(1-n(\omega_1))}{{\rm
Im}\Sigma(\omega_1,k_F)}
\end{eqnarray}
which is the same as eq.(\ref{resultcond}) obtained by the contour
integration.


\begin{thebibliography}{99}
\bibitem{Tomoi}
Unpublished notes of Tomoi Koide on branch cut in the integrals
$\ln z=\int_1^z \frac{d\zeta^{\prime}}{\zeta^{\prime}}$ where $z$ is imaginary.
\end{thebibliography}
\end{document}